\documentclass[journal,twocolumn]{IEEEtran}

\usepackage{times}
\usepackage[T1]{fontenc}
\usepackage{microtype}
\usepackage[utf8]{inputenc}
\usepackage{amsmath,amsthm,amsfonts,amsbsy}
\usepackage{newtxmath,mathrsfs,bbold,bbm,dsfont,pifont}
\usepackage{enumitem}
\usepackage{graphicx,float}
\usepackage[caption=false]{subfig}
\usepackage{array,longtable,booktabs}
\usepackage{tikz,xcolor}
\usepackage{verbatim,algpseudocode,listings}
\usepackage{csquotes}
\usepackage{hyperref}
\usepackage{lipsum}
\usepackage{cite}

\newtheorem{corollary}{Corollary}
\newtheorem{definition}{Definition}

\newtheorem{lemma}{Lemma}

\newtheorem{proposition}{Proposition}
\newtheorem{task}{Task}
\newtheorem{theorem}{Theorem}

\newcommand{\aff}{\mathrm{aff}}
\newcommand{\cone}{\mathrm{cone}}

\newcommand{\id}{\mathrm{id}}
\newcommand{\Id}{\mathrm{Id}}
\newcommand{\intr}{\mathrm{int}}

\newcommand{\spn}{\mathrm{span}}

\newcommand{\To}{\twoheadrightarrow}
\newcommand{\tr}{\mathrm{tr}}
\newcommand{\Tr}{\mathrm{Tr}}
\newcommand{\bra}[1]{\langle#1|}
\newcommand{\ket}[1]{|#1\rangle}

\newcommand{\op}[2]{\ket{#1}\bra{#2}}

\newcommand{\abb}[1]{{\textnormal{#1}}} 
\newcommand{\cc}[1]{{\overline{#1}}} 
\newcommand{\ens}[1]{{#1}} 
\newcommand{\f}[1]{{\mathcal{#1}}} 
\newcommand{\idx}[1]{{\mathtt{#1}}} 
\newcommand{\ob}[1]{{\mathscr{#1}}} 
\newcommand{\povm}[1]{{\mathbb{#1}}} 
\newcommand{\pmt}[1]{{\mathfrak{#1}}} 
\newcommand{\q}[1]{{\mathsf{#1}}} 
\newcommand{\rpl}[1]{{\tilde{#1}}} 
\newcommand{\s}[1]{{\mathcal{#1}}} 
\newcommand{\sm}[1]{{#1}} 
\newcommand{\spovm}[1]{{\mathscr{#1}}} 
\newcommand{\spa}[1]{{\mathds{#1}}} 
\newcommand{\spec}[1]{{\hat{#1}}} 
\newcommand{\sys}[1]{{#1}} 
\newcommand{\dsys}[1]{{\sys{#1}}} 

\newcommand{\AO}{\sys{A_0}}
\newcommand{\AI}{\sys{A_1}}
\newcommand{\A}{\dsys{A}}
\newcommand{\RAO}{{\rpl{\sys{A}}_\sys{0}}}
\newcommand{\RAI}{{\rpl{\sys{A}}_\sys{1}}}
\newcommand{\RA}{\rpl{\A}}
\newcommand{\APO}{{\sys{A}'_\sys{0}}}
\newcommand{\API}{{\sys{A}'_\sys{1}}}
\newcommand{\AP}{{\A'}}
\newcommand{\BO}{\sys{B_0}}
\newcommand{\BI}{\sys{B_1}}
\newcommand{\B}{\dsys{B}}
\newcommand{\RBO}{{\rpl{\sys{B}}_\sys{0}}}
\newcommand{\RBI}{{\rpl{\sys{B}}_\sys{1}}}
\newcommand{\RB}{\rpl{\B}}
\newcommand{\BPO}{{\sys{B}'_\sys{0}}}
\newcommand{\BPI}{{\sys{B}'_\sys{1}}}
\newcommand{\BP}{{\B'}}
\newcommand{\EO}{\sys{E_0}}
\newcommand{\XO}{\sys{X_0}}
\newcommand{\XI}{\sys{X_1}}
\newcommand{\X}{\dsys{X}}
\newcommand{\RXI}{{\rpl{\sys{X}}_\sys{1}}}

\definecolor{darkblue}{rgb}{0,0,0.5}
\definecolor{darkgreen}{rgb}{0,0.5,0}
\definecolor{darkred}{rgb}{0.5,0,0}
\hypersetup{
colorlinks = true,
citecolor = darkgreen, 
linkcolor = darkblue, 
urlcolor  = blue, 
filecolor = darkred 
}
\definecolor{cool_green}{rgb}{0.0,0.5,0.0}
\definecolor{cool_purple}{rgb}{0.5,0.0,0.5}

\usetikzlibrary{quantikz}
\allowdisplaybreaks

\begin{document}




\title{
Entropic and Operational Characterizations of Dynamic Quantum Resources
\author{Kaiyuan Ji and Eric Chitambar}
\thanks{Kaiyuan Ji is with the School of Electrical and Computer Engineering, Cornell University, Ithaca, New York 14850, USA (email: kj264@cornell.edu).}
\thanks{Eric Chitambar is with the Coordinated Science Laboratory and the Department of Electrical and Computer Engineering, University of Illinois at Urbana-Champaign, Urbana, Illinois 61801, USA (email: echitamb@illinois.edu).}
\thanks{An earlier version of this paper was presented in part at the 2023 IEEE International Symposium on Information Theory [DOI:10.1109/ISIT54713.2023.10206815].}
}

\maketitle


\begin{abstract}
We offer new methods for characterizing general closed and convex quantum resource theories, including dynamic ones, based on entropic concepts and operational tasks.  We propose a resource-theoretic generalization of the quantum conditional min-entropy, termed the free conditional min-entropy (FCME), in the sense that it quantifies an observer's ``subjective'' degree of uncertainty about a quantum system given that the observer's information processing is limited to free operations of the resource theory.  Using this generalized concept, we provide a complete set of entropic conditions for free convertibility between quantum states or channels in any closed and convex quantum resource theory.  We also derive an information-theoretic interpretation for the resource global robustness of a state or a channel in terms of a mutual-information-like quantity based on the FCME\@.  Apart from this entropic approach, we characterize dynamic resources by also analyzing their performance in operational tasks.  We construct operationally meaningful and complete sets of resource monotones with these tasks, which enable faithful tests of free convertibility between quantum channels.  Finally, we show that every well-defined robustness-based measure of a channel can be interpreted as an operational advantage of the channel over free channels in a communication task.
\end{abstract}


\tableofcontents


\section{Introduction}
\label{sec:introduction}

Quantum resource theories (QRTs) provide a systematic and versatile framework for studying different features of quantum mechanics~\cite{Horodecki-2013a, Coeke-2016a, Chitambar-2019a, Gour-2024c}.  Such a theory adopts the perspective of an experimenter who wishes to carry out an information-processing task in an operationally restricted environment.  The experimenter is only capable of performing a particular subset of physical operations, known as the ``free'' operations, and all operations beyond the free ones are prohibited.  This induces a partition of physical objects into two categories, one consisting of those that can be prepared using free operations, and the other consisting of those that cannot.  Since the experimenter has no means to generate objects in the second category, these objects are deemed as ``resources'' (as they are valuable and rare), in contrast to the ``free'' objects in the first category (which can be obtained at no cost).  

The rationale behind QRTs is the observation that many quantum phenomena can be captured by the same, aforementioned, formalism.  An archetypal example is the theory of entanglement~\cite{Plenio-2007a, Horodecki-2009a}, in a version of which the free operations consist of local operations and classical communication (LOCC) and whatever quantum state cannot be generated using LOCC possesses the resource of entanglement.  A variety of other quantum phenomena have been formalized and studied within the framework of QRTs~\cite{Gour-2008a, Brandao-2013a, Grudka-2014a, Veitch-2014a, Gallego-2015a, Streltsov-2017a, Rosset-2018a, Wang-2019a, Buscemi-2020a, Guff-2021a, Kaur-2021a, Wu-2021a}.  Despite their different physical natures, these phenomena have a remarkably similar structure when viewed through the lens of resource theories, and their analyses admit a similar methodology.  This has motivated a series of efforts towards understanding the \emph{general} structure of QRTs, independent of any specification of the resource~\cite{Horodecki-2013a, Brandao-2015a, Coeke-2016a, Gour-2017a, Regula-2017a, Liu-2017a, Lami-2018a, Chitambar-2019a, Liu-2019a, Takagi-2019b, Fang-2020a, Kuroiwa-2020a, Uola-2019a, Gour-2020a, Lami-2021a, Gour-2024c, George-2024a}.  Following an axiomatic approach, common structural features of different QRTs can be identified by assuming a minimal set of algebraic properties, and further valuable insights can be inferred on that basis.

While the framework of QRTs has historically been attached to ``static'' resources, i.e., resources that exist in quantum states, recent advancements have extended it to ``dynamic'' resources, which reflect properties of quantum channels~\cite{Theurer-2019a, Liu-2019b, Gour-2019b, Gour-2020a, Liu-2020a, Lu-2020a, Regula-2021a}.  A static QRT studies the manipulation of quantum states using a restricted set of channels, whereas a dynamic QRT studies the manipulation of quantum channels using a resricted set of superchannels~\cite{Chiribella-2008a, Gour-2019a}.  In this paper, we adopt a unified treatment that is inclusive enough to encompass both static and dynamic QRTs, and it has the advantage of significantly simplifying our discussion while preserving the generality.

The interplay between information theory and resource theories has long been recognized.  The connections between them are mostly centered around the following two aspects.  The first aspect is ``resource theories for information theory.''  Many core problems in information theory can be formulated in resource-theoretic terms, and additional insights have been gained from doing so~\cite{Devetak-2004a, Devetak-2008a}.  For instance, communication over (or simulation of) a noisy channel can be thought of as the task of converting the noisy channel into a noiseless one (or vice versa) when the sender and receiver are restricted to performing certain forms of encoding and decoding~\cite{Bennett-2002a, Takagi-2020a}.  Likewise, the understanding of hypothesis testing and related entropic measures has also benefited from being examined from a resource-theoretic perspective~\cite{Wang-2019a }.  The second aspect is ``information theory for resource theories.''  This is embodied by the extensive use of information-theoretic tools in the quantification of quantum resources~\cite{Liu-2017a, Regula-2017a} and derivation of conversion rates between resources~\cite{Liu-2019a}.  For instance, among the most frequently used measures of resource is the resource relative entropy, which quantifies the smallest Umegaki relative entropy between a given object and the set of free objects~\cite{Vidal-1997a, Horodecki-2013a, Winter-2016a}.  One-shot entropic measures such as the max- and min-relative entropies have also been utilized for similar purposes~\cite{Datta-2009a, Regula-2017a, Gour-2017a, Gour-2019a, Liu-2019a, George-2024a}.

In this work, we consider a new dimension of the interplay between information theory and resource theories, which is ``information theory \emph{in} resource theories.''  Intuitively, since a QRT restricts the experimenter's ability to process information, the experimenter should feel an aggravated sense of uncertainty about a given system compared to when such restrictions do not exist.  This suggests that uncertainty should be measured in a way that incorporates the operational restrictions imposed on the experimenter, and thus a generalized notion of ``entropy'' is needed in a QRT\@.  In this paper, we focus on applying this intuition to the conditional min-entropy.  The standard conditional min-entropy features its operational interpretation as a guessing probability~\cite{Konig-2009a} even in the dynamic setting~\cite{Gour-2019a}.  Here we propose a resource-theoretic generalization thereof called the ``free'' conditional min-entropy (FCME), which enjoys a similar operational relevance to that of the standard quantity except that now the experimenter can only use free operations of the underlying QRT instead of all quantum channels while making a guess.  The information-theoretic implications of the FCME go beyond uncertainty quantification amidst operational restrictions but also extend to characterizing the structure of the resource theory itself.  In particular, the FCME provides the first information-theoretic interpretations for multiple previously established and widely used resource measures, including the complete sets of resource monotones based on support functions in Refs.~\cite{Takagi-2019b, Gour-2020a, Jencova-2021a} and the resource global robustness~\cite{Harrow-2003a, Brandao-2015a} of states and channels.  We mention that a special case of the FCME in the QRT of entanglement has been studied in Ref.~\cite{Chitambar-2023a}, and its operational interpretation has been demonstrated in both state discrimination and zero-error channel simulation.

\subsection{Main contribution}
\label{sec:contribution}

The main contribution of this paper is threefold.
\begin{itemize}
	\item First, we introduce and formalize the concept of the free conditional min-entropy for both static and dynamic systems (Definitions~\ref{def:FME}--\ref{def:EFCME}).  Through an investigation of its various entropic properties (Propositions~\ref{prop:reducibility}--\ref{prop:operational}), we justify the information-theoretic meaning of the FCME as quantifying the ``subjective'' degree of uncertainty about a system from the perspective of an operationally restricted observer.
	\item Second, we find applications of the FCME in the characterization of QRTs.  Specifically, we construct a complete set of resource monotones with a clear information-theoretic meaning in terms of the FCME in a general (closed and convex) QRT (Theorem~\ref{thm:convertibility-deterministic} and Corollary~\ref{cor:convertibility-deterministic}).  We also show that the resource global robustness, of a state or a channel, can be reformulated as a mutual-information-like quantity based on the FCME (Theorem~\ref{thm:robustness-entropy} and Corollary~\ref{cor:robustness-entropy}).  This offers the first known interpretation of any resource robustness measure as a measure of information gain in a general (closed and convex) QRT\@.  Combined with the operational characterization of resource robustness measures in Refs.~\cite{Takagi-2019a, Takagi-2019b}, our result confirms the conjectured ``triangle of association'' between robustness-based measures, operational tasks, and information-theoretic quantities in general QRTs~\cite{Skrzypczyk-2019b}.
	\item Third, we also take an operational approach to characterizing quantum resources.  We provide a complete set of conditions for free convertibility between two channels in terms of the success probabilities of operational tasks (Theorems~\ref{thm:convertibility-EOP} and \ref{thm:convertibility-OCC}) and find operational interpretations of various resource robustness measures of a channel, each of which amounts to an advantage of the channel over free channels in such a task (Theorems~\ref{thm:robustness-EOP} and \ref{thm:robustness-OCC}).  In particular, our characterization goes beyond the resource global robustness or free robustness, and it applies to those against general forms of noise, thus addressing an open question in Ref.~\cite{Takagi-2019b}.
\end{itemize}

\subsection{Organization of the paper}
\label{sec:organization}

The rest of this paper is organized as follows.  In Sec.~\ref{sec:preliminaries}, we review the formalism of quantum theory with an emphasis on dynamic systems.  In Sec.~\ref{sec:QRT}, we lay out a ``top-down'' framework for general QRTs that encompasses both static and dynamic resources, with the basic components of the framework presented in Sec.~\ref{sec:axiom}, an explication of the allowed probabilistic transformations and (super)measurements in Secs.~\ref{sec:probabilistic} and \ref{sec:measurement}, and a review of resource quantification in Sec.~\ref{sec:monotones}.  In Sec.~\ref{sec:QRT}, we formally introduce the concept of the FCME, whose definition and dual formulation are given in Sec.~\ref{sec:entropy-static} in static QRTs and extended to dynamic QRTs in Sec.~\ref{sec:entropy-dynamic}.  We prove multiple properties of the FCME in Secs.~\ref{sec:entropy-properties} and \ref{sec:entropy-operational}, accompanied by detailed discussions on their information-theoretic implications.  In Sec.~\ref{sec:characterization-entropy}, we apply the FCME to characterizing general closed and convex QRTs.  This includes constructions of a complete set of entropic conditions for free convertibility (Secs.~\ref{sec:convertibility-deterministic} and \ref{sec:convertibility-probabilistic}) and an information-theoretic interpretation of the resource global robustness (Sec.~\ref{sec:robustness-entropy}).  In Sec.~\ref{sec:characterization-tasks}, we present two operational tasks, each of which can be used to faithfully test free convertibility between two given channels (Sec.~\ref{sec:convertibility-tasks}).  We also show in Sec.~\ref{sec:robustness-tasks} that our tasks can be used to estimate a broad family or robustness-based measures of dynamic resources.  In Sec.~\ref{sec:conclusion}, we summarize our results and discuss several open questions and future directions.  In Appendix~\ref{app:resource}, we provide even more general discussions on the the abstract structure of closed and convex resource theories without assuming the resources being quantum.  The arguments provided there set the foundations for the proofs of our results on characterizing quantum resources in the main text, which are contained in Appendices~\ref{app:characterization-entropy} and \ref{app:characterization-tasks}.

\begin{figure}[t]
\centering
\includegraphics[scale=0.22]{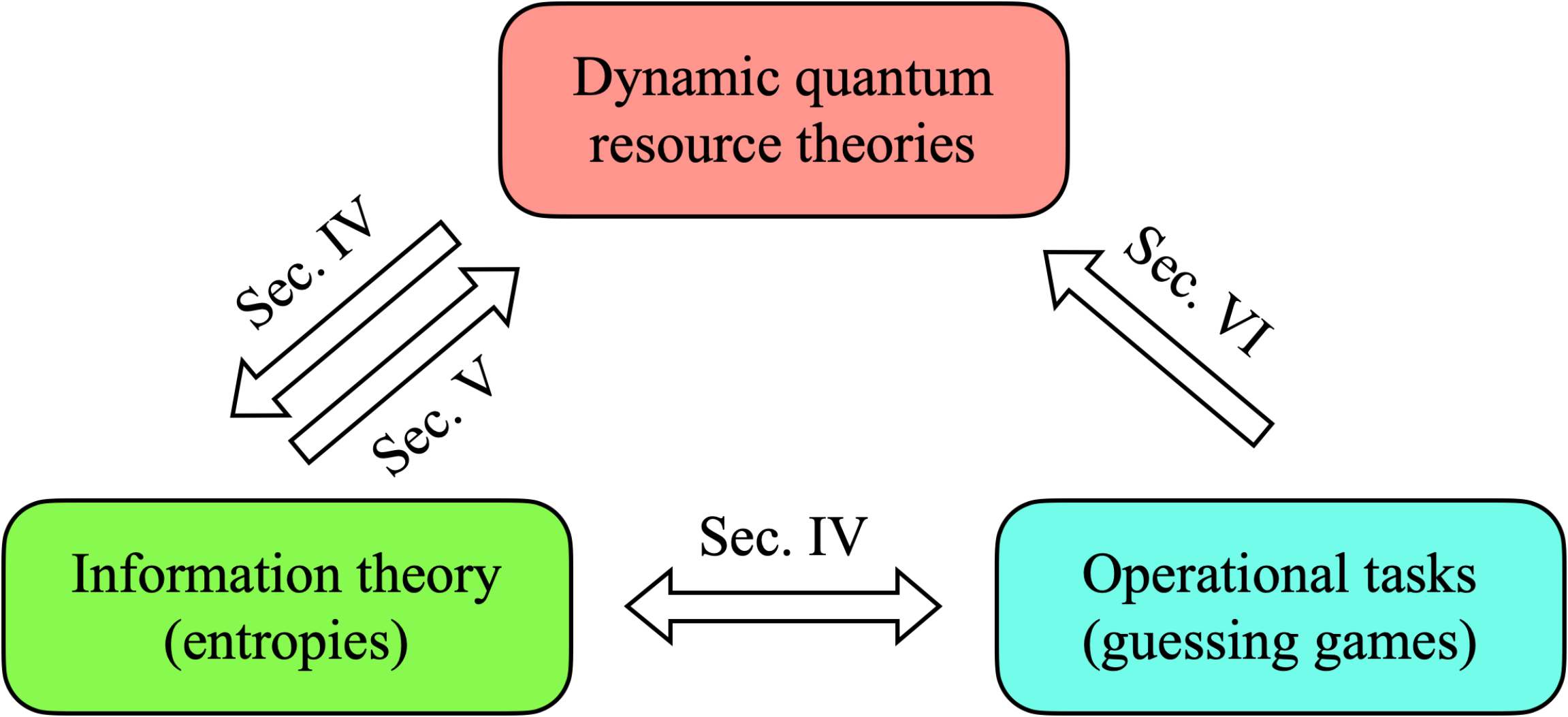}
\caption{Three-way connections established by Secs.~\ref{sec:entropy}--\ref{sec:characterization-tasks} of the paper.  Section~\ref{sec:entropy} investigates how resource-theoretic settings transform the way in which entropies are defined (the arrow pointing at the bottom left), and it also provides an operational interpretation for the FCME (the arrows at the bottom).  Section~\ref{sec:characterization-entropy} utilizes entropic concepts to characterize QRTs (the arrow pointing at the top right).  Section~\ref{sec:characterization-tasks} utilizes operational tasks to characterize QRTs (the arrow pointing at the top left).}
\label{fig:introduction}
\end{figure}

The roles of Secs.~\ref{sec:entropy}--\ref{sec:characterization-tasks} in the three-way connections between QRTs, entropies, and operational tasks are illustrated in Fig.~\ref{fig:introduction}, which echoes Ref.~\cite[Fig.~1]{Skrzypczyk-2019b}.

\section{Preliminaries}
\label{sec:preliminaries}

We start by reviewing the quantum formalism in both static and dynamic systems.  The notation of this paper is established in this section and Sec.~\ref{sec:QRT}.

\subsection{Static quantum systems}
\label{sec:system-static}

We use subscripted capital Latin letters, e.g., $\AO$, $\AI$, $\BO$, to label \emph{static quantum systems}.  Given a system $\AO$, we denote by $\spa{H}^\AO$ its associated space of Hermitian operators, and $\spa{H}_+^\AO\subset\spa{H}^\AO$ the cone of positive semidefinite operators therein.  The maximum possible rank of an operator in $\spa{H}^\AO$, also known as the dimensionality of $\AO$, is denoted by $d_\AO$.  We denote by $\#$ the trivial system, namely, the unique system of dimensionality $1$.  We denote by $\spa{R}\equiv\spa{H}^\#$ and $\spa{R}_+\equiv\spa{H}_+^\#$ the set of real numbers and the set of nonnegative real numbers, respectively.  The composition of two systems $\AO$ and $\AI$, denoted by $\AO\AI$, is identified by the composite operator space $\spa{H}^{\AO\AI}\equiv\spa{H}^\AO\otimes\spa{H}^\AI$.  A replica of $\AO$, denoted by $\RAO$, is a system operationally indistinguishable from $\AO$, with the property $d_\RAO=d_\AO$.

A \emph{quantum state} in $\AO$ is a positive semidefinite operator $\rho\in\spa{H}_+^\AO$ of unit trace.  We denote by $\q{D}^\AO\subset\spa{H}_+^\AO$ the set of states in $\AO$, and $\q{D}\equiv\bigcup_\AO\q{D}^\AO$ the set of all states, where the union is over all static systems.

The probabilistic occurrence of a state is represented by a \emph{quantum substate}, which in $\AO$ refers to a positive semidefinite operator $\rho_0\in\spa{H}_+^\AO$ of unit-bounded trace.  The substate $\rho_0$ indicates the occurrence of the state $\rho_0/\tr[\rho_0]\in\q{D}^\AO$ with probability $\tr[\rho_0]\leq1$.  We denote by $\q{D}_\leqslant^\AO$ the set of substates in $\AO$.

\subsection{Dynamic quantum systems}
\label{sec:system-dynamic}

Given two static systems $\AO$ and $\AI$, we denote by $\spa{L}^\A\equiv\spa{L}^{\AO\to\AI}$ the space of Hermiticity-preserving (HP) linear maps from $\spa{H}^\AO$ to $\spa{H}^\AI$.  Here $\A\equiv\AO\to\AI$ is a shorthand for the \emph{dynamic quantum system} evolving from $\AO$ to $\AI$.  We note that $\spa{L}^{\#\to\AI}=\spa{H}^\AI$.  Given a map $\Lambda\in\spa{L}^\A$, its Choi operator $J_\Lambda\in\spa{H}^{\AO\AI}$ is defined as
\begin{align}
\label{eq:choi-map}
	J_\Lambda^{\AO\AI}&\coloneq\left(\id^\AO\otimes\Lambda^{\RAO\to\AI}\right)\left[\spec{\phi}_+^{\AO\RAO}\right],
\end{align}
where $\id^\AO\in\spa{L}^{\AO\to\AO}$ denotes the identity map from $\AO$ to itself and $\spec{\phi}_+^{\AO\RAO}\equiv\sum_{i,j}\op{i}{j}^\AO\otimes\op{i}{j}^\RAO\in\spa{H}_+^{\AO\RAO}$ denotes the standard maximally entangled operator between $\AO$ and $\RAO$.  The bijection $\Lambda\leftrightarrow J_\Lambda$ from $\spa{L}^\A$ to $\spa{H}^{\AO\AI}$ is known as the Choi--Jamiołkowski isomorphism.

A \emph{quantum channel} from $\AO$ to $\AI$ is a linear map $\Lambda\in\spa{L}^\A$ that is completely positive (CP) and trace preserving (TP).  In other words, quantum channels are maps that preserve the set of quantum states even when acting on one part of a composite system.  According to Choi's Theorem, the CP condition of $\Lambda$ is equivalent to $J_\Lambda^{\AO\AI}\in\spa{H}_+^{\AO\AI}$, and the TP condition is equivalent to $J_\Lambda^\AO\equiv\tr_\AI[J_\Lambda^{\AO\AI}]=\spec{I}^\AO$, where $\spec{I}^\AO\in\spa{H}^\AI$ denotes the identity operator in $\AI$.  We denote by $\q{L}^\A\equiv\q{L}^{\AO\to\AI}\subset\spa{L}^\A$ the set of channels from $\AO$ to $\AI$, and $\q{L}\equiv\bigcup_{\AO,\AI}\q{L}^\A$ the set of all channels, where the union is over all input and output static systems.

Quantum channels represent deterministic operations on states.  A larger class of maps, known as \emph{quantum subchannels}, represents probabilistic operations.  A subchannel from $\AO$ to $\AI$ is a linear map $\Lambda_0\in\spa{L}^\A$ that is CP and trace nonincreasing (TNI).  Two key observations thus follow: (i) $\Lambda_0$ preserves the set of substates even when acting on one part of a composite system; and (ii) there exists another subchannel $\Lambda_1\in\spa{L}^\A$ such that $\Lambda_0+\Lambda_1\in\q{L}^\A$.  We denote by $\q{L}_\leqslant^\A$ the set of subchannels from $\AO$ to $\AI$.

\subsection{Supermaps between dynamic systems}
\label{sec:supermap}

Linear transformations between linear maps are known as \emph{supermaps}.  Given two dynamic systems $\A$ and $\B\equiv\BO\to\BI$, we denote by $\spa{S}^{\A\To\B}\equiv\spa{S}^{(\AO\to\AI)\To(\BO\to\BI)}$ the space of HP-preserving supermaps from $\spa{L}^\A$ to $\spa{L}^\B$.  We note that $\spa{S}^{(\#\to\#)\To\B}=\spa{L}^\B$ and that $\spa{S}^{(\#\to\AI)\To(\#\to\BI)}=\spa{L}^{\AI\to\BI}$.  Given a supermap $\sm{\Theta}\in\spa{S}^{\A\To\B}$, an associated linear map $\Lambda_\sm{\Theta}\in\spa{L}^{\A\B}\equiv\spa{L}^{\AO\BO\to\AI\BI}$ in the composite dynamic system $\A\B\equiv\AO\BO\to\AI\BI$ can be defined as follows~\cite{Gour-2019a}:
\begin{align}
\label{eq:lambda-supermap}
	\Lambda_\sm{\Theta}^{\A\B}&\coloneq\left(\Id^\A\otimes\sm{\Theta}^{\RA\To\B}\right)\left\{\spec{\Phi}_+^{\A\RA}\right\},
\end{align}
where $\Id^\A$ denotes the identity supermap from $\A$ to itself, $\RA\equiv\RAO\to\RAI$ denotes a replica of $\A$, and $\spec{\Phi}_+^{\A\RA}[\cdot]\equiv\tr[\spec{\phi}_+^{\AO\RAO}(\cdot)]\spec{\phi}_+^{\AI\RAI}$ denotes the standard maximally entangled map between $\A$ and $\RA$.  The bijection $\sm{\Theta}\leftrightarrow\Lambda_\sm{\Theta}$ from $\spa{S}^{\A\To\B}$ to $\spa{L}^{\A\B}$ is a dynamic analogue of the Choi--Jamiołkowski isomorphism.  Via the Choi--Jamiołkowski isomorphism $\Lambda_\sm{\Theta}\leftrightarrow J_{\Lambda_\sm{\Theta}}$, the Choi operator of the supermap $\sm{\Theta}$, denoted by $J_\sm{\Theta}\in\spa{H}^{\AO\AI\BO\BI}$, is defined as~\cite{Gour-2019a}
\begin{align}
\label{eq:choi-supermap}
	J_\sm{\Theta}^{\AO\AI\BO\BI}&\coloneq J_{\Lambda_\sm{\Theta}}^{\AO\AI\BO\BI}.
\end{align}
The bijection $\sm{\Theta}\leftrightarrow J_\sm{\Theta}$ is an isomorphism from $\spa{S}^{\A\To\B}$ to $\spa{H}^{\AO\AI\BO\BI}$.

A \emph{quantum superchannel} from $\A$ to $\B$ is a supermap $\sm{\Theta}\in\spa{S}^{\A\To\B}$ that is completely CP preserving and TP preserving.  In other words, quantum superchannels preserve the set of quantum channels even when acting on one part of a composite dynamic system.  We denote by $\q{S}^{\A\To\B}\subset\spa{S}^{\A\To\B}$ the set of superchannels from $\A$ to $\B$, and $\q{S}\equiv\bigcup_{\AO,\AI,\BO,\BI}\q{S}^{\A\To\B}$ the set of all superchannels, where union is over all quadruples of static systems.  The following lemma characterizes superchannels in both the Choi and the compositional pictures.

\begin{figure}[t]
\centering
\includegraphics[scale=0.15]{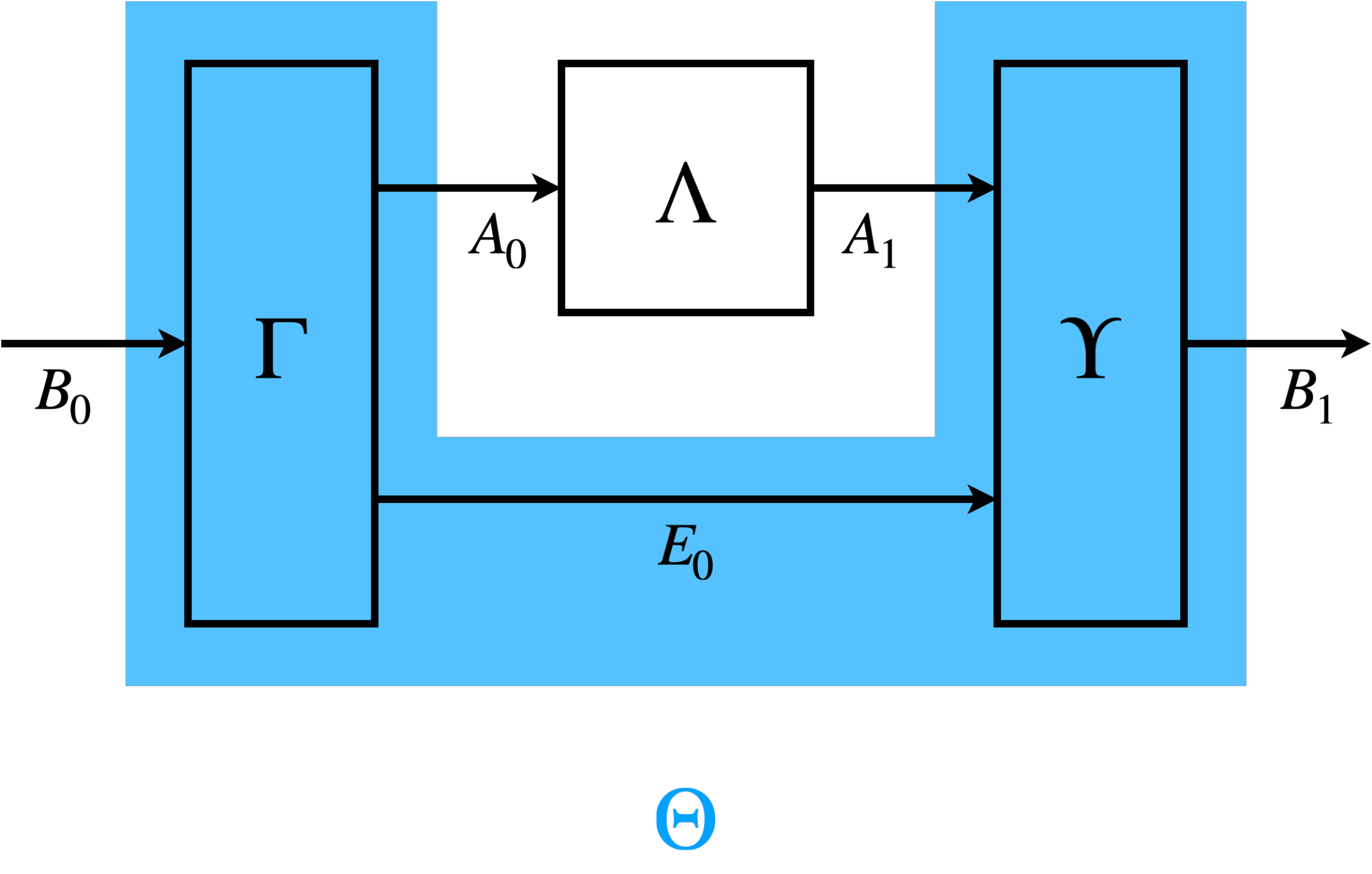}
\caption{Decomposition of a superchannel $\sm{\Theta}\in\q{S}^{\A\To\B}$ (the blue region) into a preprocessing channel $\Gamma^{\BO\to\AO\EO}$, a memory side channel $\id^\EO$, and a postprocessing channel $\Upsilon^{\AI\EO\to\BI}$, according to Eq.~\eqref{eq:superchannel-3}.  Applying $\sm{\Theta}$ to a channel $\Lambda\in\q{L}^\A$ yields $\sm{\Theta}\{\Lambda\}\in\q{L}^\B$.}
\label{fig:superchannel}
\end{figure}

\begin{lemma}[\cite{Chiribella-2008a, Gour-2019a}]
\label{lem:superchannel}
Let $\sm{\Theta}\in\spa{S}^{\A\To\B}$ be a supermap.  Then the following statements are equivalent.
\begin{enumerate}
	\item[(1)] The supermap $\sm{\Theta}$ is a superchannel.
	\item[(2)] The Choi operator $J_\sm{\Theta}\in\spa{H}_+^{\AO\AI\BO\BI}$ satisfies that
	\begin{align}
		J_\sm{\Theta}^{\AI\BO}&=\spec{I}^{\AI\BO}, \label{eq:superchannel-1}\\
		J_\sm{\Theta}^{\AO\AI\BO}&=J_\sm{\Theta}^{\AO\BO}\otimes\spec{\pi}^\AI, \label{eq:superchannel-2}
	\end{align}
	where $\spec{\pi}^\AI\equiv\frac{1}{d_\AI}\spec{I}^\AI\in\q{D}^\AI$ is the uniform state in $\AI$. 
	\item[(3)] There exists a system $\EO$ and two channels $\Gamma\in\q{L}^{\BO\to\AO\EO}$ and $\Upsilon\in\q{L}^{\AI\EO\to\BI}$ such that (see Fig.~\ref{fig:superchannel})
	\begin{align}
		\label{eq:superchannel-3}
		\sm{\Theta}^{\A\To\B}\left\{\cdot\right\}&=\Upsilon^{\AI\EO\to\BI}\circ\left(\left(\cdot\right)^\A\otimes\id^\EO\right)\circ\Gamma^{\BO\to\AO\EO}.
	\end{align}
\end{enumerate}
\end{lemma}

Quantum superchannels represent deterministic ``superoperations'' on channels.  We refer to their probabilistic relaxations as \emph{quantum subsuperchannels}.  A subsuperchannel from $\A$ to $\B$ is a supermap $\sm{\Theta}_0\in\spa{S}^{\A\To\B}$ that is completely CP preserving and completely TNI preserving~\cite{Burniston-2020a}.  We thus observe that (i) $\sm{\Theta}_0$ preserves the set of subchannels even when acting on one part of a composite dynamic system, and that (ii) there exists another subsuperchannel $\sm{\Theta}_1\in\spa{S}^{\A\To\B}$ such that $\sm{\Theta}_0+\sm{\Theta}_1\in\q{S}^{\A\To\B}$~\cite{Burniston-2020a}.  We denote by $\q{S}_\leqslant^{\A\To\B}$ the set of subsuperchannels from $\A$ to $\B$.

\section{A Top-Down Framework for Quantum Resource Theories}
\label{sec:QRT}

In this section, we formulate quantum resource theories using a ``top-down'' framework.  The framework unifies both static and dynamic QRTs, whose formal definitions we will present shortly.  We also review the definition of probabilistic free transformations and introduce the concepts of free measurements and supermeasurements.  We review resource quantification in the last part of this section, where both abstract principles and relevant examples of resource monotones are discussed.  The notation of this paper may appear somewhat clumsy, so we remind the reader to beware of nuanced fonts and symbols as we proceed (see Table~\ref{tab:notation}).

\subsection{Dynamic and static QRTs from axioms}
\label{sec:axiom}

Our framework of dynamic resource theories differs from those in some former studies, e.g., Refs.~\cite{Gour-2019b, Liu-2019b, Gour-2020a, Liu-2020a}, in that we shape the \emph{relationship} between static and dynamic resources differently.  In these former studies, a dynamic QRT \emph{is deemed as stemming from} a prespecified static QRT, and the free superoperations of the dynamic QRT are subject to the constraint that they preserve the set of free operations of the static QRT\@.  Such an approach can be understood as \emph{bottom-up}, as it starts with a static resource and extends its structure ``upwardly'' to the dynamic setting.  In contrast to the bottom-up approach, we adopt a \emph{top-down} approach, in the sense that we start with dynamic resources and consider static resources as ``downward'' reductions of the dynamic ones.  As we will argue later, the top-down approach is conceptually distinct from the bottom-up approach, and it in fact generalizes the other in a nontrivial manner.  Note that one can also study a dynamic QRT \emph{without} concerning its relationship with any static QRT, e.g., as in Refs.~\cite{Takagi-2019b,Regula-2021a,Regula-2021b}, and such an approach should be categorized as neither bottom-up nor top-down.

\begin{table}[t]
\caption{Notation of this paper}
\label{tab:notation}
\footnotesize
\renewcommand{\arraystretch}{1.2}
\begin{tabular}{|@{\;\;}m{3em}@{\;\;}@{\;\;}m{26em}@{\;\;}|}
\hline & \\[-2.2ex]
Symbol & Meaning
\\[0.5ex] \hline\hline & \\[-2.2ex]
$\spa{H}^\BI$ & Space of Hermitian operators in a system $\BI$ \\
$\spa{H}_+^\BI$ & Cone of positive semidefinite operators in $\spa{H}^\BI$ \\
$\spa{L}^\B$ & Space of HP linear maps from $\spa{H}^\BO$ to $\spa{H}^\BI$ \\
$\spa{S}^{\A\To\B}$ & Space of HP-preserving supermaps from $\spa{L}^\A$ to $\spa{L}^\B$
\\[0.5ex] \hline & \\[-2.2ex]
$\q{D}^\BI$ & Set of quantum states in $\BI$ \\
$\q{L}^\B$ & Set of quantum channels from $\BO$ to $\BI$ \\
$\q{S}^{\A\To\B}$ & Set of quantum superchannels from $\A$ to $\B$
\\[0.5ex] \hline & \\[-2.2ex]
$\q{D}_\leqslant^\BI$ & Set of quantum substates in $\BI$ \\
$\q{L}_\leqslant^\B$ & Set of quantum subchannels from $\BO$ to $\BI$ \\
$\q{S}_\leqslant^{\A\To\B}$ & Set of quantum subsuperchannels from $\A$ to $\B$
\\[0.5ex] \hline & \\[-2.2ex]
$\f{F}^\BI$ & Set of free states in $\BI$ in a static QRT $(\f{O},\f{F})$ \\
$\f{O}^{\AI\to\BI}$ & Set of free operations from $\AI$ to $\BI$ in $(\f{O},\f{F})$ \\
$\f{C}^\B$ & Set of free channels from $\BO$ to $\BI$ in a dynamic QRT $(\f{S},\f{C})$  \\
$\f{S}^{\A\To\B}$ & Set of free superoperations from $\A$ to $\B$ in $(\f{S},\f{C})$
\\[0.5ex] \hline & \\[-2.2ex]
$\f{O}_\abb{J}^{\AI\to\BI}$ & Set of Choi operators of free operations in $\f{O}^{\AI\to\BI}$ \\
$\f{S}_\abb{J}^{\A\To\B}$ & Set of Choi operators of free superoperations in $\f{S}^{\A\To\B}$
\\[0.5ex] \hline & \\[-2.2ex]
$\f{O}_\leqslant^{\AI\to\BI}$ & Set of probabilistic free operations from $\AI$ to $\BI$ in $(\f{O},\f{F})$ \\
$\f{S}_\leqslant^{\A\To\B}$ & Set of probabilistic free superoperations from $\A$ to $\B$ in $(\f{S},\f{C})$
\\[0.5ex] \hline & \\[-2.2ex]
$\f{O}_\abb{M}^\AI$ & Set of free measurements on $\AI$ in $(\f{O},\f{F})$ \\
$\f{S}_\abb{M}^\A$ & Set of free supermeasurements on $\A$ in $(\f{S},\f{C})$
\\[0.5ex] \hline
\end{tabular}
\end{table}

In what follows, we formally define a dynamic QRT and its induced static QRT in the top-down framework.  A summary of the notation is provided in Table~\ref{tab:notation} (see Secs.~\ref{sec:probabilistic} and \ref{sec:measurement} for missing definitions).  We also summarize in Fig.~\ref{fig:QRT} the relationships between different components of the QRT and the standard quantum theory.

\begin{definition}
\label{def:QRT-dynamic}
Let $\f{S}\subseteq\q{S}$ be a subset of superchannels.  For any two dynamic systems $\A$ and $\B$, define the following components:
\begin{enumerate}
	\item[(1)] a set of superchannels $\f{S}^{\A\To\B}\coloneq\f{S}\cap\q{S}^{\A\To\B}$;
	\item[(2)] a set of channels $\f{C}^\B\coloneq\f{S}^{(\#\to\#)\To\B}\subseteq\q{L}^\B$.
\end{enumerate}
Denote $\f{C}\equiv\bigcup_{\BO,\BI}\f{C}^\B\subseteq\q{L}$.  Then the tuple $(\f{S},\f{C})$ is called a \textbf{dynamic quantum resource theory} whenever the following conditions hold:
\begin{enumerate}
	\item[(1)] for any dynamic system $\B$, it holds that $\Id^\B\in\f{S}^{\B\To\B}$;
	\item[(2)] for any three dynamic systems $\A$, $\B$, and $\dsys{C}$, if $\sm{\Theta}\in\f{S}^{\A\To\B}$ and $\sm{\Theta}'\in\f{S}^{\B\To\dsys{C}}$, then $\sm{\Theta}'\circ\sm{\Theta}\in\f{S}^{\A\To\dsys{C}}$.
\end{enumerate}
In this case, superchannels in $\f{S}$ are called \textbf{(deterministic) free superoperations}, and channels in $\f{C}$ are called \textbf{free channels}.  We say that $(\f{S},\f{C})$ is \textbf{closed} and \textbf{convex} whenever $\f{S}^{\A\To\B}$ is a closed and convex set for all $\A$ and $\B$.
\end{definition}

Briefly, a dynamic QRT is completely determined by a subset of superchannels which contains the identity supermaps and which is closed under composition.

Following the top-down approach, every static QRT is a ``subtheory'' of a dynamic QRT, only regarding dynamic systems with a trivial input.

\begin{definition}
\label{def:QRT-static}
Let $(\f{S},\f{C})$ be a dynamic QRT\@.  For any two static systems $\AI$ and $\BI$, define the following components:
\begin{enumerate}
	\item[(1)] a set of channels $\f{O}^{\AI\to\BI}\coloneq\f{S}^{(\#\to\AI)\To(\#\to\BI)}\subseteq\q{L}^{\AI\to\BI}$;
	\item[(2)] a set of states $\f{F}^\BI\coloneq\f{O}^{\#\to\BI}\subseteq\q{D}^\BI$.
\end{enumerate}
Denote $\f{O}\equiv\bigcup_{\AI,\BI}\f{O}^{\AI\to\BI}\subseteq\q{L}$ and $\f{F}\equiv\bigcup_\BI\f{F}^\BI\subseteq\q{D}$.  Then the tuple $(\f{O},\f{F})$ is called the \textbf{static quantum resource theory} reduced from $(\f{S},\f{C})$.  In this case, channels in $\f{O}$ are called \textbf{(deterministic) free operations}, and states in $\f{F}$ are called \textbf{free states}.  We say that $(\f{O},\f{F})$ is \textbf{closed} and \textbf{convex} whenever $\f{O}^{\AI\to\BI}$ is a closed and convex set for all $\AI$ and $\BI$.
\end{definition}

We denote by $\f{S}_\abb{J}^{\A\To\B}\coloneq\{J_\sm{\Theta}\colon\sm{\Theta}\in\f{S}^{\A\To\B}\}\subset\spa{H}_+^{\AO\AI\BO\BI}$ the set of Choi operators of free superoperations from $\A$ to $\B$, and $\f{O}_\abb{J}^{\AI\to\BI}\coloneq\f{S}_\abb{J}^{(\#\to\AI)\To(\#\to\BI)}\subset\spa{H}_+^{\AI\BI}$ the set of Choi operators of free operations from $\AI$ to $\BI$.

\begin{figure}[t]
\[\begin{tikzcd}
&&\substack{\displaystyle\f{S}^{\A\To\B} \\\\ \abb{(free superoperations)}} \\
\\
\\
\substack{\displaystyle\f{C}^\B \\\\ \abb{(free channels)}} &&\substack{\displaystyle\q{S}^{\A\To\B} \\\\ \abb{(quantum superchannels)}} &&\substack{\displaystyle\f{O}^{\AI\to\BI} \\\\ \abb{(free operations)}} \\
\\
\\
\substack{\displaystyle\q{L}^\B \\\\ \abb{(quantum channels)}} &&\substack{\displaystyle\f{F}^\BI \\\\ \abb{(free states)}} &&\substack{\displaystyle\q{L}^{\AI\to\BI} \\\\ \abb{(quantum channels)}} \\
\\
\\
&&\substack{\displaystyle\q{D}^\BI \\\\ \abb{(quantum states)}}
\arrow["{\subseteq}"{description}, color={rgb,255:red,184;green,46;blue,46}, shorten <=4pt, shorten >=4pt, from=7-3, to=10-3]
\arrow["{\BO=\#}"{description}, color={rgb,255:red,44;green,44;blue,175}, shorten <=5pt, shorten >=5pt, from=7-1, to=10-3]
\arrow["{\AI=\#}"{description}, color={rgb,255:red,35;green,139;blue,35}, shorten <=5pt, shorten >=5pt, from=7-5, to=10-3]
\arrow["{\AO=\AI=\#}"{description, pos=0.7}, color={rgb,255:red,35;green,139;blue,35}, shorten <=5pt, shorten >=5pt, from=4-3, to=7-1]
\arrow["{\AO=\BO=\#}"{description, pos=0.7}, color={rgb,255:red,44;green,44;blue,175}, shorten <=5pt, shorten >=5pt, from=4-3, to=7-5]
\arrow["{\BO=\#}"{description, pos=0.3}, color={rgb,255:red,44;green,44;blue,175}, shorten <=5pt, shorten >=5pt, from=4-1, to=7-3]
\arrow["{\AO=\AI=\#}"{description}, color={rgb,255:red,35;green,139;blue,35}, shorten <=5pt, shorten >=5pt, from=1-3, to=4-1]
\arrow["{\AI=\#}"{description, pos=0.3}, color={rgb,255:red,35;green,139;blue,35}, shorten <=5pt, shorten >=5pt, from=4-5, to=7-3]
\arrow["{\AO=\BO=\#}"{description}, color={rgb,255:red,44;green,44;blue,175}, shorten <=5pt, shorten >=5pt, from=1-3, to=4-5]
\arrow["{\subseteq}"{description}, color={rgb,255:red,184;green,46;blue,46}, shorten <=4pt, shorten >=4pt, from=4-1, to=7-1]
\arrow["{\subseteq}"{description}, color={rgb,255:red,184;green,46;blue,46}, shorten <=4pt, shorten >=4pt, from=4-5, to=7-5]
\arrow["{\subseteq}"{description}, color={rgb,255:red,184;green,46;blue,46}, shorten <=4pt, shorten >=4pt, from=1-3, to=4-3]
\end{tikzcd}\]
\caption{Relationships between components of a dynamic QRT $(\f{S},\f{C})$, its reduced static QRT $(\f{O},\f{F})$, and the standard quantum theory $(\q{S},\q{L},\q{D})$ in a top-down framework.  Green arrows represent transformation-to-object reduction.  Blue arrows represent dynamic-to-static reduction.  Red arrows represent the relationship ``being a subset of.''  Note that the set of free channels $\f{C}\equiv\bigcup_{\BO,\BI}\f{C}^\B$ and the set of free operations $\f{O}\equiv\bigcup_{\AI,\BI}\f{O}^{\AI\to\BI}$ do not necessarily coincide.}
\label{fig:QRT}
\end{figure}

So far we have established the bulk of our framework of resource theories, encompassing both static and dynamic QRTs, in a top-down fashion (from the top level of superoperations down to the bottom level of states).  The dynamic-to-static reduction implies that any property possessed by a dynamic QRT is inherited by its reduced static QRT\@.

We now comment on the conceptual similarity and distinction between the top-down and the bottom-up approaches.  The similarity between them lies in that Definition~\ref{def:QRT-static} coincides with the conventional definition of a static QRT~\cite{Chitambar-2019a}, with the latter specified solely by a subset of channels $\f{O}\subseteq\q{L}$ containing the identity maps and closed under composition.  Despite this, there is a remarkable distinction between the top-down and the bottom-up approaches regarding what constitutes a dynamic QRT\@.  In fact, Definition~\ref{def:QRT-dynamic} is a strict generalization of the bottom-up approach to dynamic QRTs~\cite{Gour-2019b, Liu-2019b, Gour-2020a, Liu-2020a}.  As mentioned before, a dynamic QRT in the bottom-up sense is extended from some pre-existing static QRT, whose free operations automatically serve as the free channels in the dynamic QRT\@.  However, not distinguishing between \emph{free channels} and \emph{free operations} implicitly requires the identity maps to be free channels in every dynamic QRT, a condition known to be violated in many situations.  One (counter)example is dynamic quantum coherence, for which the identity maps are naturally free operations~\cite{Baumgratz-2014a, Streltsov-2017a} but not necessarily free channels~\cite{Saxena-2020a}.  We remark that this discrepancy, while contradicting the bottom-up approach, can be naturally addressed in the top-down framework by viewing the static theories of coherence~\cite{Baumgratz-2014a, Streltsov-2017a} as the reduced static QRTs of the corresponding dynamic theories (namely, MIO from MISC, and DIO from DISC)~\cite{Saxena-2020a}.  Apart from quantum coherence, the resource theories of quantum memories~\cite{Rosset-2018a} and of communication~\cite{Devetak-2008a, Takagi-2020a} are likewise incompatible with the bottom-up approach, as the identity maps are maximally resourceful (and thus not free) channels in these theories.

We conclude this subsection by pointing out a \emph{conjugation invariance principle} of resource theories which is generally regarded true in this paper.  We believe that any ``physically motivated'' QRT should keep its structure invariant under complex conjugation with respect to a prespecified computational basis.  This specifically means that the sets $\f{S}$, $\f{C}$, $\f{O}$, and $\f{F}$ are all closed under complex conjugation with respect to the said basis.  We remark that this invariance principle is empirically observed in virtually all known resource theories of physical interest.

\subsection{Probabilistic free transformations}
\label{sec:probabilistic}

In a static QRT $(\f{O},\f{F})$, the set of free operations prescribes what \emph{deterministic} transformations (i.e., CPTP maps) are allowed between states.  We now discuss how this prescription would affect the realizability of \emph{probabilistic} transformations in $(\f{O},\f{F})$.  Recall that a probabilistic state transformation is represented by a subchannel.  Since a subchannel is not a complete physical process, to decide its realizability, we must put the subchannel back in a complete physical process (i.e., a channel) where it could have belonged.  This complete process is represented by a quantum instrument (i.e., a q-to-qc channel), which involves an ancillary classical ``flag'' system indicating whether the intended subchannel is carried out successfully.

\begin{definition}[\cite{Chitambar-2019a}]
\label{def:probabilistic-static}
A subchannel $\Psi_0\in\q{L}_\leqslant^{\AI\to\BI}$ is called a \textbf{probabilistic free operation} in a static QRT $(\f{O},\f{F})$, denoted by $\Psi_0\in\f{O}_\leqslant^{\AI\to\BI}$, whenever there exists a classical system $\XI$, a subchannel $\Psi_1\in\q{L}_\leqslant^{\AI\to\BI}$, and a free operation $\Psi\in\f{O}^{\AI\to\BI\XI}$ such that
\begin{align}
	&\Psi^{\AI\to\BI\XI}\left[\cdot\right] \notag\\
	&\quad=\Psi_0^{\AI\to\BI}\left[\cdot\right]\otimes\op{0}{0}^\XI+\Psi_1^{\AI\to\BI}\left[\cdot\right]\otimes\op{1}{1}^\XI.
\end{align}
\end{definition}

Probabilistic free operations defined as such have been studied in the QRT of coherence~\cite{Fang-2018a}, and a related but slightly different treatment has also been considered in the QRT of thermodynamics~\cite{Alhambra-2016a} where the classical flag system is present in both the input and the output systems (since classical flags cannot be freely generated in their context but can be reused).  The intuition behind Definition~\ref{def:probabilistic-static} is that the only operational means of carrying out a subchannel $\Psi_0$ at no cost of resource is to implement a free quantum instrument $\{\Psi_0,\Psi_1\}$ with some supplementary subchannel $\Psi_1$ and then hope for the outcome $0$ to occur.  We observe that the set $\f{O}_\leqslant^{\AI\to\BI}$ is closed and convex if $(\f{O},\f{F})$ is a closed and convex static QRT\@.

Now we move forward to the dynamic setting, in which probabilistic transformations are between subchannels and in the form of subsuperchannels.  Probabilistic free transformations in a dynamic QRT are defined following the same idea as in a static QRT\@.

\begin{definition}
\label{def:probabilistic-dynamic}
A subsuperchannel $\sm{\Theta}_0\in\q{S}_\leqslant^{\A\To\B}$ is called a \textbf{probabilistic free superoperation} in a dynamic QRT $(\f{S},\f{C})$, denoted by $\sm{\Theta}_0\in\f{S}_\leqslant^{\A\To\B}$, whenever there exists a classical system $\XI$, a subsuperchannel $\sm{\Theta}_1\in\q{S}_\leqslant^{\A\To\B}$, and a free superoperation $\sm{\Theta}\in\f{S}^{\A\To\B^\times}$, where $\B^\times\equiv\BO\to\BI\XI$, such that
\begin{align}
	&\left(\sm{\Theta}\left\{\cdot\right\}\right)^{\B^\times}\left[\cdot\right] \notag\\
	&\quad=\left(\sm{\Theta}_0\left\{\cdot\right\}\right)^\B\left[\cdot\right]\otimes\op{0}{0}^\XI+\left(\sm{\Theta}_1\left\{\cdot\right\}\right)^\B\left[\cdot\right]\otimes\op{1}{1}^\XI.
\end{align}
\end{definition}

The intuition behind Definition~\ref{def:probabilistic-dynamic} is that a subsuperchannel is considered to be freely realizable probabilistically if and only if it is a component of some free ``superinstrument''~\cite{Burniston-2020a}.  We observe that the set $\f{S}_\leqslant^{\A\To\B}$ is closed and convex if $(\f{S},\f{C})$ is a closed and convex dynamic QRT\@.

\subsection{Free measurements and supermeasurements}
\label{sec:measurement}

In quantum theory, measurements are processes that feed on a quantum state as input and generate a classical signal (known as the outcome) as output.  Being a q-to-c channel, a quantum measurement is nonetheless most commonly represented by a \emph{positive operator-valued measure (POVM)}, which on a system $\AI$ is a collection of positive semidefinite operators $\povm{M}\equiv\{M_m\}_m$ with $M_m\in\spa{H}_+^\AI$ for all $m$ and satisfying $\sum_{m}M_m=\spec{I}$.  The probability $p_m$ of obtaining an outcome $m$ when measuring a state $\rho\in\q{D}^\AI$ according to the POVM $\povm{M}$ is dictated by the Born rule as follows:
\begin{align}
\label{eq:born}
	p_m&\coloneq\tr\left[M_m\rho\right]\quad\forall m.
\end{align}
It can be conveniently verified that $\{p_m\}_m$ in Eq.~\eqref{eq:born} is a valid probability distribution, as $p_m\geq0$ for all $m$ and $\sum_{m}p_m=1$.

Now we consider a static QRT $(\f{O},\f{F})$.  Since quantum measurements are q-to-c channels, the separation between the free operations $\f{O}$ and the resourceful operations $\q{L}\setminus\f{O}$ implicitly entails a classification of measurements into ``free'' measurements and ``resourceful'' measurements.  This idea is made explicit as follows.

\begin{definition}
\label{def:measurement-free}
A POVM $\povm{M}\equiv\{M_m\}_m$ on a system $\AI$ is called a \textbf{free measurement} in a static QRT $(\f{O},\f{F})$, denoted by $\povm{M}\in\f{O}_\abb{M}^\AI$, whenever there exits a classical system $X_1$ such that
\begin{align}
	\sum_{m}M_m^\AI\otimes\op{m}{m}^\XI&\in\f{O}_\abb{J}^{\AI\to\XI}.
\end{align}
\end{definition}

Provided the conjugation invariance principle (introduced in Sec.~\ref{sec:axiom}), Definition~\ref{def:measurement-free} is simply saying that a POVM is considered to be a free measurement if and only if its corresponding q-to-c channel is a free operation.

Having identified the free measurements of a static QRT, we naturally wonder if the same idea applies to the dynamic setting.  Therefore, our next aim is to identify the ``free supermeasurements,'' i.e., measurements on channels that are operationally free in a dynamic QRT\@.  But before that, we need to clarify what a ``measurement on channels'' refers to.

Both the terms ``supermeasurement'' and ``measurement on channels'' are initially coined by the authors of Ref.~\cite{Burniston-2020a} to refer to probabilistic transformations of channels (i.e., what we refer to as subsuperchannels in this paper).  Here, we borrow these two terms (and use them interchangeably) to describe a rather different concept than in Ref.~\cite{Burniston-2020a}, and this tweak of nomenclature appears sensible in our context.

Recall that measurements on states represent physical processes that turn a quantum state into a classical state (i.e., a probability distribution).  Analogously, by ``measurements on channels'' we in particular mean physical processes that turn a quantum channel into a classical channel (i.e., a conditional probability distribution).  One way of understanding such a process is to view it as a q-to-c superchannel.  An alternative but equally insightful way is to represent the process in terms of a collection of positive semidefinite operators satisfying a certain set of constraints, i.e., in a way similar to (and generalizing) the POVM representation of a measurement on states.  We refer to this operator representation of a supermeasurement as a \emph{superPOVM}.

\begin{definition}
\label{def:superPOVM}
A \textbf{superPOVM} on a dynamic system $\A$ is a collection of positive semidefinite operators $\spovm{M}\equiv\{\mu_{x_1|x_0}\}_{x_0,x_1}$ with $\mu_{x_1|x_0}\in\spa{H}_+^{\AO\AI}$ for all $x_0,x_1$ and satisfying that
\begin{align}
	\sum_{x_1}\mu_{x_1|x_0}^\AI&=\spec{I}^\AI\quad\forall x_0, \label{eq:superPOVM-1}\\
	\sum_{x_1}\mu_{x_1|x_0}^{\AO\AI}&=\sum_{x_1}\mu_{x_1|x_0}^\AO\otimes\spec{\pi}^\AI\quad\forall x_0. \label{eq:superPOVM-2}
\end{align}
\end{definition}

Equations~\eqref{eq:superPOVM-1} and \eqref{eq:superPOVM-2} are adaptations of Eqs.~\eqref{eq:superchannel-1} and \eqref{eq:superchannel-2} with the systems $\BO$ and $\BI$ (here $\XO$ and $\XI$) classical.

Following the superPOVM formalism, we introduce what we call the \emph{superBorn rule}, which elucidates the syntax of obtaining a classical channel $\{p_{x_1|x_0}\}_{x_0,x_1}$ by applying a superPOVM $\spovm{M}$ on $\A$ to a quantum channel $\Lambda\in\q{L}^\A$, as follows:
\begin{align}
\label{eq:superborn}
	p_{x_1|x_0}&\coloneq\tr\left[\mu_{x_1|x_0}J_\Lambda\right]\quad\forall x_0,x_1.
\end{align}
The following proposition legitimates the superBorn rule, showing that $\{p_{x_1|x_0}\}_{x_0,x_1}$ is a valid conditional probability distribution.

\begin{proposition}
\label{prop:superborn}
The probability $p_{x_1|x_0}$ (defined in Eq.~\eqref{eq:superborn}) of obtaining an outcome $x_1$ from $x_0$ by measuring a channel $\Lambda\in\q{L}^\A$ according to a superPOVM $\spovm{M}\equiv\{\mu_{x_1|x_0}\}_{x_0,x_1}$ on $\A$ has the following properties.
\begin{enumerate}
	\item[(1)] Positivity: $p_{x_1|x_0}\geq0$ for all $x_0,x_1$.
	\item[(2)] Normalization: $\sum_{x_1}p_{x_1|x_0}=1$ for all $x_0$.
\end{enumerate}
\end{proposition}

\begin{proof}
The positivity of $p_{x_1|x_0}$ for all $x_0,x_1$ is a direct consequence of the positivity of $\mu_{x_1|x_0}$ and $J_\Lambda$.  The following proves the normalization property:
\begin{align}
	\sum_{x_1}p_{x_1|x_0}&=\sum_{x_1}\tr\left[\mu_{x_1|x_0}J_\Lambda\right] \\
	&=\tr\left[\left(\sum_{x_1}\mu_{x_1|x_0}^\AO\otimes\spec{\pi}^\AI\right)J_\Lambda^{\AO\AI}\right] \\
	&=\frac{1}{d_\AI}\tr\left[\sum_{x_1}\mu_{x_1|x_0}^\AO J_\Lambda^\AO\right] \\
	&=\frac{1}{d_\AI}\tr\left[\sum_{x_1}\mu_{x_1|x_0}^\AO\right] \\
	&=\frac{1}{d_\AI}\tr\left[\sum_{x_1}\mu_{x_1|x_0}^\AI\right] \\
	&=1\quad\forall x_0.
\end{align}
\end{proof}

\begin{figure}[t]
\centering
\includegraphics[scale=0.15]{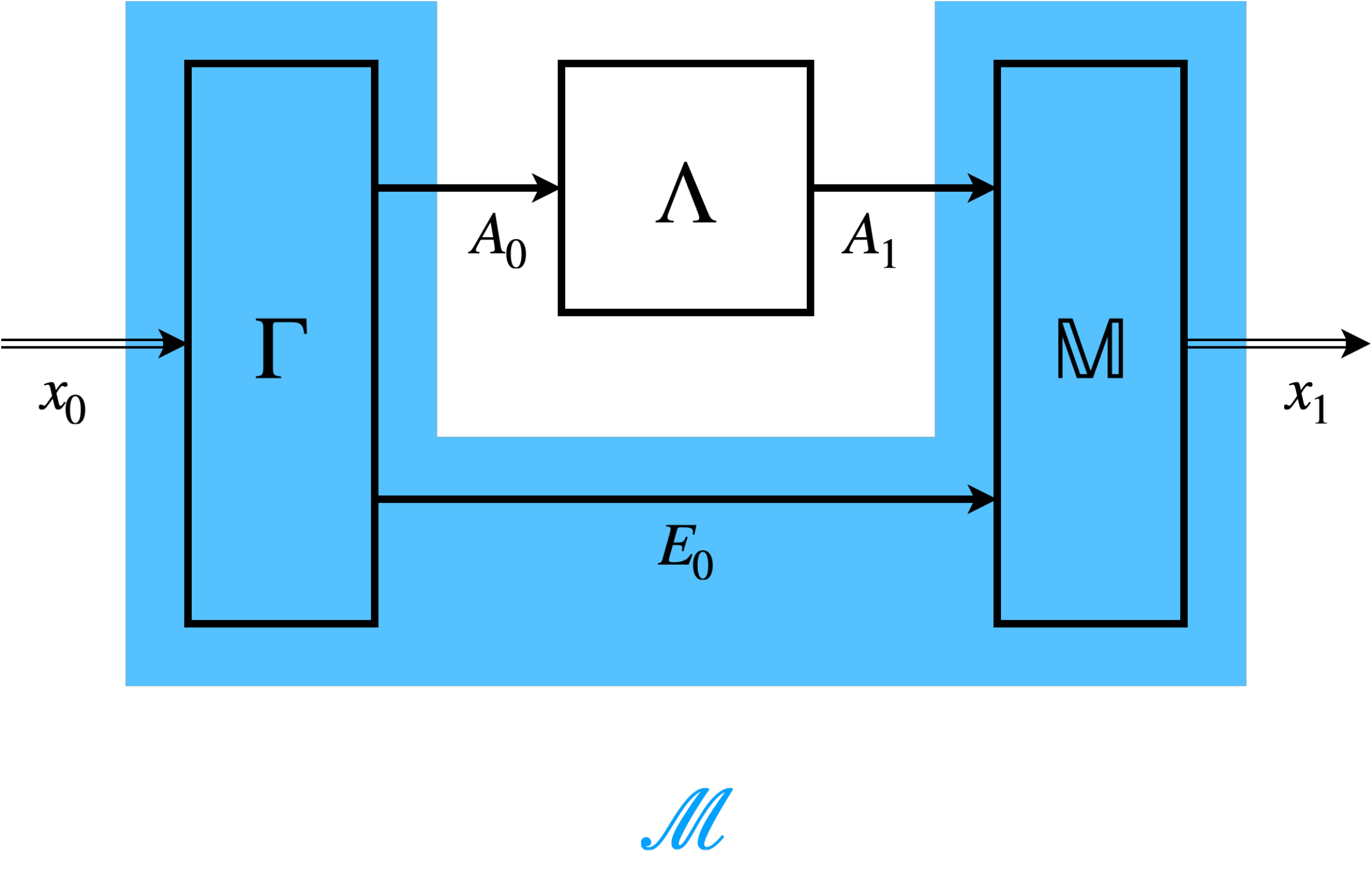}
\caption{Decomposition of a supermeasurement $\spovm{M}\equiv\{\mu_{x_1|x_0}\}_{x_0,x_1}$ (the blue region) into a preprocessing c-to-q channel $\Gamma^{\XO\to\AO\EO}$, a memory side channel $\id^\EO$, and a postprocessing POVM $\povm{M}\equiv\{M_{x_1}\}_{x_1}$.  Applying $\spovm{M}$ to a channel $\Lambda\in\q{L}^\A$ yields a conditional probability distribution $\{p_{x_1|x_0}\}_{x_0,x_1}$.}
\label{fig:superPOVM}
\end{figure}

Definition~\ref{def:superPOVM} and Eq.~\eqref{eq:superborn} can also be understood from a compositional perspective (see Fig.~\ref{fig:superPOVM}).  As a q-to-c superchannel, every supermeasurement on $\A$ can be decomposed into a preprocessing c-to-q channel $\Gamma^{\XO\to\AO\EO}$, a memory side channel $\id^\EO$, and a postprocessing POVM $\povm{M}\equiv\{M_{x_1}\}_{x_1}$ on $\AI\EO$ (see Lemma~\ref{lem:superchannel}).  By the Born rule in Eq.~\eqref{eq:born}, the probability of obtaining an outcome $x_1$ from $x_0$ when applying the supermeasurement to a channel $\Lambda\in\q{L}^\A$ is given by
\begin{align}
	&p_{x_1|x_0} \notag\\
	&\quad=\tr\left[M_{x_1}^{\AI\EO}\left(\Lambda^\A\otimes\id^\EO\right)\circ\Gamma^{\XO\to\AO\EO}\left[\op{x_0}{x_0}^\XO\right]\right] \\
	&\quad=\tr\left[\left(\phi_+^{\AO\RAO}\otimes M_{x_1}^{\AI\EO}\right)\left(J_\Lambda^{\AO\AI}\otimes\Gamma^{\XO\to\RAO\EO}\left[\op{x_0}{x_0}^\XO\right]\right)\right] \\
	&\quad=\tr\left[\mu_{x_1|x_0}J_\Lambda\right]\quad\forall x_0,x_1, \label{eq:superPOVM-3}
\end{align}
where
\begin{align}
	\mu_{x_1|x_0}^{\AO\AI}&\coloneq\tr_{\RAO\EO}\left[\left(\phi_+^{\AO\RAO}\otimes M_{x_1}^{\AI\EO}\right)\Gamma^{\XO\to\RAO\EO}\left[\op{x_0}{x_0}^\XO\right]\right] \notag\\
	&\quad\forall x_0,x_1. \label{eq:superPOVM-4}
\end{align}
We note that Eq.~\eqref{eq:superPOVM-3} is consistent with the superBorn rule in Eq.~\eqref{eq:superborn} and that the collection of operators $\spovm{M}\equiv\{\mu_{x_1|x_0}\}_{x_0,x_1}$ defined in Eq.~\eqref{eq:superPOVM-4} is a valid superPOVM according to Definition~\ref{def:superPOVM} since
\begin{align}
	\sum_{x_1}\mu_{x_1|x_0}^\AI&=\spec{I}^\AI\quad\forall x_0, \\
	\sum_{x_1}\mu_{x_1|x_0}^{\AO\AI}&=d_\AI\left(\tr_\EO\circ\Gamma^{\XO\to\RAO\EO}\left[\op{x_0}{x_0}^\XO\right]\right)^\top\otimes\spec{\pi}^\AI\quad\forall x_0.
\end{align}

Now we are ready to generalize Definition~\ref{def:measurement-free} to dynamic QRTs.

\begin{definition}
\label{def:supermeasurement-free}
A superPOVM $\spovm{M}\coloneq\{\mu_{x_1|x_0}\}_{x_0,x_1}$ on a dynamic system $\A$ is called a \textbf{free supermeasurement} in a dynamic QRT $(\f{S},\f{C})$, denoted by $\spovm{M}\in\f{S}_\abb{M}^\A$, whenever there exists a dynamic classical system $\X\equiv\XO\to\XI$ such that
\begin{align}
	\sum_{x_0,x_1}\mu_{x_1|x_0}^{\AO\AI}\otimes\op{x_0}{x_0}^\XO\otimes\op{x_1}{x_1}^\XI&\in\f{S}_\abb{J}^{\A\To\X}.
\end{align}
\end{definition}

Definition~\ref{def:supermeasurement-free} essentially means that a supermeasurement is considered to be free in a dynamic QRT if and only if its corresponding q-to-c superchannel is a free superoperation.

\subsection{Resource monotones, robustness measures, and operational tasks}
\label{sec:monotones}

Comparing the resourcefulness of different objects is a primary theme of resource theories.  (Here ``objects'' generally refer to states in a static QRT or channels in a dynamic QRT, and ``transformations'' refer to operations in a static QRT or superoperations in a dynamic QRT\@.  For abstract discussions of general resource theories in terms of objects and transformations, see Appendix~\ref{app:resource}.)  A handy way of making such comparison is via resource quantification.  This involves defining a real-valued function on the set of objects, so that comparing the resourcefulness of these objects can be turned into comparing their respective function values.  Any function $f$ meaningful for resource quantification must be monotonically nonincreasing under free transformations, i.e., satisfying $f\circ\ob{T}\leq f$ for every free transformation $\ob{T}$.  This comes from the intuition that an object should not turn up more resourceful undergoing a free transformation, and it essentially enforces that the total preorder (between objects) entailed by $f$ should not contradict the partial preorder entailed by convertibility via free transformations.

Various forms of resource monotones have been studied to serve different needs.  Common types of monotones include: distance-based monotones, entropy-based monotones, monotones based on geometry of the object space, and monotones based on operational tasks~\cite{Chitambar-2019a}.  A significant portion of this paper will be devoted to establishing quantitative connections between distinct types of resource monotones.

One prominent types of geometrically inspired monotones is the \emph{resource robustness measures}.  Let $\f{F}$ denote the set of free objects in a resource theory.  Let $\s{K}$ be a subset of all objects.  Then the \emph{resource robustness} of an object $\ob{Q}$ against the subset $\s{K}$ is defined as~\cite{Chitambar-2019a}
\begin{align}
\label{eq:robustness}
	R^{\s{K},\f{F}}(\ob{Q})&\coloneq\min_{\scriptsize\left\{\begin{array}{c}
		r\in\spa{R}_+, \\
		\ob{K}\in\s{K}\colon \\
		\frac{\ob{Q}+r\ob{K}}{1+r}\in\f{F}
	\end{array}\right\}}r.
\end{align}
The resource robustness $R^{\s{K},\f{F}}(\ob{Q})$ quantifies the minimum amount of admixing noise needed from $\s{K}$ to dissolve all the resource contained in $\ob{Q}$.  It is shown in Appendix~\ref{app:robustness} that the robustness measure $R^{\s{K},\f{F}}(\ob{Q})$ is a resource monotone with respect to $\ob{Q}$ if $\s{K}$ is contractive under every free transformation.  Typical examples of resource robustness measures as a monotone include the \emph{resource global robustness}~\cite{Harrow-2003a, Brandao-2015a}, for which $\s{K}$ is fixed as the entire set of objects, and the \emph{resource free robustness}~\cite{Vidal-1999a}, for which $\s{K}$ is fixed as the set of free objects $\f{F}$.  Another frequently considered resource robustness measure, despite not always being a monotone~\cite{Harrow-2003a}, is the \emph{resource random robustness}~\cite{Vidal-1999a}, for which $\s{K}$ contains only the maximally mixed object (e.g., the uniform state in a static QRT or the uniform channel in a dynamic QRT).

Another important class of resource monotones are those emerging in \emph{operational tasks}.  Basically, an operational task is a scenario where: (i) a set of rules, parametrized by $\ob{G}$, is announced; (ii) the performer of the task operates within the rules and obtains a result $i$; and (iii) the performer is rewarded with a real-valued score $s_i(\ob{G})$ based on the result.  For our purpose, we also assume the following about the performer's operational capacity for performing such a task: (i) the performer has access to a single use of an object $\ob{Q}$, which serves as a resource to assist their completion of the task; and (ii) the performer has the freedom to choose their strategy $\ob{T}$, within a strategy set $\f{T}$, for how to utilize their resource in response to the rules $\ob{G}$.  Let $p_i(\ob{T}[\ob{Q}];\ob{G})$ denote the occurrence probability of a result $i$ given that a strategy $\ob{T}\in\f{T}$ has been adopted on the resource $\ob{Q}$.  Then the performer's expected score for the task under the optimal strategy is given by 
\begin{align}
	S^\f{T}(\ob{Q};\ob{G})&\coloneq\max_{\ob{T}\in\f{T}}\sum_{i}s_i(\ob{G})p_i(\ob{T}\left[\ob{Q}\right];\ob{G}).
\end{align}
We note that when the strategy set $\f{T}$ is the set of free transformations applied to $\ob{Q}$, the expected score $S^\f{T}(\ob{Q};\ob{G})$ becomes a resource monotone with respect to $\ob{Q}$ for any $\ob{G}$.  To see this, for any $\ob{T}\in\f{T}$ and $\ob{G}$, we have that
\begin{align}
	S^\f{T}(\ob{T}\left[\ob{Q}\right];\ob{G})&=\max_{\ob{T}'\in\f{T}}\sum_{i}s_i(\ob{G})p_i(\ob{T}'\circ\ob{T}\left[\ob{Q}\right];\ob{G}) \\
	&\leq\max_{\ob{T}'\in\f{T}}\sum_{i}s_i(\ob{G})p_i(\ob{T}'\left[\ob{Q}\right];\ob{G}) \label{eq:monotones-1}\\
	&=S^\f{T}(\ob{Q};\ob{G})\quad\forall\ob{Q}, \label{eq:monotones-2}
\end{align}
where Eq.~\eqref{eq:monotones-1} uses the fact that the set of free transformations is closed under composition.  A special rewarding rule is to only consider success and failure, which corresponds to a binary result $i\in\{\abb{succ},\abb{fail}\}$ with the score function $s_\abb{succ}(\ob{G})\coloneq1$ and $s_\abb{fail}(\ob{G})\coloneq0$ for any $\ob{G}$.  Accordingly, the maximum expected score is given by the maximum success probability:
\begin{align}
	P^\f{T}(\ob{Q};\ob{G})&\coloneq\max_{\ob{T}\in\f{T}}p_\abb{succ}(\ob{T}\left[\ob{Q}\right];\ob{G}).
\end{align}

We readily notice that every resource monotone $f$ by definition implies a necessary condition for the existence of a (deterministic) free transformation that converts one object $\ob{Q}$ to anther object $\ob{Q}'$, as the free convertibility from $\ob{Q}$ to $\ob{Q}'$ implies $f(\ob{Q})\geq f(\ob{Q}')$.  We say that a family of resource monotones $\{f_w\}_{w\in\idx{W}}$ constitutes a \emph{complete set of resource monotones} whenever it collectively also provides a sufficient condition for such free convertibility.  For such a family, given any two objects $\ob{Q}$ and $\ob{Q}'$, there exists a free transformation $\ob{T}$ such that $\ob{T}[\ob{Q}]=\ob{Q}'$ if and only if $f_w(\ob{Q})\geq f_w(\ob{Q}')$ for all $w\in\idx{W}$.

Lastly, we can generalize the notion the resource monotones by taking into account probabilistic transformations.  We say that a function $f$ is a \emph{resource monotone in the probabilistic sense} whenever it is monotonically nonincreasing under probabilistic free transformations, i.e., whenever $f\circ\ob{T}\leq f$ for every probabilistic free transformation $\ob{T}$, defined, e.g., in the spirit of Definitions~\ref{def:probabilistic-static} and \ref{def:probabilistic-dynamic}.  A family of such monotones is said to be \emph{complete} if it provides both necessary and sufficient conditions for the convertibility from one probabilistic object (e.g., a substate or subchannel) to another via some probabilistic free transformation.  We note that monotones in the probabilistic sense are stronger than those in the deterministic sense.

\section{Free Conditional Min-Entropy}
\label{sec:entropy}

In this section, we generalize an important entropic measure in quantum information theory, namely the quantum (conditional) min-entropy, to scenarios where variable operational restrictions may apply.  We start by reviewing the definitions of the quantum (conditional) min-entropy in both the static~\cite{Renner-2005a, Konig-2009a} and the dynamic~\cite{Gour-2019a} settings.  Then we propose their resource-theoretic generalizations, namely the \emph{free (conditional) min-entropy} and its dynamic extension, as measures of quantum uncertainty in operationally restricted scenarios.

From now on, we always assume a \emph{closed} and \emph{convex} structure of any QRT considered in this paper.  Minimally, we only need to assume that the set $\f{S}^{\A\To\B}$ is closed and convex for all $\AO$, $\AI$, $\BO$, and $\BI$, and then the closedness and convexity of $\f{C}^\B$, $\f{O}^{\AI\to\BI}$, and $\f{F}^\BI$ would follow as a consequence of top-down reduction.  We mention that closedness and convexity are empirical and reasonable assumptions to make, as a variety of concrete QRTs are known to satisfy them both.  Even though there exist QRTs that are not closed or not convex, it may still be insightful to consider their closures or convex expansions as approximations.

Our approach to the generalized entropic concepts relies heavily on \emph{convex conic programming}~\cite{Boyd-2004a} (see Appendix~\ref{app:program} for a review of the basics), and this explains why closedness and convexity are technically critical to this paper.  Given a closed and convex set $\s{U}\subseteq\spa{U}$ in an inner-product vector space $\spa{U}$, the \emph{cone} generated by $\spa{U}$ is defined as
\begin{align}
	\cone(\s{U})&\coloneq\left\{r\gamma\colon\gamma\in\s{U},\;r\in\spa{R}_+\right\}\subseteq\spa{U},
\end{align}
which contains all nonnegative real multiples of the elements in $\s{U}$.  The \emph{dual cone} of $\s{U}$ is defined as
\begin{align}
	\label{eq:dual-cone}
	\cone^*(\s{U})&\coloneq\left\{\alpha\in\spa{U}\colon\left\langle\alpha,\gamma\right\rangle\geq0\;\;\forall\gamma\in\s{U}\right\},
\end{align}
which contains all elements whose inner product with every element in $\s{U}$ is nonnegative.

\subsection{Conditional min-entropy: static and dynamic}
\label{sec:entropy-quantum}

An entropy is a measure of uncertainty, quantifying an observer's lack of knowledge about a physical system.  A conditional entropy quantifies such lack of knowledge about one system (the \emph{concerned} system) given that the observer has operational access to another system (the \emph{conditioned} system).  Due to the potential correlation between the two systems, the observer's operation on the conditioned system may add to their prior knowledge about the concerned system, thus resulting in a reduced lack of knowledge about the concerned system.

Entropies and conditional entropies are very inclusive concepts, as they can be characterized as any functions subject to a few information-theoretic axioms~\cite{Gour-2021a, Brandsen-2021a, Gour-2024a, Gour-2024b}.  Our discussions focus on one particular entropy function (as well as its conditional variant), namely the quantum (conditional) min-entropy.  As shown in a recent work~\cite{Gour-2024a}, the quantum (conditional) min-entropy can be portrayed as the most ``conservative'' way of quantifying (conditional) uncertainty, in the sense that it gives the smallest value among all plausible entropic functions under certain axioms.

Given a state $\rho\in\q{D}^\BI$, the \emph{(quantum) min-entropy} of the system $\BI$ is defined as~\cite{Renner-2005a}
\begin{align}
\label{eq:QME}
	H_{\min}(\BI)_\rho&\coloneq-\log\min_{\scriptsize\left\{\begin{array}{c}
		r\in\spa{R}\colon \\
		r\spec{I}^\BI-\rho^\BI\in\spa{H}_+^\BI
	\end{array}\right\}}r.
\end{align}
Given a bipartite state $\rho\in\q{D}^{\AI\BI}$, the \emph{(quantum) conditional min-entropy} of the system $\BI$ conditioned on $\AI$ is defined as~\cite{Renner-2005a}
\begin{align}
\label{eq:QCME}
	H_{\min}(\BI|\AI)_\rho&\coloneq-\log\min_{\scriptsize\left\{\begin{array}{c}
		\gamma\in\spa{H}^\AI\colon \\
		\gamma^\AI\otimes\spec{I}^\BI-\rho^{\AI\BI}\in\spa{H}_+^{\AI\BI}
	\end{array}\right\}}\tr\left[\gamma\right].
\end{align}
The conditional min-entropy reduces to the min-entropy if the conditioned system $\AI$ and the concerned system $\BI$ are decoupled, and this is consistent with the intuition that observing $\AI$ cannot provide any information about $\BI$ if the two systems are independent.  Entropic properties possessed by the min- and conditional min-entropies include monotonicity~\cite{Gour-2019a, Gour-2021a, Brandsen-2021a}, additivity, and others~\cite{Renner-2005a}.  Besides, the conditional min-entropy has an operational interpretation as a guessing probability when the concerned system $\BI$ is classical~\cite{Konig-2009a}.  We will elaborate on these properties in Secs.~\ref{sec:entropy-properties} and \ref{sec:entropy-operational} as we examine resource-theoretic generalizations of these entropies.

The definitions of min- and conditional min-entropies can be extended into dynamic systems as well~\cite{Gour-2019a, Gour-2021b, Gour-2024b}, quantifying the noisiness of dynamic processes.  Given a channel $\Lambda\in\q{L}^\B$, the \emph{extended (quantum) min-entropy} of the dynamic system $\B$ is defined as~\cite{Gour-2019a}
\begin{align}
\label{eq:EQME}
	H_{\min}(\B)_\Lambda&\coloneq H_{\min}(\BI|\BO)_{\frac{1}{d_\BO}J_\Lambda}.
\end{align}
Given a bipartite channel $\Lambda\in\q{L}^{\A\B}$, the \emph{extended (quantum) conditional min-entropy} of the dynamic system $\B$ conditioned on $\A$ is defined as~\cite{Gour-2019a}
\begin{align}
\label{eq:EQCME}
	&H_{\min}(\B|\A)_\Lambda\coloneq\log\left(d_\AO d_\BO\right) \notag\\
	&\quad-\log\min_{\scriptsize\left\{\begin{array}{c}
		\gamma\in\spa{H}^{\AO\AI\BO}\colon \\
		\gamma^{\AO\AI\BO}\otimes\spec{I}^\BI-J_\Lambda^{\AO\AI\BO\BI}\in\spa{H}_+^{\AO\AI\BO\BI}, \\
		\gamma^{\AO\BO}=\spec{\pi}^\AO\otimes\gamma^\BO
	\end{array}\right\}}\tr\left[\gamma\right].
\end{align}
The extended entropies generalize the unextended ones in the sense that the former reduce to the latter if defined on a replacement channel.  Also, the conditional-to-unconditional reduction still holds for the extended entropies if the channel $\Lambda\in\q{L}^{\A\B}$ is a product channel between $\A$ and $\B$.  For more discussions about properties of these entropies, see Refs.~\cite{Gour-2019a, Gour-2021b}.

The above entropies are defined in terms of semidefinite programs, which are a special type of convex conic program.  The dual formulations of these entropies are postponed to Secs.~\ref{sec:entropy-static} and \ref{sec:entropy-dynamic}, where we introduce generalizations thereof in more detail.

\subsection{FCME of static systems}
\label{sec:entropy-static}

Recall that an entropy quantifies an observer's lack of knowledge about a system.  Following this interpretation, while the entropy is a property of the system, it depends on the observer as well.  Consequently, for observers with different operational properties, entropies should be defined differently.  In the conventional context, the dependence of entropies on the observer is rarely alluded to because the observer is always tacitly assumed to be a ``quantum'' agent, meaning that the observer's admissible operations are only subject to quantum theory, as reflected by the quantum min-entropies in Sec.~\ref{sec:entropy-quantum} (which becomes clear when their dual formulations are presented).  Now, we generalize the concept of entropies in the sense that, we drop the tacit assumption of the observer being precisely quantum, and we rather consider a general observer whose admissible operations are restricted to a \emph{variable} subset of channels, denoted by $\f{O}\subseteq\q{L}$.

For the subset $\f{O}$ to be operationally reasonable, we naturally assume the following conditions satisfied by $\f{O}$.  First, it contains the identity maps; namely, ``inaction'' is always admissible.  Second, it is closed under composition; namely, successive admissible operations are admissible as a whole.  An extra condition is its closedness and convexity.  As such, we note that the operational restrictions on the observer are precisely described by the closed and convex static QRT $(\f{O},\f{F})$, with the free operations $\f{O}$ specified above and the free states $\f{F}$ reduced from $\f{O}$.  We refer to the entropies defined pertaining to this resource-theoretic observer as \emph{free entropies}.  We remark that our analysis of free entropies subsumes that of quantum entropies, as the latter corresponds to the special case in which $(\f{O},\f{F})$ equals the standard quantum theory $(\q{L},\q{D})$.

In the presence of operational restrictions described by a static QRT $(\f{O},\f{F})$, we propose the following adjusted definition of the min-entropy.

\begin{definition}
\label{def:FME}
Given a substate $\rho\in\q{D}_\leqslant^\BI$, the \textbf{free min-entropy (FME)} of the system $\BI$ is defined as
\begin{align}
\label{eq:FME}
	H_{\min}^\f{F}(\BI)_\rho&\coloneq-\log\min_{\scriptsize\left\{\begin{array}{c}
		r\in\spa{R}\colon \\
		r\spec{I}^\BI-\rho^\BI\in\cone^*(\f{F}^\BI)
	\end{array}\right\}}r,
\end{align}
where $\cone^*(\f{F}^\BI)$ is the dual cone of the set of free states in $\BI$.
\end{definition}

The FME is defined in terms of a convex conic program.  The quantum min-entropy defined in Eq.~\eqref{eq:QME} is a special case of the FME for when $\f{F}=\q{D}$, i.e., when all states are free states.  In general, due to the relation $\cone(\f{F}^\BI)\subseteq\spa{H}_+^\BI\subseteq\cone^*(\f{F}^\BI)$, the feasible region of the program on the right-hand side of Eq.~\eqref{eq:FME} has at least one interior point, and hence by Slater's condition, strong duality holds~\cite{Boyd-2004a}.  As a result, the FME can be equivalently expressed in its dual form as follows (see the derivation in Appendix~\ref{app:FME-dual}):
\begin{align}
	\label{eq:FME-dual}
	2^{-H_{\min}^\f{F}(\BI)_\rho}&=\max_{\tau\in\f{F}^\BI}\tr\left[\tau\rho\right].
\end{align}

Analogously, we define the resource-theoretic generalization of the conditional min-entropy as follows (also see Ref.~\cite{George-2024a}).

\begin{definition}
\label{def:FCME}
Given a bipartite substate $\rho\in\q{D}_\leqslant^{\AI\BI}$, the \textbf{free conditional min-entropy} of the system $\BI$ conditioned on $\AI$ is defined as
\begin{align}
\label{eq:FCME}
	&H_{\min}^\f{O}(\BI|\AI)_\rho \notag\\
	&\quad\coloneq-\log\min_{\scriptsize\left\{\begin{array}{c}
		\gamma\in\spa{H}^\AI\colon \\
		\gamma^\AI\otimes\spec{I}^\BI-\rho^{\AI\BI}\in\cone^*(\f{O}_\abb{J}^{\AI\to\BI})
	\end{array}\right\}}\tr\left[\gamma\right],
\end{align}
where $\cone^*(\f{O}_\abb{J}^{\AI\to\BI})$ is the dual cone of the set of Choi operators of free operations.
\end{definition}

The program on the right-hand side of Eq.~\eqref{eq:FCME} is strictly feasible due to the relation $\cone(\f{O}_\abb{J}^{\AI\to\BI})\subseteq\spa{H}_+^{\AI\BI}\subseteq\cone^*(\f{O}_\abb{J}^{\AI\to\BI})$, and thus by Slater's condition, strong duality holds.  The dual form of the FCME is given by (see the derivation in Appendix~\ref{app:FCME-dual})
\begin{align}
	&2^{-H_{\min}^\f{O}(\BI|\AI)_\rho} \notag\\
	&\quad=\max_{\Psi\in\f{O}^{\AI\to\BI}}\tr\left[J_\Psi\rho\right] \label{eq:FCME-dual-1}\\
	&\quad=\max_{\Psi\in\f{O}^{\AI\to\BI}}\tr\left[\spec{\phi}_+^{\RBI\BI}\left(\Psi^{\AI\to\RBI}\otimes\id^\BI\right)\left[\rho^{\AI\BI}\right]\right]. \label{eq:FCME-dual-2}
\end{align}
Comparing the dual form of the FCME and that of the quantum conditional min-entropy~\cite[Eq.~(32)]{Konig-2009a}, we can notice that the latter is a special case of the former for when $\f{O}=\q{L}$, i.e., when no operational restrictions are imposed.

We note that the FME and the FCME are defined by replacing the positive semidefinite cones in the definitions of the quantum min- and conditional min-entropies with cones relevant to the underlying QRT (see Definitions~\ref{def:FME} and \ref{def:FCME}).  This makes them fundamentally different from quantities defined by optimizing the quantum min- or conditional min-entropy over a certain set relevant to the QRT, such as those appearing in Refs.~\cite{Gour-2017a,Gour-2018a,Gour-2019a}.

\subsection{FCME of dynamic systems}
\label{sec:entropy-dynamic}

Applying the same brand of resource-theoretic generalization to the extended (conditional) min-entropy, we now define free entropies that measure the uncertainty about a dynamic system from an operationally restricted observer's perspective.  The operational restrictions are described by a dynamic QRT $(\f{S},\f{C})$ with $\f{S}$ the free superoperations and $\f{C}$ the free channels.  As before, we assume that $(\f{S},\f{C})$ is closed and convex.

\begin{definition}
\label{def:EFME}
Given a subchannel $\Lambda\in\q{L}_\leqslant^\B$, the \textbf{extended free min-entropy (EFME)} of the dynamic system $\B$ is defined as
\begin{align}
\label{eq:EFME}
	H_{\min}^\f{C}(\B)_\Lambda&\coloneq H_{\min}^\f{C}(\BI|\BO)_{\frac{1}{d_\BO}J_\Lambda},
\end{align}
where $H_{\min}^\f{C}(\cdot|\cdot)_{(\cdot)}$ follows Definition~\ref{def:FCME} by replacing $\f{O}$ with the set of free channels $\f{C}$.
\end{definition}

The extended quantum min-entropy defined in Eq.~\eqref{eq:EQME} is a special case of the EFME for when $\f{C}=\q{L}$, i.e., when all channels are free channels.  Following Eqs.~\eqref{eq:FCME-dual-1} and \eqref{eq:FCME-dual-2}, the dual form of the EFME is given by
\begin{align}
	d_\BO2^{-H_{\min}^\f{C}(\B)_\Lambda}&=\max_{\Psi\in\f{C}^\B}\tr\left[J_\Psi J_\Lambda\right] \label{eq:EFME-dual-1}\\
	&=\max_{\Psi\in\f{C}^\B}\tr\left[\spec{\phi}_+^{\RBI\BI}\left(\Psi^\RB\otimes\Lambda^\B\right)\left[\spec{\phi}_+^{\RBO\BO}\right]\right]. \label{eq:EFME-dual-2}
\end{align}

Finally, we generalize the extended conditional min-entropy.

\begin{definition}
\label{def:EFCME}
Given a bipartite subchannel $\Lambda\in\q{L}_\leqslant^{\A\B}$, the \textbf{extended free conditional min-entropy (EFCME)} of the dynamic system $\B$ conditioned on $\A$ is defined as
\begin{align}
\label{eq:EFCME}
	&H_{\min}^\f{S}(\B|\A)_\Lambda\coloneq\log\left(d_\AO d_\BO\right) \notag\\
	&\quad-\log\min_{\scriptsize\left\{\begin{array}{c}
		\gamma\in\spa{H}^{\AO\AI\BO}\colon \\
		\gamma^{\AO\AI\BO}\otimes\spec{I}^\BI-J_\Lambda^{\AO\AI\BO\BI}\in\cone^*(\f{S}_\abb{J}^{\A\To\B}), \\
		\gamma^{\AO\BO}=\spec{\pi}^\AO\otimes\gamma^\BO
	\end{array}\right\}}\tr\left[\gamma\right],
\end{align}
where $\cone^*(\f{S}_\abb{J}^{\A\To\B})$ is the dual cone of the set of Choi operators of free superoperations.
\end{definition}

The program on the right-hand side of Eq.~\eqref{eq:EFCME} is strictly feasible due to the relation $\cone(\f{S}_\abb{J}^{\A\To\B})\subseteq\spa{H}_+^{\AO\AI\BO\BI}\subseteq\cone^*(\f{S}_\abb{J}^{\A\To\B})$, and thus strong duality follows.  The dual form of the EFCME is given by (see the derivation in Appendix~\ref{app:EFCME-dual})
\begin{align}
	&d_\BO2^{-H_{\min}^\f{S}(\B|\A)_\Lambda} \notag\\
	&\quad=\max_{\sm{\Theta}\in\f{S}^{\A\To\B}}\tr\left[J_\sm{\Theta}J_\Lambda\right] \label{eq:EFCME-dual-1}\\
	&\quad=\max_{\sm{\Theta}\in\f{S}^{\A\To\B}}\tr\left[\spec{\phi}_+^{\RBI\BI}\left(\sm{\Theta}^{\A\To\RB}\otimes\Id^\B\right)\left\{\Lambda^{\A\B}\right\}\left[\spec{\phi}_+^{\RBO\BO}\right]\right]. \label{eq:EFCME-dual-2}
\end{align}
Comparing the dual form of the EFCME and that of the extended quantum conditional min-entropy~\cite[Eq.~(63)]{Gour-2019a}, we can notice that the latter is a special case of the former for when $\f{S}=\q{S}$, i.e., when all superchannels are free for channel transformations.  An similar formula to Eq.~\eqref{eq:EFCME-dual-1} can also be found in Ref.~\cite[Sec.~III-B]{Jencova-2021a}.

We summarize in Fig.~\ref{fig:entropy} the relationships between the quantum and free min-entropies introduced so far.

\begin{figure}[t]
\[\begin{tikzcd}
&& {\substack{\displaystyle H_{\min}^\f{S}(\B|\A)_\Lambda \\\\ \abb{(EFCME)}}} \\
\\
\\
{\substack{\displaystyle H_{\min}^\f{C}(\B)_\Lambda \\\\ {\abb{(EFME)}}}} && {\substack{\displaystyle H_{\min}(\B|\A)_\Lambda \\\\ \abb{(Ext.~Q.~Cond.~Min-Ent.)}}} && {\substack{\displaystyle H_{\min}^\f{O}(\BI|\AI)_\rho \\\\ \abb{(FCME)}}} \\
\\
\\
{\substack{\displaystyle H_{\min}(\B)_\Lambda \\\\ \abb{(Ext.~Q.~Min-Ent.)}}} && {\substack{\displaystyle H_{\min}^\f{F}(\BI)_\rho \\\\ \abb{(FME)}}} && {\substack{\displaystyle H_{\min}(\BI|\AI)_\rho \\\\ \abb{(Q.~Cond.~Min-Ent.)}}} \\
\\
\\
&& {\substack{\displaystyle H_{\min}(\BI)_\rho \\\\ \abb{(Q.~Min-Ent.)}}}
\arrow["{\f{S}=\q{S}}"{description}, color={rgb,255:red,184;green,46;blue,46}, shorten <=5pt, shorten >=5pt, from=7-3, to=10-3]
\arrow["{\BO=\#}"{description}, color={rgb,255:red,44;green,44;blue,175}, shorten <=5pt, shorten >=5pt, from=7-1, to=10-3]
\arrow["{\AI=\#}"{description}, color={rgb,255:red,35;green,139;blue,35}, shorten <=5pt, shorten >=5pt, from=7-5, to=10-3]
\arrow["{\AO=\AI=\#}"{description, pos=0.7}, color={rgb,255:red,35;green,139;blue,35}, shorten <=5pt, shorten >=5pt, from=4-3, to=7-1]
\arrow["{\AO=\BO=\#}"{description, pos=0.7}, color={rgb,255:red,44;green,44;blue,175}, shorten <=5pt, shorten >=5pt, from=4-3, to=7-5]
\arrow["{\BO=\#}"{description, pos=0.3}, color={rgb,255:red,44;green,44;blue,175}, shorten <=5pt, shorten >=5pt, from=4-1, to=7-3]
\arrow["{\AO=\AI=\#}"{description}, color={rgb,255:red,35;green,139;blue,35}, shorten <=5pt, shorten >=5pt, from=1-3, to=4-1]
\arrow["{\AI=\#}"{description, pos=0.3}, color={rgb,255:red,35;green,139;blue,35}, shorten <=5pt, shorten >=5pt, from=4-5, to=7-3]
\arrow["{\AO=\BO=\#}"{description}, color={rgb,255:red,44;green,44;blue,175}, shorten <=5pt, shorten >=5pt, from=1-3, to=4-5]
\arrow["{\f{S}=\q{S}}"{description}, color={rgb,255:red,184;green,46;blue,46}, shorten <=5pt, shorten >=5pt, from=4-1, to=7-1]
\arrow["{\f{S}=\q{S}}"{description}, color={rgb,255:red,184;green,46;blue,46}, shorten <=5pt, shorten >=5pt, from=4-5, to=7-5]
\arrow["{\f{S}=\q{S}}"{description}, color={rgb,255:red,184;green,46;blue,46}, shorten <=5pt, shorten >=5pt, from=1-3, to=4-3]
\end{tikzcd}\]
\caption{Relationships between the free min-entropies and the quantum min-entropies.  Green arrows represent conditional-to-unconditional reduction.  Blue arrows represent extended-to-unextended reduction.  Red arrows represent free-to-quantum reduction.  We note the parallel structure between this figure and Fig.~\ref{fig:QRT}.}
\label{fig:entropy}
\end{figure}

\subsection{Entropic properties}
\label{sec:entropy-properties}

We prove several properties of the free min-entropies and compare them with the corresponding properties of quantum entropies.  We find similarities in between, justifying the free min-entropies as a plausible way of quantifying uncertainty in operationally restricted scenarios.  In the meantime, we notice several aspects in which the free min-entropies diverge from quantum entropies, showing exotic but interesting behaviours.  Let $(\f{S},\f{C})$ be a closed and convex dynamic QRT whose reduced static QRT is $(\f{O},\f{F})$.

\begin{proposition}
\label{prop:reducibility}
The free min-entropies have the following \textbf{reducibility} properties.
\begin{enumerate}
	\item[(1)] $H_{\min}^\f{F}(\BI)_\rho=H_{\min}^\f{O}(\BI|\#)_\rho$ for any substate $\rho\in\q{D}_\leqslant^\BI$.
	\item[(2)] $H_{\min}^\f{F}(\BI)_\rho=H_{\min}^\f{C}(\#\to\BI)_\Lambda$ for any preparation subchannel $\Lambda[\cdot]\coloneq(\cdot)\rho\in\q{L}_\leqslant^{\#\to\BI}$.
	\item[(3)] $H_{\min}^\f{C}(\B)_\Lambda=H_{\min}^\f{S}(\B|\#\to\#)_\Lambda$ for any subchannel $\Lambda\in\q{L}_\leqslant^\B$.
	\item[(4)] $H_{\min}^\f{O}(\BI|\AI)_\rho=H_{\min}^\f{S}((\#\to\BI)|(\#\to\AI))_\Lambda$ for any preparation subchannel $\Lambda[\cdot]\coloneq(\cdot)\rho\in\q{L}_\leqslant^{\#\to\AI\BI}$.
\end{enumerate}
\end{proposition}

Proposition~\ref{prop:reducibility} formally explicates the relationships in Fig.~\ref{fig:entropy} between the free min-entropies.  It is self-evident by the definitions of free min-entropies and following the top-down framework of resource theories (see Fig.~\ref{fig:QRT}).  Such reducibility implies that any property possessed by the extended free min-entropies is also possessed by the unextended free min-entropies, so that no separate proof for the latter is needed.

Monotonicity under noisy operations is a defining property that qualifies an entropy as a valid uncertainty measure~\cite{Gour-2019a, Brandsen-2021a, Gour-2021a, Gour-2021b}.  The following proposition shows that the free min-entropies satisfy monotonicity within the operational restrictions of the resource theory.

\begin{proposition}
\label{prop:monotonicity}
The free min-entropies have the following \textbf{monotonicity} properties.
\begin{enumerate}
	\item[(1)] $H_{\min}^\f{F}(\BPI)_{\Psi[\rho]}\geq H_{\min}^\f{F}(\BI)_\rho$ for any substate $\rho\in\q{D}_\leqslant^\BI$ and subchannel $\Psi\in\q{L}_\leqslant^{\BI\to\BPI}$ such that $\Psi^\dagger\in\f{O}^{\BPI\to\BI}$.
	\item[(2)] $H_{\min}^\f{O}(\BI|\API)_{(\Psi\otimes\id)[\rho]}\geq H_{\min}^\f{O}(\BI|\AI)_\rho$ for any substate $\rho\in\q{D}_\leqslant^{\AI\BI}$ and free operation $\Psi\in\f{O}^{\AI\to\API}$.
	\item[(3)] $H_{\min}^\f{C}(\BP)_{\sm{\Theta}\{\Lambda\}}\geq H_{\min}^\f{C}(\B)_\Lambda$ for any subchannel $\Lambda\in\q{L}_\leqslant^\B$ and subsuperchannel $\sm{\Theta}\in\q{S}_\leqslant^{\B\To\BP}$ such that $\sm{\Theta}^\dagger\in\f{S}^{\BP\To\B}$.
	\item[(4)] $H_{\min}^\f{S}(\B|\AP)_{(\sm{\Theta}\otimes\Id)\{\Lambda\}}\geq H_{\min}^\f{S}(\B|\A)_\Lambda$ for any subchannel $\Lambda\in\q{L}_\leqslant^{\A\B}$ and free superoperation $\sm{\Theta}\in\f{S}^{\A\To\AP}$.
\end{enumerate}
Here $(\cdot)^\dagger$ denotes adjunction of a map or supermap, as defined in Eq.~\eqref{aeq:adjoint} and further explained in Appendix~\ref{app:transformation}.
\end{proposition}

The proof of Proposition~\ref{prop:monotonicity} is in Appendix~\ref{app:monotoncity}.  Let us provide some intuitions behind these monotonicity properties.  Proposition~\ref{prop:monotonicity}(1) indicates that the FME is nondecreasing under bistochastic operations whose ``time-reversal'' is a free operation.  Here the time-reversal $\Psi^\dagger$ of $\Psi$ should be interpreted as the input-output inversion of $\Psi$ from a temporally backward agent's perspective~\cite{Chiribella-2021a}.  To give a concrete example, the random unitary channel $\Psi\coloneq\sum_{i}p_i\ob{U}_i$ with the unitary channel $\ob{U}_i^\dagger\in\f{O}^{\BPI\to\BI}$ being free for all $i$ is such a channel with a free time-reversal $\Psi^\dagger=\sum_{i}p_i\ob{U}_i^\dagger\in\f{O}^{\BPI\to\BI}$, and so it should not decrease the FME according to Proposition~\ref{prop:monotonicity}(1).  In particular, the example of the random unitary channel is consistent with the intuition that a random mixture of several ``realizably reversible'' operations must not reduce the uncertainty about a system~\cite{Gour-2021a}.  Here each $\ob{U}_i$ is ``realizably reversible'' in the sense that its inverse map is a free operation in the QRT\@.  Next we shift our attention to Proposition~\ref{prop:monotonicity}(2), which asserts that the FCME is nondecreasing under (deterministic) free operations applied to the conditioned system.  Colloquially, this means that letting the conditioned system $\AI$ undergo free processing cannot make the system $\AI$ more informative about the concerned system $\BI$, and this is reasonable in the resource-theoretic scenario.  Proposition~\ref{prop:monotonicity}(3) and (4) are dynamic generalizations of (1) and (2), and they admit analogous intuitive interpretations.

Additivity on decoupled systems is another defining property of entropies~\cite{Gour-2019a, Brandsen-2021a, Gour-2021a, Gour-2021b}.  However, unlike the quantum entropies, the free min-entropies fail to satisfy additivity in general, although they do satisfy subadditivity on decoupled systems.

\begin{proposition}
\label{prop:subadditivity}
If the set of free superoperations $\f{S}$ is closed under tensor product, then the free min-entropies have the following \textbf{subadditivity} properties.
\begin{enumerate}
	\item[(1)] $H_{\min}^\f{F}(\BI\BPI)_{\rho\otimes\tau}\leq H_{\min}^\f{F}(\BI)_\rho+H_{\min}^\f{F}(\BPI)_\tau$ for any two substates $\rho\in\q{D}_\leqslant^\BI$ and $\tau\in\q{D}_\leqslant^\BPI$.
	\item[(2)] $H_{\min}^\f{O}(\BI\BPI|\AI\API)_{\rho\otimes\tau}\leq H_{\min}^\f{O}(\BI|\AI)_\rho+H_{\min}^\f{O}(\BPI|\API)_\tau$ for any two substates $\rho\in\q{D}_\leqslant^{\AI\BI}$ and $\tau\in\q{D}_\leqslant^{\API\BPI}$.
	\item[(3)] $H_{\min}^\f{C}(\B\BP)_{\Lambda\otimes\Psi}\leq H_{\min}^\f{C}(\B)_\Lambda+H_{\min}^\f{C}(\BP)_\Psi$ for any two subchannels $\Lambda\in\q{L}_\leqslant^\B$ and $\Psi\in\q{L}_\leqslant^\BP$.
	\item[(4)] $H_{\min}^\f{S}(\B\BP|\A\AP)_{\Lambda\otimes\Psi}\leq H_{\min}^\f{S}(\B|\A)_\Lambda+H_{\min}^\f{S}(\BP|\AP)_\Psi$ for any two subchannels $\Lambda\in\q{L}_\leqslant^{\A\B}$ and $\Psi\in\q{L}_\leqslant^{\AP\BP}$.
\end{enumerate}
\end{proposition}

The proof of Proposition~\ref{prop:subadditivity} is in Appendix~\ref{app:subadditivity}.  Toy examples of QRTs in which the free min-entropies violate additivity can be explicitly constructed and are omitted here.  In such QRTs, for two systems decoupled, the observer's overall uncertainty about the composite system can be strictly less than their accumulative uncertainty about the individual systems.  Weird as it may seem, we will elaborate on this point later, showing that such an anomaly on decoupled systems is quite where the merit of the free min-entropies lies.

The following proposition provides upper and lower bounds on the free min-entropies, and it continues to demonstrate the exotic behaviours of the free min-entropies on decoupled systems.

\begin{proposition}
\label{prop:bounds}
If the discarding superchannel in $\A$ is a free superoperation, i.e., $\Tr\{\cdot\}\equiv\tr\circ(\cdot)[\spec{\pi}]\in\f{S}^{\A\To(\#\to\#)}$, then the free min-entropies have the following \textbf{bounds}.
\begin{enumerate}
	\item[(1)] $H_{\min}(\BI)_\omega\leq H_{\min}^\f{O}(\BI|\AI)_{\rho\otimes\omega}\leq H_{\min}^\f{F}(\BI)_\omega$ for any state $\rho\in\q{D}^\AI$ and substate $\omega\in\q{D}_\leqslant^\BI$.  The rightmost inequality becomes equality if $\rho\in\f{F}^\AI$.
	\item[(2)] $H_{\min}(\B)_\Omega\leq H_{\min}^\f{S}(\B|\A)_{\Lambda\otimes\Omega}\leq H_{\min}^\f{C}(\B)_\Omega$ for any channel $\Lambda\in\q{L}^\A$ and subchannel $\Omega\in\q{L}_\leqslant^\B$.  The rightmost inequality becomes equality if $\Lambda\in\f{C}^\A$.
\end{enumerate}
If the uniform channel in $\B$ is a free channel, i.e., $\spec{\Pi}[\cdot]\equiv\tr[\cdot]\spec{\pi}\in\f{C}^\B$, then the FME and the EFME have the following \textbf{bounds}.
\begin{enumerate}
	\item[(3)] $0\leq H_{\min}^\f{F}(\BI)_\rho\leq\log d_\BI$ for any state $\rho\in\q{D}^\BI$.  The leftmost inequality becomes equality if $\rho$ is a free pure state, and the rightmost inequality becomes equality if $\rho=\spec{\pi}$.
	\item[(4)] $-\log\min\{d_\BO,d_\BI\}\leq H_{\min}^\f{C}(\B)_\Lambda\leq\log d_\BI$ for any channel $\Lambda\in\q{L}^\B$.  The leftmost inequality becomes equality if $\Lambda$ is a free isometric channel, and the rightmost inequality becomes equality if $\Lambda=\spec{\Pi}$.
\end{enumerate}
\end{proposition}

The proof of Proposition~\ref{prop:bounds} is in Appendix~\ref{app:bounds}.  According to Proposition~\ref{prop:bounds}(1), the FME is bounded from below by the quantum min-entropy, which is consistent with the intuition that operational restrictions weaken the observer's ability to comprehend information and thus entails a severer lack of knowledge.  It is helpful to think of $H_{\min}(\BI)_\omega$ as the ``intrinsic'' degree of uncertainty about $\BI$ and $H_{\min}^\f{F}(\BI)_\omega-H_{\min}(\BI)_\omega$ as the amount of information about $\BI$ that is in principle available but beyond the observer's restricted comprehension.  For example, from a classical observer's point of view (in which case the free states are incoherent states, i.e., states that are diagonal under the computational basis), the maximally coherent qubit $\op{+}{+}\coloneq\frac{1}{2}(\ket{0}+\ket{1})(\bra{0}+\bra{1})$ and the uniform qubit $\spec{\pi}\coloneq\frac{1}{2}(\op{0}{0}+\op{1}{1})$ have the same value of FME, not because they possess the same amount of intrinsic uncertainty, but because they \emph{appear} equally uncertain to the observer who is  incapable of comprehending the additional, ``coherent'' information stored in the maximally coherent qubit $\op{+}{+}$.  Apart from the above, Proposition~\ref{prop:bounds}(1) also manifests how the FCME behaves differently from the quantum conditional min-entropy when the conditioned system $\AI$ and the concerned system $\BI$ are decoupled.  For such decoupled systems, the FCME is bounded from above by the FME, and this bound is reached if $\AI$ has no resource.  We will show in Sec.~\ref{sec:convertibility-deterministic} that if $\AI$ is in a resourceful state, one can always find a state $\omega$ in $\BI$ such that the rightmost inequality in (1) is strict.  This implies that when access to resourceful operations is denied, access to a resourceful state in $\AI$ may still be advantageous to comprehending information about $\BI$, despite $\AI$ and $\BI$ being independent.  An operational interpretation of this is that the resource in $\AI$ can be consumed to simulate resourceful operations that enable the observer to reduce the uncertainty about $\BI$.  On the other hand, as indicated by the leftmost inequality in (1), mining static resource from an independent system in this way can at best overcome the adverse effect of restricted comprehension, but it can never surmount the intrinsic uncertainty about $\BI$.  Proposition~\ref{prop:bounds}(2) is the dynamic generalization of (1), and the above discussions apply analogously.

\subsection{Operational interpretations as guessing probabilities}
\label{sec:entropy-operational}

Finding a simple operational interpretation for a mathematical quantity is a crucial step towards establishing its physical significance.  We demonstrate the operational interpretations of the FCME and the EFCME as guessing probabilities when the concerned system is classical.  Such interpretations have been discovered for the quantum conditional min-entropy~\cite{Konig-2009a} and its extended version~\cite{Gour-2019a}, and they are regarded as emblematic features of conditional min-entropies.

We consider the task of \emph{state discrimination} within a given ensemble.  Representing a stochastic quantum source, a \emph{state ensemble} in a system $\AI$ is a collection of substates $\ens{\varrho}\equiv\{\rho_{x_1}\}_{x_1}$ with $\rho_{x_1}\in\q{D}_\leqslant^\AI$ for all $x_1$ and satisfying $\sum_{x_1}\tr[\rho_{x_1}]=1$.  An alternative representation of the ensemble $\ens{\varrho}$ is in terms of a qc state in $\AI\XI$ (with $\XI$ a classical system) as follows:
\begin{align}
\label{eq:ensemble}
	\rho^{\AI\XI}&\equiv\sum_{x_1}\rho_{x_1}^\AI\otimes\op{x_1}{x_1}^\XI.
\end{align}
In a state discrimination task, an observer Alice is given a random state from the ensemble $\ens{\varrho}$ and wants to distinguish which state she actually gets, i.e., to guess the actual label $x_1$.  The most general strategy Alice can adopt is to measure the state she holds according to a POVM $\povm{M}\equiv\{M_{x_1'}\}_{x_1'}$ and submit the measurement outcome $x_1'$ as her guess at $x_1$.  If her operational restrictions are described by a closed and convex static QRT $(\f{O},\f{F})$, then her maximum probability of making a correct guess is given by
\begin{align}
\label{eq:probability-static}
	P_\abb{guess}^\f{O}(\ens{\varrho})&\coloneq\max_{\povm{M}\in\f{O}_\abb{M}^\AI}\sum_{x_1}\tr\left[M_{x_1}\rho_{x_1}\right],
\end{align}
where the maximum is over all free measurements on $\AI$.

\begin{figure}[t]
\centering
\includegraphics[scale=0.15]{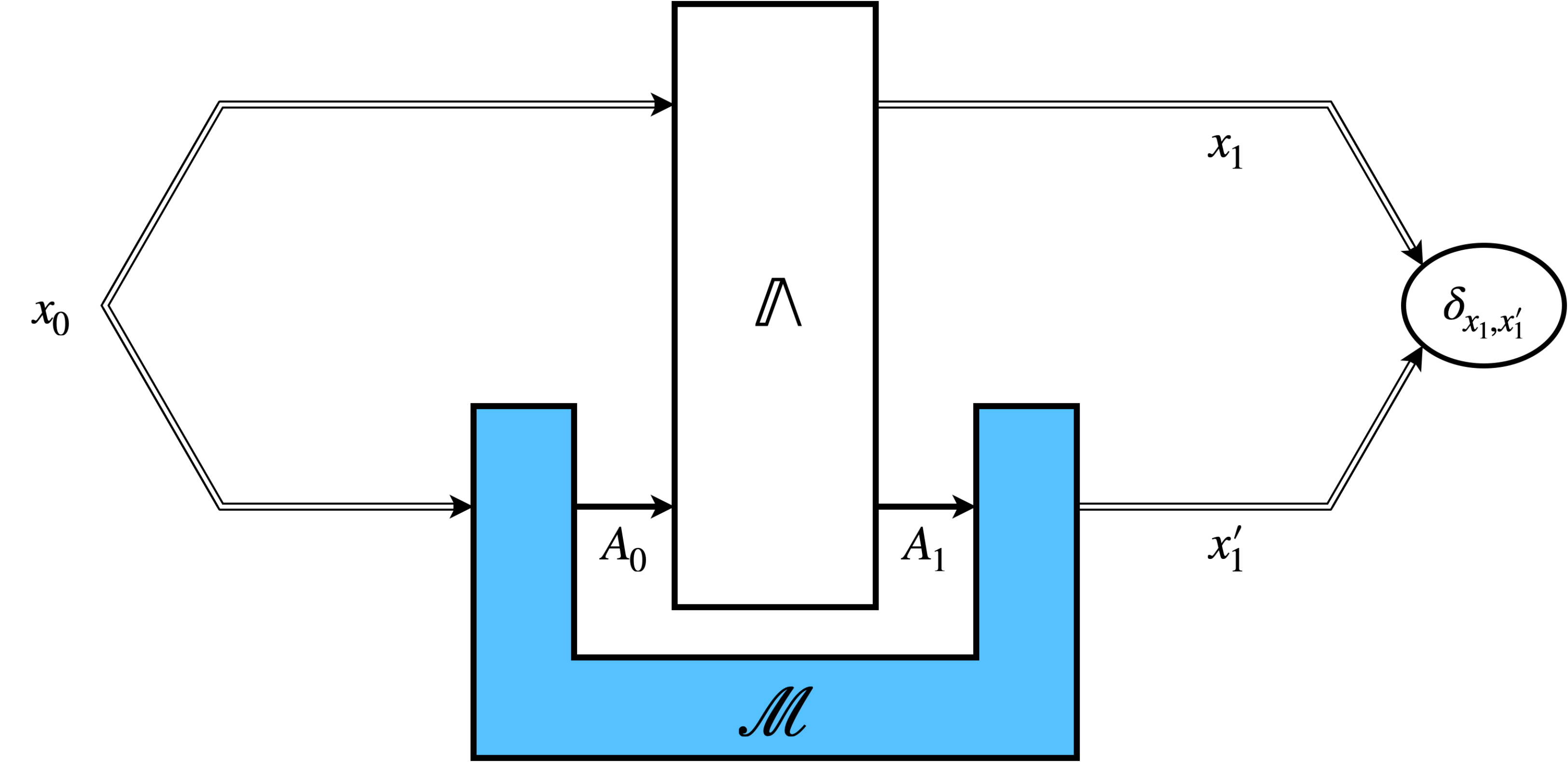}
\caption{Subchannel discrimination within a multi-instrument $\povm{\Lambda}\equiv\{\Lambda_{x_1|x_0}\}_{x_0,x_1}$.  Solid arrows stand for quantum systems and hollow arrows for classical systems.  Alice's goal is to make a correct guess at the actual instrument outcome $x_1$ given the actual value of a randomly generated control signal $x_0$.  As indicated by the blue region, her strategy is described by a free supermeasurement $\spovm{M}\equiv\{\mu_{x_1'|x_0}\}_{x_0,x_1'}$ applied to the quantum part of $\povm{\Lambda}$.  She submits the supermeasurement outcome $x_1'$ as her guess at $x_1$.  She succeeds whenever $x_1'=x_1$.}
\label{fig:SD}
\end{figure}

Before relating the guessing probability $P_\abb{guess}^\f{O}(\ens{\varrho})$ to the FCME, we consider another task which generalizes state discrimination, called \emph{subchannel discrimination}.  In this task, Alice bets on a quantum instrument device that has a classical control.  Such a device is represented by a \emph{multi-instrument}, which in a dynamic system $\A$ is a collection of subchannels $\povm{\Lambda}\equiv\{\Lambda_{x_1|x_0}\}_{x_0,x_1}$ with $\Lambda_{x_1|x_0}\in\q{L}_\leqslant^\A$ for all $x_0,x_1$ and $\sum_{x_1}\Lambda_{x_1|x_0}$ being TP for all $x_0$.  The device functions by executing the quantum instrument $\{\Lambda_{x_1|x_0}\}_{x_1}$ conditioned on a control signal $x_0$.  An alternative representation of the multi-instrument $\povm{\Lambda}$ is in terms of a qc-to-qc channel in $\A\X$ (with $\X\equiv\XO\to\XI$ a dynamic classical system) as follows:
\begin{align}
	\label{eq:multi}
	\Lambda^{\A\X}\left[\cdot\right]&\equiv\sum_{x_0}\Lambda_{x_1|x_0}^\A\left[\left(\spec{I}^\AO\otimes\bra{x_0}^\XO\right)\left(\cdot\right)^{\AO\XO}\left(\spec{I}^\AO\otimes\ket{x_0}^\XO\right)\right] \notag\\
	&\quad\otimes\op{x_1}{x_1}^\XI.
\end{align}
In a subchannel discrimination task, Alice holds a multi-instrument $\povm{\Lambda}$ but can only interact with its quantum input and quantum output.  Her goal is to guess the outcome $x_1$ of the multi-instrument after being revealed that a control signal $x_0$ has been fed into the device.  Since the quantum part of the multi-instrument can be manipulated with appropriate pre-, post-, and parallel processing, Alice's most general strategy is described by a supermeasurement $\spovm{M}\equiv\{\mu_{x_1'|x_0}\}_{x_0,x_1'}$ applied to the dynamic system $\A$ whose outcome $x_1'$ is submitted as her guess at $x_1$.  We show in Fig.~\ref{fig:SD} an illustration of her strategy in the task.  If Alice's operational restrictions are described by a closed and convex dynamic QRT $(\f{S},\f{C})$, then by the superBorn rule in Eq.~\eqref{eq:superborn}, on a uniformly randomly generated control signal $x_0$, her maximum average probability of making a correct guess at $x_1$ is given by
\begin{align}
\label{eq:probability-dynamic}
	P_\abb{guess}^\f{S}(\povm{\Lambda})&\coloneq\frac{1}{d_\XO}\max_{\spovm{M}\in\f{S}_\abb{M}^\A}\sum_{x_0,x_1}\tr\left[\mu_{x_1|x_0}J_{\Lambda_{x_1|x_0}}\right],
\end{align}
where the maximum is over all free supermeasurements on $\A$.

The following proposition establishes the operational interpretations of the FCME and the EFCME as Alice's guessing probabilities in the state discrimination and subchannel discrimination tasks.

\begin{proposition}
\label{prop:operational}
The FCME and the EFCME can be interpreted as the following \textbf{guessing probabilities}.
\begin{enumerate}
	\item[(1)] $H_{\min}^\f{O}(\XI|\AI)_\rho=-\log P_\abb{guess}^\f{O}(\ens{\varrho})$ for any state ensemble $\ens{\varrho}\equiv\{\rho_{x_1}\}_{x_1}$ in $\AI$, where $\rho\in\q{D}^{\AI\XI}$ is defined in Eq.~\eqref{eq:ensemble}.
	\item[(2)] $H_{\min}^\f{S}(\X|\A)_\Lambda=-\log P_\abb{guess}^\f{S}(\povm{\Lambda})$ for any multi-instrument $\povm{\Lambda}\equiv\{\Lambda_{x_1|x_0}\}_{x_0,x_1}$ in $\A$, where $\Lambda\in\q{L}^{\A\X}$ is defined in Eq.~\eqref{eq:multi}.
\end{enumerate}
\end{proposition}

The proof of Proposition~\ref{prop:operational} is in Appendix~\ref{app:operational}.

\section{Resource Characterization with the FCME}
\label{sec:characterization-entropy}

In this section, we apply the free min-entropies to an information-theoretic characterization of general closed and convex QRTs.  Our characterization consists of three parts and is mainly discussed in the dynamic setting.  First, we derive a complete set of entropic conditions for deciding deterministic convertibility between channels via free superoperations.  Second, we generalize the convertibility conditions from the deterministic sense to the probabilistic sense.  Third, we provide an information-theoretic interpretation for the resource global robustness of channels as a mutual-information-like quantity induced by the free min-entropies.  Due to the top-down framework of resource theories, our results apply to general closed and convex static QRTs as well.  Essentially, our characterization reveals a new angle underpinning the inseparable relationship between resource theories and quantum information theory.

\subsection{Deterministic free convertibility}
\label{sec:convertibility-deterministic}

It is of fundamental concern in any resource theory to develop a systematic approach for deciding whether a give object is convertible to another via (deterministic) free transformations.  Numerous works have been devoted to characterizing deterministic free convertibility in terms of complete sets of resource monotones in different types of QRTs~\cite{Chefles-2009a, Buscemi-2012a, Buscemi-2014a, Jencova-2016a, Buscemi-2016a, Gour-2018a,  Rosset-2018a, Gour-2019a, Skrzypczyk-2019a, Skrzypczyk-2019b, Buscemi-2020a, Lipka-Bartosik-2020a, Jencova-2021a, Lipka-Bartosik-2021a, Ji-2024a}.  Recently, such complete monotones have been developed in general (closed and convex) resource theories of states and measurements beyond quantum~\cite{Takagi-2019b} and in general (closed and convex) dynamic QRTs~\cite{Gour-2020a, Jencova-2021a}.  In the dynamic setting, however, the physical significance of the monotones in Refs.~\cite{Gour-2020a, Jencova-2021a} was unclear.  In what follows, we close this gap by proving that these monotones are essentially entropic conditions in terms of the \emph{free} min-entropies.  We also derive simplifications of these monotones.

Let $(\f{S},\f{C})$ be a closed and convex dynamic QRT\@.  We say that a POVM $\povm{N}\equiv\{N_n\}_{n\in\idx{N}}$ on a system $\BO$ is \emph{informationally complete} whenever $\spn(\{N_n\}_n)=\spa{H}^\BO$.

\begin{theorem}
\label{thm:convertibility-deterministic}
Let $\Lambda\in\q{L}^\A$ and $\Psi\in\q{L}^\B$ be two channels.  Let $\povm{N}\equiv\{N_n\}_{n\in\idx{N}}$ be an arbitrary informationally complete POVM on $\BO$ with linearly independent POVM elements.  Then the following statements are equivalent.
\begin{enumerate}
	\item[(1)] There exists a free superoperation $\sm{\Theta}\in\f{S}^{\A\To\B}$ such that $\sm{\Theta}\{\Lambda\}=\Psi$.
	\item[(2)] For every channel $\Omega\in\q{L}^\B$,
	\begin{align}
		\label{eq:convertibility-deterministic}
		H_{\min}^\f{S}(\RB|\A)_{\Lambda^\A\otimes\Omega^\RB}&\leq H_{\min}^\f{S}(\RB|\B)_{\Psi^\B\otimes\Omega^\RB}.
	\end{align}
	\item[(3)] Equation~\eqref{eq:convertibility-deterministic} holds for every measure-and-prepare channel $\Omega\in\q{L}^\B$ of the form
	\begin{align}
		\label{eq:convertibility-measure-and-prepare}
		\Omega\left[\cdot\right]&=\sum_{n}\tr\left[N_n\left(\cdot\right)\right]\omega_n,
	\end{align}
	where $\omega_n\in\q{D}^\BI$ is a variable state for all $n$.
	\item[(4)] It holds that $g(\Lambda,\Psi)\geq1$, where
	\begin{subequations}
	\label{eq:convertibility-program}
	\begin{align}
		g(\Lambda,\Psi)&\coloneq\max r \\
		\textnormal{s.t.:}&\quad r\in\spa{R}, \\
		&\quad\alpha\in\cone(\f{S}_\abb{J}^{\A\To\B}), \\
		&\quad\tr_{\AO\AI\BO}\left[\alpha^{\AO\AI\BO\BI}\left(J_\Lambda^{\AO\AI}\otimes N_n^\BO\otimes\spec{I}^\BI\right)\right] \notag\\
		&\quad\quad\geq r\tr_\BO\left[(J_\Psi^\top)^{\BO\BI}\left(N_n^\BO\otimes\spec{I}^\BI\right)\right]\quad\forall n\in\idx{N}, \label{eq:convertibility-program-1}\\
		&\quad\alpha^{\AI\BO}=\spec{I}^{\AI\BO}, \label{eq:convertibility-program-2}\\
		&\quad\alpha^{\AO\AI\BO}=\alpha^{\AO\BO}\otimes\spec{\pi}^\AI. \label{eq:convertibility-program-3}
	\end{align}
	\end{subequations}
\end{enumerate}
\end{theorem}

The proof of Theorem~\ref{thm:convertibility-deterministic} is in Appendix~\ref{app:convertibility-deterministic}.

The set of entropic conditions in Theorem~\ref{thm:convertibility-deterministic}(2) is mathematically equivalent to the complete set of resource monotones provided in Ref.~\cite[Theorem~4]{Gour-2020a} and slightly strengthens that in Ref.~\cite[Theorem~3, $\varepsilon=0$]{Jencova-2021a}.  And yet, the information-theoretic interpretation of these known monotones is not evident until they are recast, in Theorem~\ref{thm:convertibility-deterministic}(2), in terms of the extended free conditional min-entropy (see Definition~\ref{def:EFCME}).  The equivalence between Theorem~\ref{thm:convertibility-deterministic}(1) and (2) indicates that a channel $\Lambda$ is freely convertible to another channel $\Psi$ if and only if $\Lambda$ always discloses at least as much information as $\Psi$ does about an independent dynamic system $\RB$ (a replica of $\B$), regardless of the channel process that $\RB$ bears.  Note that this is yet another demonstration, as well as an application, of the anomaly of the free min-entropies on decoupled systems.

A technical contribution of Theorem~\ref{thm:convertibility-deterministic} beyond Refs.~\cite{Gour-2020a, Jencova-2021a} is the simplification from (2) to (3), through which the size of the set of convertibility conditions is significantly reduced.  This simplification is inspired by Ref.~\cite{Gour-2018a}, utilizing the fact that the set of measure-and-prepare channels spans the entire space of TP linear maps.  Furthermore, Theorem~\ref{thm:convertibility-deterministic}(4) translates the infinitude of entropic conditions in (2) or (3) into a single convex conic optimization problem, which is solvable in principle given the mathematical representation of the channels $\Lambda$ and $\Psi$ and of the QRT $(\f{S},\f{C})$, and which can be efficiently solved (with respect to system dimensionalities) if so is the membership problem of $\f{S}$.

The following corollary is a reduction of Theorem~\ref{thm:convertibility-deterministic} to the static setting.  It is equivalent to the observation made around Ref.~\cite[Eq.~(120)]{Takagi-2019b}, but now phrased in terms of the free conditional min-entropy (see Definition~\ref{def:FCME}).  Let $(\f{O},\f{F})$ be a closed and convex static QRT\@.

\begin{corollary}
\label{cor:convertibility-deterministic}
Let $\rho\in\q{D}^\AI$ and $\tau\in\q{D}^\BI$ be two states.  Then there exists a free operation $\Psi\in\f{O}^{\AI\to\BI}$ such that $\Psi[\rho]=\tau$ if and only if
\begin{align}
	H_{\min}^\f{O}(\RBI|\AI)_{\rho^\AI\otimes\omega^\RBI}&\leq H_{\min}^\f{O}(\RBI|\BI)_{\tau^\BI\otimes\omega^\RBI}
\end{align}
for every state $\omega\in\q{D}^\BI$.
\end{corollary}

\subsection{Probabilistic free convertibility}
\label{sec:convertibility-probabilistic}

Apart from deterministic free convertibility, probabilistic convertibility between objects using free transformations is also a topic of theoretical and practical concern.  Recently, probabilistic convertibility between states via resource-nongenerating CP maps has been investigated in Refs.~\cite{Regula-2022a, Regula-2022b}, but the suggested criteria do not apply to conversion tasks with a specified success probability.  The study of probabilistic convertibility between channels or subchannels has also been incomplete, though some resource manipulation tasks such as resource distillation have been explored in related settings~\cite{Regula-2021b}.  We complement these prior works with a complete set of conditions for convertibility between subchannels via probabilistic free superoperations, expressed in terms of the extended free conditional min-entropy.  It is worth noting that our definition of probabilistic free transformations follows Definitions~\ref{def:probabilistic-static} and \ref{def:probabilistic-dynamic} and is thus conceptually different from the resource-nongenerating probabilistic transformations considered in Refs.~\cite{Regula-2021b, Regula-2022a, Regula-2022b}.  Let $(\f{S},\f{C})$ be a closed and convex dynamic QRT\@.

\begin{theorem}
\label{thm:convertibility-probabilistic}
Let $\Lambda_0\in\q{L}_\leqslant^\A$ and $\Psi_0\in\q{L}_\leqslant^\B$ be two subchannels.  Let $\Lambda_1\in\q{L}_\leqslant^\A$ and $\Psi_1\in\q{L}_\leqslant^\B$ be two arbitrary subchannels such that $\Lambda_0+\Lambda_1\in\q{L}^\A$ and $\Psi_0+\Psi_1\in\q{L}^\B$.  Then there exists a probabilistic free superoperation $\sm{\Theta}_0\in\f{S}_\leqslant^{\A\To\B}$ such that $\sm{\Theta}_0\{\Lambda_0\}=\Psi_0$ if and only if there exists a classical system $\XI$ (thereby $\B^\times\equiv\BO\to\BI\XI$) such that
\begin{align}
	\label{eq:convertibility-probabilistic}
	H_{\min}^\f{S}(\RB^\times|\A)_{\Lambda_0^\A\otimes\Omega_0^{\RB^\times}+\Lambda_1^\A\otimes\frac{2}{3}\spec{\Pi}^{\RB^\times}}&\leq H_{\min}^\f{S}(\RB^\times|\B)_{\Psi_0^\B\otimes\Omega_0^{\RB^\times}+\Psi_1^\B\otimes\frac{2}{3}\spec{\Pi}^{\RB^\times}}
\end{align}
for every subchannel $\Omega_0\in\q{L}_\leqslant^{\B^\times}$ of the form
\begin{align}
	\label{eq:convertibility-QC}
	\Omega_0^{\B^\times}\left[\cdot\right]&\coloneq\frac{2}{3}\Gamma_0^\B\left[\cdot\right]\otimes\op{0}{0}^\XI+\frac{1}{3}\spec{\Pi}^\B\left[\cdot\right]\otimes\op{1}{1}^\XI,
\end{align}
where $\Gamma_0\in\q{L}_\leqslant^\B$ is a variable subchannel.
\end{theorem}

The proof of Theorem~\ref{thm:convertibility-probabilistic} is detailed in Appendix~\ref{app:convertibility-probabilistic}, utilizing the fact that $\f{S}_\leqslant^{\A\To\B}$ is closed and convex given the same properties of $\f{S}^{\A\To\B}$.  Note that the only requirement for the subchannels $\Lambda_1$ and $\Psi_1$ in Theorem~\ref{thm:convertibility-probabilistic} is that they supplement $\Lambda_0$ and $\Psi_0$ to compose \emph{some} channels $\Lambda_0+\Lambda_1$ and $\Psi_0+\Psi_1$, respectively.  Given $\Lambda_0$ and $\Psi_0$, eligible choices of $\Lambda_1$ and $\Psi_1$ are not unique, with the following as an example:
\begin{align}
	\Lambda_1\left[\cdot\right]&\coloneq\tr\left[\left(\spec{I}-(J_{\Lambda_0}^\top)^\AO\right)\left(\cdot\right)\right]\spec{\pi}, \\
	\Psi_1\left[\cdot\right]&\coloneq\tr\left[\left(\spec{I}-(J_{\Psi_0}^\top)^\BO\right)\left(\cdot\right)\right]\spec{\pi}.
\end{align}

Since substates are just rescaled states, a probabilistic conversion task in a static QRT is typically specified by a source state, a target state, and a success probability in which the conversion is realized.  The following corollary tailors Theorem~\ref{thm:convertibility-probabilistic} for a closed and convex static QRT $(\f{O},\f{F})$.

\begin{corollary}
\label{cor:convertibility-probabilistic}
Let $\rho\in\q{D}^\AI$ and $\tau\in\q{D}^\BI$ be two states and $p\in[0,1]$ be a probability.  Then there exists a probabilistic free operation $\Psi_0\in\f{O}_\leqslant^{\AI\to\BI}$ such that $\Psi_0[\rho]=p\tau$ if and only if there exists a classical system $\XI$ such that
\begin{align}
	&H_{\min}^\f{O}(\RBI\XI|\AI)_{\rho^\AI\otimes\omega_0^{\RBI\XI}} \notag\\
	&\quad\leq H_{\min}^\f{O}(\RBI\XI|\BI)_{\tau^\BI\otimes\left(p\omega_0^{\RBI\XI}+\frac{2\left(1-p\right)}{3}\spec{\pi}^{\RBI\XI}\right)}
\end{align}
for every substate $\omega_0\in\q{D}_\leqslant^{\RBI\XI}$ of the form
\begin{align}
	\omega_0^{\RBI\XI}&\coloneq\frac{2}{3}\tau_0^\RBI\otimes\op{0}{0}^\XI+\frac{1}{3}\spec{\pi}^\RBI\otimes\op{1}{1}^\XI,
\end{align}
where $\tau_0\in\q{D}_\leqslant^\BI$ is a variable substate.  
\end{corollary}

The reduction from Theorem~\ref{thm:convertibility-probabilistic} to Corollary~\ref{cor:convertibility-probabilistic} is by choosing $\tau_1\coloneq(1-p)\tau_0$ as the supplementary substate for $p\tau_0$.

\subsection{Resource global robustness}
\label{sec:robustness-entropy}

Resource robustness measures are functions that quantify an object's resistance to noise in terms of remaining resourceful, as discussed in Sec.~\ref{sec:monotones}.  One representative of such measures is the \emph{resource global robustness}~\cite{Harrow-2003a, Brandao-2015a}, which quantifies the minimum amount of ``global'' noise (within the object set) that an object can be mixed with before losing its identity as a resourceful object.  Following Eq.~\eqref{eq:robustness}, the resource global robustness of a state $\rho\in\q{D}^\AI$ in a static QRT $(\f{O},\f{F})$ is defined as
\begin{align}
	R_\abb{glob}^\f{F}(\rho)&\coloneq R^{\q{D},\f{F}}(\rho) \\
	&=\min_{\scriptsize\left\{\begin{array}{c}
		r\in\spa{R}_+, \\
		\tau\in\q{D}^\AI\colon \\
		\frac{\rho+r\tau}{1+r}\in\f{F}^\AI
	\end{array}\right\}}r.
\end{align}
The resource global robustness of a channel $\Lambda\in\q{L}^\AI$ in a dynamic QRT $(\f{S},\f{C})$ is defined as
\begin{align}
	R_\abb{glob}^\f{C}(\Lambda)&\coloneq R^{\q{L},\f{C}}(\Lambda) \\
	&=\min_{\scriptsize\left\{\begin{array}{c}
		r\in\spa{R}_+, \\
		\Psi\in\q{L}^\A\colon \\
		\frac{\Lambda+r\Psi}{1+r}\in\f{C}^\A
	\end{array}\right\}}r.
\end{align}
In what follows, we show that the resource global robustness, of either states or channels, has an entropic expression in terms of the free min-entropies, provided the closedness and convexity of the underlying QRT\@.  Remarkably, this expression has a clear meaning from an information-theoretic perspective.

Let $(\f{O},\f{F})$ be a closed and convex static QRT\@.  Having the concepts of FME and FCME established, given a bipartite state $\rho\in\q{D}^{\AI\BI}$, we can naturally define the \emph{free min-mutual information} between the two systems $\AI$ and $\BI$ as
\begin{align}
	\label{eq:mutual-static}
	I_{\min}^\f{O}(\AI;\BI)_\rho&\coloneq H_{\min}^\f{F}(\BI)_\rho-H_{\min}^\f{O}(\BI|\AI)_\rho,
\end{align}
where $H_{\min}^\f{F}(\BI)_\rho$ is defined on the marginal state $\rho^\BI\in\q{D}^\BI$.  Let $(\f{S},\f{C})$ be a closed and convex dynamic QRT\@.  Likewise, given a bipartite channel $\Lambda\in\q{L}^{\A\B}$, we define the \emph{extended free min-mutual information} between the two dynamic systems $\A$ and $\B$ as
\begin{align}
	\label{eq:mutual-dynamic}
	I_{\min}^\f{S}(\A;\B)_\Lambda&\coloneq H_{\min}^\f{C}(\B)_\Lambda-H_{\min}^\f{S}(\B|\A)_\Lambda,
\end{align}
where $H_{\min}^\f{C}(\B)_\Lambda$ is defined on the marginal channel:
\begin{align}
	\Lambda^\B\left[\cdot\right]&\equiv\Tr_\A\left\{\Lambda^{\A\B}\right\}\left[\cdot\right] \\
	&=\tr_\AI\circ\Lambda^{\A\B}\left[\spec{\pi}^\AO\otimes\left(\cdot\right)^\BO\right].
\end{align}
Note that both Eqs.~\eqref{eq:mutual-static} and \eqref{eq:mutual-dynamic} are asymmetric between the two systems involved.

The following theorem establishes a close connection between the resource global robustness of channels and the extended free min-mutual information.

\begin{theorem}
\label{thm:robustness-entropy}
Let $\Lambda\in\q{L}^\A$ be a channel.  Then
\begin{align}
	\label{eq:robustness-entropy}
	\log\left(1+R_\abb{glob}^\f{C}(\Lambda)\right)&=\sup_{\Omega\in\q{L}}I_{\min}^\f{S}(\A;\B)_{\Lambda^\A\otimes\Omega^\B},
\end{align}
where the supremum is over all dynamic systems and all channels therein.
\end{theorem}

The proof of Theorem~\ref{thm:robustness-entropy} is detailed in Appendix~\ref{app:robustness-entropy}, employing the convex conic programming formulation of the resource global robustness.  In the proof, we first establish an inequality between the two sides of Eq.~\eqref{eq:robustness-entropy}, and the inequality is then shown to be equalized by any optimizer of the dual program.  Also, the proof is assisted by a technique employed in Refs.~\cite{Piani-2015a, Lipka-Bartosik-2020a}.

From an information-theoretic perspective, the right-hand side of Eq.~\eqref{eq:robustness-entropy} represents the supremum amount of information that $\Lambda$ can potentially provide about any system $\B$ independent of it.  This quantity shall then be referred to as the ``independent free min-informativeness'' of $\Lambda$, and by Theorem~\ref{thm:robustness-entropy}, it coincides with the logarithmic resource global robustness of $\Lambda$.  Note that since the system $\B$ in Eq.~\eqref{eq:robustness-entropy} is also optimized over, its dimensionality can be unbounded in general.  However, as implied in the proof, the supremum in Eq.~\eqref{eq:robustness-entropy} is actually approachable by a q-to-qc channel (i.e., a quantum instrument) whose classical output system is the only system unbounded.

In the static QRT $(\f{O},\f{F})$, the resource global robustness of states has a similar information-theoretic interpretation.

\begin{corollary}
\label{cor:robustness-entropy}
Let $\rho\in\q{D}^\AI$ be a state.  Then
\begin{align}
	\log\left(1+R_\abb{glob}^\f{F}(\rho)\right)&=\max_{\omega\in\q{D}^\AI}I_{\min}^\f{O}(\AI;\RAI)_{\rho^\AI\otimes\omega^\RAI}.
\end{align}
\end{corollary}

Note that the unboundedness of the other system $\RAI$ is not demanded in the static setting.

We now comment on the connections between our entropic characterization of the resource global robustness and some prior results.  First, due to the equivalence between the logarithmic resource global robustness of a state $\rho\in\q{D}^\AI$ and its resource max-relative entropy~\cite{Datta-2009a}, Corollary~\ref{cor:robustness-entropy} immediately gives that
\begin{align}
\label{eq:max}
	\min_{\sigma\in\f{F}^\AI}D_{\max}(\rho\|\sigma)&=\max_{\omega\in\q{D}^\AI}I_{\min}^\f{O}(\AI;\RAI)_{\rho\otimes\omega},
\end{align}
where
\begin{align}
	D_{\max}(\rho\|\sigma)&\coloneq\log\min_{\scriptsize\left\{\begin{array}{c}
		r\in\spa{R}\colon \\
		r\sigma^\AI-\rho^\AI\in\spa{H}_+^\AI
	\end{array}\right\}}r
\end{align}
is the max-relative entropy between two states $\rho,\sigma\in\q{D}^\AI$~\cite{Datta-2009b}.  This relates two information-theoretically inspired resource monotones with distinct nature.  More generally, a dynamic version of Eq.~\eqref{eq:max} is implied by Theorem~\ref{thm:robustness-entropy}:
\begin{align}
	\min_{\Psi\in\f{C}^\A}D_{\max}(\Lambda\|\Psi)&=\sup_{\Omega\in\q{L}}I_{\min}^\f{S}(\A;\B)_{\Lambda^\A\otimes\Omega^\B},
\end{align}
which follows from the fact that the max-relative entropy between two channels~\cite{Leditzky-2018a} is equal to that between their normalized Choi operators~\cite{Wilde-2020a}.  Second, in general closed and convex QRTs, it has been shown that the resource global robustness of a state or a channel has an operational interpretation as the advantage it provides over free states or free channels in certain discrimination tasks~\cite{Takagi-2019a, Takagi-2019b}.  The information-theoretic interpretation of the resource global robustness, as we provide in Theorem~\ref{thm:robustness-entropy} or Corollary~\ref{cor:robustness-entropy}, thereby complements the known operational interpretation and addresses the final missing piece of the ``triangle of associations,'' priorly conjectured in Ref.~\cite{Skrzypczyk-2019b}, about the potential equivalence between robustness-based measures, operational tasks, and information-theoretic quantities.

\section{Resource Characterization with Operational Tasks}
\label{sec:characterization-tasks}

In this section, we characterize general closed and convex QRTs through new designs of operational tasks.  Compared to the entropy-based method presented in Sec.~\ref{sec:characterization-entropy}, operational tasks often have the advantage of relying on the physical presence, rather than the mathematical representation, of the resources being characterized.  For instance, employing Theorem~\ref{thm:convertibility-deterministic} to test free convertibility between two channels requires access to complete and efficient representation of these channels and of the underlying QRT\@.  But operational tasks do not require such representation; instead, they expect the performer of the task to hold physical instances of these channels and is operationally restricted by the QRT\@.  In other words, operational tasks translate mathematical criteria into experimentally observable measures, and in doing so they bypass drawbacks such as incomplete tomographic information or computational hardness.

In what follows, we propose two different operational tasks and use them to characterize general closed and convex dynamic QRTs.  These tasks share similar features in design with the nontransient guessing games considered in Ref.~\cite{Ji-2024a}.  As discussed in Sec.~\ref{sec:monotones}, each task is specified by a set of variable parameters denoted by $\ob{G}$, and the performer of the task, named Alice, receives a rewarding score $s_i(\ob{G})\in\spa{R}$ if her action leads to a result $i$.  Note that the task itself is defined in a resource-independent manner, whereas we assume that Alice, the performer of the task, has an operational capacity limited by a closed and convex dynamic QRT $(\f{S},\f{C})$.  In addition, we assume that Alice can access a single use of a channel $\Lambda$, which serves as a dynamic resource that she can utilize to enhance her performance in the task.  Then without loss of generality, Alice's maximum expected score for the task is given by
\begin{align}
	S^\f{S}(\Lambda;\ob{G})&=\max_{\sm{\Theta}\in\f{S}}\sum_{i}s_i(\ob{G})p_i(\sm{\Theta}\{\Lambda\};\ob{G}),
\end{align}
where $p_i(\cdot;\ob{G})$ denotes the occurrence probability of a result $i$ given Alice's actual action in the task.  We establish results in the following two regards.  First, we have shown in Eq.~\eqref{eq:monotones-2} that $S^\f{S}(\Lambda;\ob{G})$ is a resource monotone with respect to $\Lambda$, and we here show that $S^\f{S}(\Lambda;\ob{G})$ constitutes a complete set of resource monotones when the parameters $\ob{G}$ vary within a certain scope.  Second, we investigate the operational advantage the resource $\Lambda$ brings, over when this resource is absent, in terms of the ratio
\begin{align}
\frac{S^\f{S}(\Lambda;\ob{G})}{\max_{\Psi\in\f{C}}S^\f{S}(\Psi;\ob{G})}
\end{align}
and show that such advantage is intimately related to various resource robustness measures of $\Lambda$.

\subsection{Operational tasks as free convertibility tests}
\label{sec:convertibility-tasks}

We first introduce the operational tasks and show how they give rise to complete sets of resource monotones.

We motivate the first operational task by considering the following scenario.  Let Alice be given a POVM $\povm{M}\equiv\{M_m\}_{m\in\idx{M}}$.  She wishes to predict the outcome of $\povm{M}$ given that she can use $\povm{M}$ to measure whatever state she wants to.  Generally, her strategy for such outcome prediction is described by a state ensemble $\ens{\varrho}\equiv\{\rho_{m'}\}_{m'\in\idx{M}}$, which means that with probability $\tr[\rho_{m'}]$, she measures the state $\rho_{m'}/\tr[\rho_{m'}]$ according to $\povm{M}$ and predicts the measurement outcome to be $m'$.  Her success probability under this strategy is given by $\sum_{m\in\idx{M}}\tr[M_m\rho_m]$.

Now we generalize the above scenario to a distributed setting.  Let $\povm{M}$ be a distributed POVM, shared between Alice and another party, called the referee (he/him).  We also let Alice and the referee share a maximally entangled state $\spec{\varphi}_+$ beforehand.  Now Alice's goal is to predict the outcome of $\povm{M}$ given that she can feed any state into her input system of $\povm{M}$, also given that the referee will feed his share of $\spec{\varphi}_+$ into his input system of $\povm{M}$.  Formally, we recapitulate this distributed scenario as the following task.

\begin{figure}[t]
\centering
\includegraphics[scale=0.15]{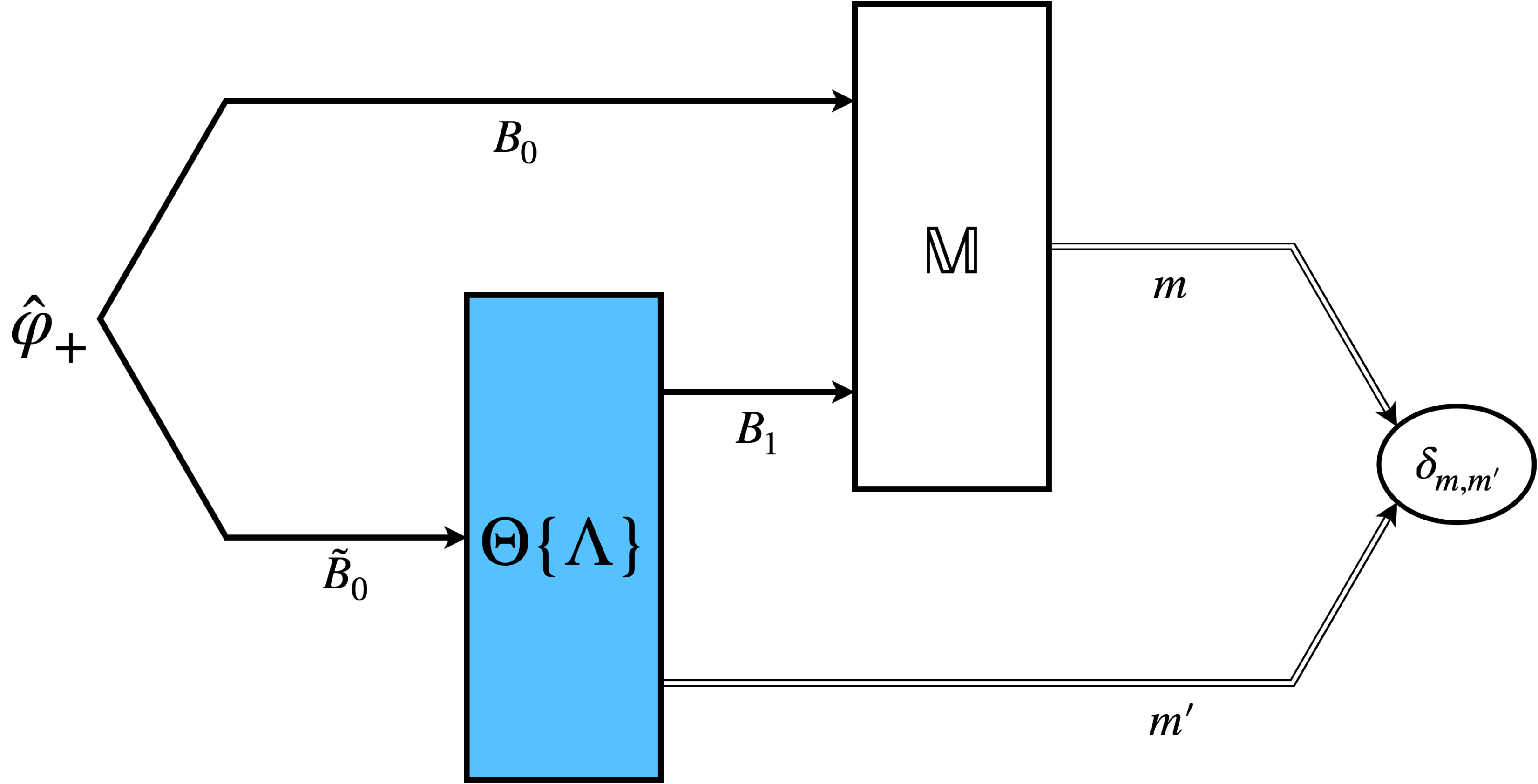}
\caption{Entanglement-assisted outcome prediction about a distributed POVM $\povm{M}\equiv\{M_m\}_m$.  Solid arrows stand for quantum systems and hollow arrows for classical systems.  Alice's goal is to make a correct prediction of the measurement outcome $m$.  As indicated by the blue region, her strategy is described by a free superoperation $\sm{\Theta}$ applied to her channel $\Lambda$ and using $\sm{\Theta}\{\Lambda\}$ for generating the submitted state in $\BI$ and her guess $m'$ at $m$.  She succeeds whenever $m'=m$.}
\label{fig:EOP}
\end{figure}

\begin{task}
\label{def:EOP}
An \textbf{entanglement-assisted outcome prediction (EOP)} task for Alice is specified by a distributed POVM $\povm{M}\equiv\{M_m\}_{m\in\idx{M}}$ between two systems $\BO$ and $\BI$.  It has the following steps.
\begin{enumerate}
	\item Alice and the referee share a maximally entangled state $\spec{\varphi}_+\equiv\frac{1}{d_\BO}\spec{\phi}_+\in\q{D}^{\BO\RBO}$, with the referee's system being $\BO$.
	\item Alice submits a state in $\BI$ and an index $m'\in\idx{M}$ to the referee.
	\item The referee measures $\BO\BI$ according to $\povm{M}$ and obtains an outcome $m\in\idx{M}$.  Alice succeeds whenever $m=m'$. 
\end{enumerate}
\end{task}

As mentioned before, we assume that Alice is operationally restricted by a closed and convex dynamic QRT $(\f{S},\f{C})$ and that she has one-shot access to a channel $\Lambda\in\q{L}^\A$.  This implies that her strategy in the EOP task can in general be described as (i) transforming $\Lambda$ into a q-to-qc channel $\sm{\Theta}\{\Lambda\}$ using a free superoperation $\sm{\Theta}\in\f{S}^{\A\To\B^\times}$, where $\B^\times\equiv\BO\to\BI\XI$, and (ii) feeding her share of $\spec{\varphi}_+$ into $\sm{\Theta}\{\Lambda\}$ to generate the system $\BI$ and the index $m'$ that she needs to submit in Step~2 of Task~\ref{def:EOP}.  We show in Fig.~\ref{fig:EOP} an illustration of both Alice's and the referee's moves in the task.  It follows that Alice's maximum probability of making a correct prediction is given by
\begin{align}
\label{eq:EOP}
	P_\abb{EOP}^\f{S}(\Lambda;\povm{M})&\coloneq\max_{\sm{\Theta}\in\f{S}^{\A\To\B^\times}}\sum_{m}\tr\left[\left(M_m^{\BO\BI}\otimes\op{m}{m}^\XI\right)\right. \notag\\
	&\quad\times\left.\left(\id^\BO\otimes\left(\sm{\Theta}\left\{\Lambda\right\}\right)^{\RBO\to\BI\XI}\right)\left[\spec{\varphi}_+^{\BO\RBO}\right]\right].
\end{align}

The following theorem asserts that through variation of the POVM $\povm{M}$, the probability $P_\abb{EOP}^\f{S}(\Lambda;\povm{M})$ forms a complete set of resource monotones with respect to $\Lambda$.  In other words, the task of entanglement-assisted outcome prediction provides a scheme for faithfully testing free convertibility between channels in any closed and convex dynamic QRT\@.

\begin{theorem}
\label{thm:convertibility-EOP}
Let $\Lambda\in\q{L}^\A$ and $\Psi\in\q{L}^\B$ be two channels.  Then there exists a free superoperation $\sm{\Theta}\in\f{S}^{\A\To\B}$ such that $\sm{\Theta}\{\Lambda\}=\Psi$ if and only if $P_\abb{EOP}^\f{S}(\Lambda;\povm{M})\geq P_\abb{EOP}^\f{S}(\Psi;\povm{M})$ for every POVM $\povm{M}\equiv\{M_m\}_{m\in\idx{M}}$ on $\BO\BI$.
\end{theorem}

The proof of Theorem~\ref{thm:convertibility-EOP} is in Appendix~\ref{app:convertibility-EOP}.  We comment that in Theorem~\ref{thm:convertibility-EOP}, the outcome set $\idx{M}$ of $\povm{M}$ is also variable, and thus its size can be (countably) unbounded in general.

We now introduce the second operational task, which has a simpler structure compared to the EOP task (Task~\ref{def:EOP}).  As shown in Fig.~\ref{fig:EOP}, implementing the EOP task involves both entanglement distribution and (variable) distributed measurements.  These experimental requirements, which can be challenging in practice, can however be bypassed by new design of tasks.

We consider a scenario of one-shot classical communication as follows.  Let $n\in\idx{N}$ be a classical index generated by a stochastic source.  The index $n$ is encoded according to a state ensemble $\ens{\upsilon}\equiv\{\nu_n\}_{n\in\idx{N}}$, transmitted through a quantum channel $\Lambda$, and decoded using a POVM $\povm{N}\equiv\{N_{n'}\}_{n'\in\idx{N}}$.  Here $\tr[\nu_n]$ is equal to the probability of the index $n$ being generated for each $n\in\idx{N}$, and the outcome $n'$ of $\povm{N}$ is the recovered index from the transmission.  The probability of the transmission being faithful, i.e., of $n'=n$, is given by $\sum_{n\in\idx{N}}\tr[N_n\Lambda[\nu_n]]$.  We recapitulate this task as a parametrized task performed by the holder (Alice) of the quantum channel $\Lambda$.

\begin{figure}[t]
\centering
\includegraphics[scale=0.15]{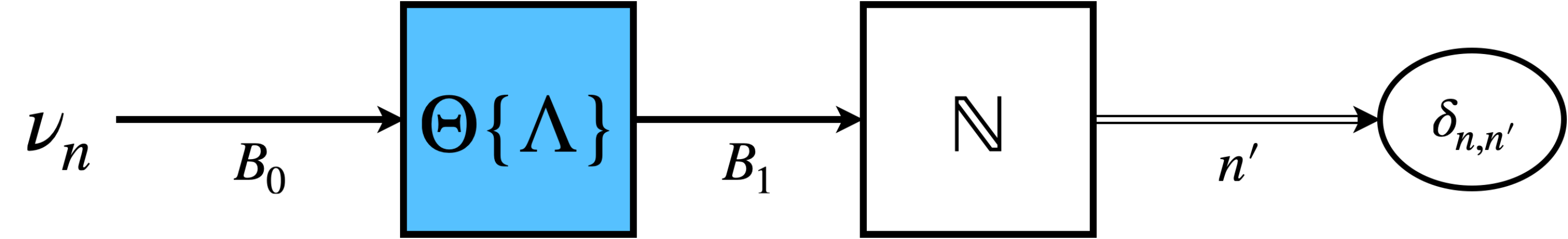}
\caption{One-shot classical communication over an encoding ensemble $\ens{\upsilon}\equiv\{\nu_n\}_n$ and a decoding POVM $\povm{N}\equiv\{N_{n'}\}_{n'}$.  Solid arrows stand for quantum systems and hollow arrows for classical systems.  Alice's goal is to faithfully transmit the classical index $n$ from the left end to the right end.  As indicated by the blue region, her strategy is described by a free superoperation $\sm{\Theta}$ applied to her channel $\Lambda$ and using $\sm{\Theta}\{\Lambda\}$ for data transmission.  She succeeds whenever $n'=n$.}
\label{fig:OCC}
\end{figure}

\begin{task}
\label{def:OCC}
A \textbf{one-shot classical communication (OCC)} task for Alice is specified by a state ensemble $\ens{\upsilon}\equiv\{\nu_n\}_{n\in\idx{N}}$ in a system $\BO$ and a POVM $\povm{N}\equiv\{N_{n'}\}_{n'\in\idx{N}}$ on a system $\BI$.  It has the following steps.
\begin{enumerate}
	\item An index $n\in\idx{N}$ is generated with probability $\tr[\nu_n]$.  The state $\nu_n/\tr[\nu_n]\in\q{D}^\BO$ is given to Alice without letting her know $n$.
	\item Alice submits a state in $\BI$.
	\item The system $\BI$ is measured according to the POVM $\povm{N}$, producing an outcome $n'$.  Alice succeeds whenever $n'=n$.
\end{enumerate}
\end{task}

As before, we assume that Alice is operationally restricted by a closed and convex QRT $(\f{S},\f{C})$ and has one-shot access to a channel $\Lambda\in\q{L}^\A$.  Then her most general strategy for the OCC task is to transform $\Lambda$ with a free superoperation $\sm{\Theta}\in\f{S}^{\A\To\B}$ and use $\sm{\Theta}\{\Lambda\}$ to conduct data transmission in Step~2 of Task~\ref{def:OCC}.  We show in Fig.~\ref{fig:OCC} an illustration of the task and Alice's strategy.  Her maximum probability of conducting a faithful transmission is given by
\begin{align}
\label{eq:OCC}
	P_\abb{OCC}^\f{S}(\Lambda;\ens{\upsilon},\povm{N})&\coloneq\max_{\sm{\Theta}\in\f{S}^{\A\To\B}}\sum_{n}\tr\left[N_n\sm{\Theta}\left\{\Lambda\right\}\left[\nu_n\right]\right].
\end{align}

The following theorem asserts that if the POVM $\povm{N}$ is informationally complete, then through variation of the state ensemble $\ens{\upsilon}$, the probability $P_\abb{OCC}^\f{S}(\Lambda;\ens{\upsilon},\povm{N})$ forms a complete set of resource monotones with respect to $\Lambda$.  In other words, the task of one-shot classical communication can faithfully test free convertibility using a fixed measurement device and a variable source device.

\begin{theorem}
\label{thm:convertibility-OCC}
Let $\Lambda\in\q{L}^\A$ and $\Psi\in\q{L}^\B$ be two channels.  Let $\povm{N}\equiv\{N_{n'}\}_{n'\in\idx{N}}$ be an arbitrary informationally complete POVM on $\BI$.  Then there exists a free superoperation $\sm{\Theta}\in\f{S}^{\A\To\B}$ such that $\sm{\Theta}\{\Lambda\}=\Psi$ if and only if $P_\abb{OCC}^\f{S}(\Lambda;\ens{\upsilon},\povm{N})\geq P_\abb{OCC}^\f{S}(\Psi;\ens{\upsilon},\povm{N})$ for every state ensemble $\ens{\upsilon}\equiv\{\nu_n\}_{n\in\idx{N}}$ in $\BO$.
\end{theorem}

The proof of Theorem~\ref{thm:convertibility-OCC} is in Appendix~\ref{app:convertibility-OCC}.  Theorem~\ref{thm:convertibility-OCC} indicates that a channel's abilities to transmit classical data under different encoding schemes completely reflect its free convertibility properties.

\subsection{Relation to resource robustness measures}
\label{sec:robustness-tasks}

Apart from testing free convertibility between two channels, we can also use these operational tasks to characterize the resourcefulness of a single channel $\Lambda$.  This is done by investigating the extent to which the channel $\Lambda$ can enhance Alice's performance in these tasks.  For example, in the EOP task (Task~\ref{def:EOP}), Alice's maximum probability of correct prediction, with the assistance of $\Lambda$, is given by $P_\abb{EOP}^\f{S}(\Lambda;\povm{M})$, as defined in Eq.~\eqref{eq:EOP}.  Whereas if the assistance of $\Lambda$ is removed, her maximum prediction probability would be
\begin{align}
\label{eq:EOP-free}
	P_\abb{EOP}^\f{C}(\povm{M})&\coloneq\max_{\Psi\in\f{C}}P_\abb{EOP}^\f{S}(\Psi;\povm{M}).
\end{align}
Theorem~\ref{thm:convertibility-EOP} implies that if $\Lambda$ is resourceful, then there exists a POVM $\povm{M}$ such that $P_\abb{EOP}^\f{S}(\Lambda;\povm{M})/P_\abb{EOP}^\f{C}(\povm{M})>1$, namely that $\Lambda$ provides a strict operational advantage in the particular task specified by $\povm{M}$.  The following theorem asserts that this advantage, at its supremum over all POVMs, is exactly quantified by the resource global robustness of $\Lambda$.

\begin{theorem}
\label{thm:robustness-EOP}
Let $\Lambda\in\q{L}^\A$ be a channel.  Then
\begin{align}
	1+R_\abb{glob}^\f{C}(\Lambda)&=\sup_\povm{M}\frac{P_\abb{EOP}^\f{S}(\Lambda;\povm{M})}{P_\abb{EOP}^\f{C}(\povm{M})},
\end{align}
where the supremum is over all POVMs on $\AO\AI$.
\end{theorem}

The proof of Theorem~\ref{thm:robustness-EOP} is detailed in Appendix~\ref{app:robustness-EOP}, following a similar idea to that of Theorem~\ref{thm:robustness-entropy}.

A main difference between Theorem~\ref{thm:robustness-EOP} and the operational characterization of the resource global robustness of channels in Ref.~\cite{Takagi-2019b} (and related results in concrete QRTs, e.g., that in Ref.~\cite{Takagi-2020a}) is that, we here assume Alice to have the operational freedom to manipulate her channel with free superoperations during a task, a possibility not incorporated in Refs.~\cite{Takagi-2019b, Takagi-2020a}.

For the OCC task (Task~\ref{def:OCC}), the operational advantage of a channel $\Lambda$ can be similarly formulated.  Although the advantage in the OCC task is not directly linked with resource robustness measures, it can be characterized using the free min-mutual information measure introduced in Sec.~\ref{sec:robustness-entropy}.  Let $\q{L}_\abb{MP}\subset\q{L}$ denote the set of all measure-and-prepare channels (i.e., entanglement-breaking channels) between arbitrary systems.  Also define
\begin{align}
\label{eq:OCC-free}
	P_\abb{OCC}^\f{C}(\ens{\upsilon},\povm{N})&\coloneq\max_{\Psi\in\f{C}}P_\abb{OCC}^\f{S}(\Psi;\ens{\upsilon},\povm{N}),
\end{align}
which represents Alice's optimal performance without the assistance of $\Lambda$ in the OCC task.

\begin{theorem}
\label{thm:entropy-OCC}
Let $\Lambda\in\q{L}^\A$ be a channel.  Then
\begin{align}
	\sup_{\Omega\in\q{L}_\abb{MP}}I_{\min}^\f{S}(\A;\B)_{\Lambda^\A\otimes\Omega^\B}&=\log\sup_{\ens{\upsilon},\povm{N}}\frac{P_\abb{OCC}^\f{S}(\Lambda;\ens{\upsilon},\povm{N})}{P_\abb{OCC}^\f{C}(\ens{\upsilon},\povm{N})},
\end{align}
where the supremum on the left-hand side is over all dynamic systems and all measure-and-prepare channels therein, and the supremum on the right-hand side is over all pairs of static systems with all state ensembles in the first system and all POVMs on the second system.
\end{theorem}

The proof of Theorem~\ref{thm:entropy-OCC} is in Appendix~\ref{app:entropy-OCC}.

Finally, we consider a modification of the OCC task, which makes it versatile enough to characterize resource robustness measures beyond the resource global robustness.  We show that \emph{every} well-defined resource robustness measure $R^{\s{K},\f{C}}(\Lambda)$ of a channel $\Lambda$ can be interpreted as an operational advantage of $\Lambda$ in a tailored version of the modified task, addressing an open problem raised in Ref.~\cite{Takagi-2019b}.  Here $R^{\s{K},\f{C}}(\Lambda)$ is the resource robustness of $\Lambda\in\q{L}^\A$ against a subset of channels $\s{K}\subseteq\q{L}$ (thereby $\s{K}^\A\equiv\s{K}\cap\q{L}^\A$), defined as
\begin{align}
	R^{\s{K},\f{C}}(\Lambda)&\coloneq\min_{\scriptsize\left\{\begin{array}{c}
		r\in\spa{R}_+, \\
		\Psi\in\s{K}^\A\colon \\
		\frac{\Lambda+r\Psi}{1+r}\in\f{C}^\A
	\end{array}\right\}}r.
\end{align}
The modified task has almost the same steps as the original OCC task, except that in Step~3 of Task~\ref{def:OCC}, instead of recording whether Alice succeeds (i.e., whether $n'=n$), it rewards Alice with a score $s_{n,n'}\in[-1,1]$ based on the index $n$ and Alice's recovery $n'$ thereof.  Under the assumption that Alice is operationally restricted by a closed and convex QRT $(\f{S},\f{C})$ and has one-shot access to a channel $\Lambda\in\q{L}^\A$, her maximum expected score is given by
\begin{align}
\label{eq:OCC-score}
	S_\abb{OCC}^\f{S}(\Lambda;s,\ens{\upsilon},\povm{N})&\coloneq\max_{\sm{\Theta}\in\f{S}^{\A\To\B}}\sum_{n,n'}s_{n,n'}\tr\left[N_{n'}\sm{\Theta}\left\{\Lambda\right\}\left[\nu_n\right]\right].
\end{align}
If Alice's freedom to manipulate her channel $\Lambda$ using free superoperations is deprived, her maximum expected score would be
\begin{align}
\label{eq:OCC-Id}
	S_\abb{OCC}^{\{\Id\}}(\Lambda;s,\ens{\upsilon},\povm{N})&\coloneq\sum_{n,n'}s_{n,n'}\tr\left[N_{n'}\Lambda\left[\nu_n\right]\right].
\end{align}
For any subset of channels $\s{K}\subseteq\q{L}$, we define
\begin{align}
\label{eq:OCC-K}
	\bar{S}_\abb{OCC}^\s{K}(s,\ens{\upsilon},\povm{N})&\coloneq\min_{\Psi\in\s{K}}S_\abb{OCC}^{\{\Id\}}(\Psi;s,\ens{\upsilon},\povm{N}).
\end{align}
Note that the right-hand side of Eq.~\eqref{eq:OCC-K} is a minimization instead of a maximization.  Also define
\begin{align}
	S_\abb{OCC}^\f{C}(s,\ens{\upsilon},\povm{N})&\coloneq\max_{\Psi\in\f{C}}S_\abb{OCC}^{\{\Id\}}(\Psi;s,\ens{\upsilon},\povm{N}) \label{eq:OCC-Id-free-score}\\
	&=\max_{\Psi\in\f{C}}S_\abb{OCC}^\f{S}(\Psi;s,\ens{\upsilon},\povm{N}), \label{eq:OCC-free-score}
\end{align}
where Eq.~\eqref{eq:OCC-free-score} follows from the fact that $\f{C}$ is contractive under any free superoperation.  We say that a state ensemble $\ens{\upsilon}\equiv\{\nu_n\}_{n\in\idx{N}}$ in a system $\AO$ is \emph{tomographically complete} whenever $\spn(\{\nu_n\}_n)=\spa{H}^\AO$.

\begin{theorem}
\label{thm:robustness-OCC}
Let $\s{K}\subseteq\q{L}$ be a subset of channels satisfying the following conditions:
\begin{enumerate}
	\item[(1)] the set $\s{K}^\A$ is closed and convex;
	\item[(2)] the value of $R^{\s{K},\f{C}}(\Psi)$ is bounded for all $\Psi\in\q{L}^\A$.
\end{enumerate}
Let $\Lambda\in\q{L}^\A$ be a channel.  Let $\ens{\upsilon}\equiv\{\nu_n\}_{n\in\idx{N}}$ be an arbitrary tomographically complete state ensemble in $\AO$, and let $\povm{N}\equiv\{N_{n'}\}_{n'\in\idx{N}}$ be an arbitrary informationally complete POVM on $\AI$.  Then
\begin{align}
\label{eq:robustness-OCC-1}
	1+R^{\s{K},\f{C}}(\Lambda)&=\max_{\scriptsize\left\{\begin{array}{c}
		s\in[-1,1]^{\idx{N}\times\idx{N}}\colon \\
		\bar{S}_\abb{OCC}^\s{K}(s,\ens{\upsilon},\povm{N})\geq0
	\end{array}\right\}}\frac{S_\abb{OCC}^{\{\Id\}}(\Lambda;s,\ens{\upsilon},\povm{N})}{S_\abb{OCC}^\f{C}(s,\ens{\upsilon},\povm{N})}.
\end{align}
If $\s{K}$ is contractive under all free superoperations, then
\begin{align}
\label{eq:robustness-OCC-2}
	1+R^{\s{K},\f{C}}(\Lambda)&=\max_{\scriptsize\left\{\begin{array}{c}
		s\in[-1,1]^{\idx{N}\times\idx{N}}\colon \\
		\bar{S}_\abb{OCC}^\s{K}(s,\ens{\upsilon},\povm{N})\geq0
	\end{array}\right\}}\frac{S_\abb{OCC}^\f{S}(\Lambda;s,\ens{\upsilon},\povm{N})}{S_\abb{OCC}^\f{C}(s,\ens{\upsilon},\povm{N})}.
\end{align}
\end{theorem}

The proof of Theorem~\ref{thm:robustness-OCC} is in Appendix~\ref{app:robustness-OCC}.

Our characterization in Theorem~\ref{thm:robustness-OCC}, while applying to a more general class of robustness-based measures, follows a proof idea reminiscent of those in Refs.~\cite{Uola-2020b, Yuan-2021a}.  Apart from its generality, Theorem~\ref{thm:robustness-OCC} differs from Refs.~\cite{Uola-2020b, Yuan-2021a} in several other aspects.  First, in Eq.~\eqref{eq:robustness-OCC-2}, Alice is additionally assumed the operational freedom to apply free superoperations to her channel during a task, a situation beyond the consideration of Refs.~\cite{Uola-2020b, Yuan-2021a}.  Second, the parameter optimizations in Eqs.~\eqref{eq:robustness-OCC-1} and \eqref{eq:robustness-OCC-2} are only with respect to the score function while keeping the source and measurement devices fixed.  This enables an experimentally friendly implementation of robustness estimation: one only needs to optimize the classical postprocessing based on the data from a single set of quantum hardware.

As a corollary of Theorem~\ref{thm:robustness-OCC}, both the \emph{resource free robustness} and the \emph{resource random robustness} can be characterized with the modified task.  The resource free robustness of a channel $\Lambda\in\q{L}^\A$ in $(\f{S},\f{C})$ is defined as
\begin{align}
	R_\abb{free}^\f{C}(\Lambda)&\coloneq R^{\f{C},\f{C}}(\Lambda) \\
	&=\min_{\scriptsize\left\{\begin{array}{c}
		r\in\spa{R}_+, \\
		\Psi\in\f{C}^\A\colon \\
		\frac{\Lambda+r\Psi}{1+r}\in\f{C}^\A
	\end{array}\right\}}r,
\end{align}
and the resource random robustness of $\Lambda$ is defined as
\begin{align}
	R_\abb{rand}^\f{C}(\Lambda)&\coloneq R^{\{\spec{\Pi}\},\f{C}}(\Lambda) \\
	&=\min_{\scriptsize\left\{\begin{array}{c}
		r\in\spa{R}_+\colon \\
		\frac{\Lambda+r\spec{\Pi}}{1+r}\in\f{C}^\A
	\end{array}\right\}}r,
\end{align}
where $\spec{\Pi}\in\q{L}^\A$ is the uniform channel in $\A$.

\begin{corollary}
\label{cor:robustness-OCC}
Let $\Lambda\in\q{L}^\A$ be a channel.  Let $\ens{\upsilon}\equiv\{\nu_n\}_{n\in\idx{N}}$ be an arbitrary tomographically complete state ensemble in $\AO$, and let $\povm{N}\equiv\{N_{n'}\}_{n'\in\idx{N}}$ be an arbitrary informationally complete POVM on $\AI$.  If the value of $R_\abb{free}^\f{C}(\Psi)$ is bounded for all $\Psi\in\q{L}^\A$, then
\begin{align}
	1+R_\abb{free}^\f{C}(\Lambda)&=\max_{\scriptsize\left\{\begin{array}{c}
		s\in[-1,1]^{\idx{N}\times\idx{N}}\colon \\
		\bar{S}_\abb{OCC}^\s{L}(s,\ens{\upsilon},\povm{N})\geq0
	\end{array}\right\}}\frac{S_\abb{OCC}^{\{\Id\}}(\Lambda;s,\ens{\upsilon},\povm{N})}{S_\abb{OCC}^\f{C}(s,\ens{\upsilon},\povm{N})} \label{eq:robustness-OCC-3}\\
	&=\max_{\scriptsize\left\{\begin{array}{c}
		s\in[-1,1]^{\idx{N}\times\idx{N}}\colon \\
		\bar{S}_\abb{OCC}^\s{L}(s,\ens{\upsilon},\povm{N})\geq0
	\end{array}\right\}}\frac{S_\abb{OCC}^\f{S}(\Lambda;s,\ens{\upsilon},\povm{N})}{S_\abb{OCC}^\f{C}(s,\ens{\upsilon},\povm{N})}.
\end{align}
If the value of $R_\abb{rand}^\f{C}(\Psi)$ is bounded for all $\Psi\in\q{L}^\A$, then
\begin{align}
	1+R_\abb{rand}^\f{C}(\Lambda)&=\max_{\scriptsize\left\{\begin{array}{c}
		s\in[-1,1]^{\idx{N}\times\idx{N}}\colon \\
		\sum_{n,n'}s_{n,n'}\tr\left[N_{n'}\right]\tr\left[\nu_n\right]\geq0
	\end{array}\right\}}\frac{S_\abb{OCC}^{\{\Id\}}(\Lambda;s,\ens{\upsilon},\povm{N})}{S_\abb{OCC}^\f{C}(s,\ens{\upsilon},\povm{N})}.
\end{align}
\end{corollary}

Note that the condition $\bar{S}_\abb{OCC}^\s{L}(s,\ens{\upsilon},\povm{N})\geq0$ in Eq.~\eqref{eq:robustness-OCC-3} essentially means that the variable function $s$ is restricted to those under which the modified task realizes a resource witness~\cite{Chitambar-2019a}.

\section{Conclusions and Discussions}
\label{sec:conclusion}

\subsection{Summary of results}
\label{sec:summary}

In this work, we have studied the manipulation and quantification of general quantum resources using two distinct approaches.  The first approach is based on a generalization of entropic concepts.  Specifically, we have interpreted entropies as quantifying the uncertainty about a given system from a given observer's perspective.  This motivates the idea of taking into account not only the property of the system, but also that of the observer, when defining entropies.  From this standpoint, all conventional definitions of entropies correspond to the special case of a ``completely unrestricted'' observer in the standard quantum theory.  A more general possibility is the observer being restricted, whose admissible operations are captured by a closed and convex quantum resource theory.  Since the observer's ability to access information is potentially weakened, the measures of uncertainty are expected to alter accordingly and depend on the QRT\@.  We have demonstrated this intuition by proposing a resource-theoretic generalization of the (conditional) min-entropy, termed the free (conditional) min-entropy, to quantify the ``subjective'' degree of uncertainty from the perspective of such an operationally restricted observer (Definitions~\ref{def:FME} and \ref{def:FCME}).  We have also extended the definitions of the free (conditional) min-entropy to dynamic systems (Definitions~\ref{def:EFME} and \ref{def:EFCME}), in a way consistent with the unextended definitions (Fig.~\ref{fig:entropy}) thanks to a top-down framework of QRTs (Fig.~\ref{fig:QRT}).  We have further justified the information-theoretic relevance of these free min-entropies by proving multiple properties thereof (Propositions~\ref{prop:reducibility}--\ref{prop:operational}), including monotonicity under free (super)operations, while pointing out and making sense of aspects in which the free entropies contrasts standard entropies, such as a lack of additivity.

We have then found applications of the free min-entropies in characterizing the QRT in which they are defined.  Specifically, we have constructed a complete set of entropic conditions for deterministic and probabilistic free convertibility between two given (sub)channels or (sub)states (Theorems~\ref{thm:convertibility-deterministic} and \ref{thm:convertibility-probabilistic} and Corollaries~\ref{cor:convertibility-deterministic} and \ref{cor:convertibility-probabilistic}).  We have also established an information-theoretic interpretation of the resource global robustness of a state or a channel in terms of a mutual-information-like quantity based on the free min-entropies.

The second approach that we have taken to characterize quantum resources is based on operational tasks.  We have considered two operational tasks (Tasks~\ref{def:EOP} and \ref{def:OCC}) and derived complete sets of resource monotones in terms of the success probabilities therein (Theorems~\ref{thm:convertibility-EOP} and \ref{thm:convertibility-OCC}).  We have also found operational interpretations for a broad family of resource robustness measures, for many of which no such interpretations were known before (Theorems~\ref{thm:robustness-EOP} and \ref{thm:robustness-OCC}).

\subsection{Open questions and outlook}
\label{sec:question}

We list several directions for future exploration.  First, as we have presented a resource-theoretic generalization of the min-entropies, a natural question is whether such a generalization can be found for other entropies, divergences, or distinguishability measures.  What makes the min-entropies relatively easy to generalize is their clear operational meaning in the one-shot regime.  Therefore, we would at least expect that other information measures which arise in one-shot operational tasks can be generalized following a similar recipe, may the task be hypothesis testing~\cite{Helstrom-1967a, Holevo-1972a, Dupuis-2013a, Regula-2024a}, reverse test~\cite{Matsumoto-2018a}, or conversion between ordered pairs of states~\cite{Wang-2019a}.  Once generalizations of one-shot information measures are put in place, we would be curious about whether any fundamental results in the standard quantum information theory (e.g., the quantum Stein's lemma~\cite{Hiai-1991a, Ogawa-2000a} and the quantum asymptotic equipartition property~\cite{Tomamichel-2009a}) can be suitably generalized to the resource-theoretic setting~\cite{Brandao-2020a, George-2024a}.  The hope is that success in generalizing these results would provide meaningful resource-theoretic generalizations for information measures whose operational interpretation only manifests in the asymptotic regime, such as the Umegaki relative entropy~\cite{Hiai-1991a, Ogawa-2000a} and the conditional von Neumann entropy~\cite{Horodecki-2005a, Tomamichel-2009a}.

Apart from the above constructive approach based on operational interpretations of entropies, it would also be insightful to develop a fully axiomatic approach for resource-theoretic generalization of entropies~\cite{Matsumoto-2010a, Gour-2021a, Gour-2021c, Gour-2024a}.  What is a minimal set of information-theoretic principles that a ``free entropy'' should satisfy, that exclusively determines a free entropy?  To answer this question systematically would certainly justify any possible operational proposal of a free entropy from an information-theoretic standpoint.  We hope that the rudimentary discussions in Sec.~\ref{sec:entropy-properties} of this paper can be found instructive in this regard.

Another potential connection worthy of attention is with the formulation of entropies in generalized probabilistic theories (GPTs)~\cite{Barrett-2007a, Barnum-2010a, Short-2010a, Kimura-2010a}.  Both specified by general (instead of self-dual) cones, the structures of GPTs and QRTs bear somewhat of resemblance, and presumably the formulation and analysis of entropies in these theories should have certain points in common as well.  We hope that the respective attempts in both theories can inspire each other, and we leave such potential connections for future discovery.

\section*{Acknowledgements}
\label{sec:acknowledgements}

We thank Francesco Buscemi, Ian George, and Mark M. Wilde for helpful comments on the presentation of this paper.  This work is supported by NSF Award No.~2112890.  E.C. is supported by the U.S. Department of Energy Office of Science National Quantum Information Science Research Centers.



\section*{Appendix}

Appendices~\ref{app:program}, \ref{app:transformation}, and \ref{app:resource} are presented in the general context of topological vector spaces.  Appendices~\ref{app:dual}, \ref{app:properties}, \ref{app:characterization-entropy}, and \ref{app:characterization-tasks} apply specifically to quantum theory and resonate directly with the main text.

\appendices

\section{Convex conic programming}
\label{app:program}

Let $\spa{U}$ be a finite-dimensional vector space over $\spa{R}$.  We denote by $\spa{U}^*$ the \emph{dual vector space} of $\spa{U}$.  We denote by $\langle\alpha,\gamma\rangle\in\spa{R}$ the \emph{inner product} between an \emph{element} $\gamma\in\spa{U}$ and a \emph{functional} $\alpha\in\spa{U}^*$.

Let $\s{U}\subseteq\spa{U}$ be a closed and convex set.  The \emph{cone} generated by $\s{U}$, denoted by $\cone(\s{U})\subseteq\spa{U}$, is defined as
\begin{align}
	\cone(\s{U})&\coloneq\left\{r\gamma\colon\gamma\in\s{U},\;r\in\spa{R}_+\right\}.
\end{align}
The \emph{dual cone} of $\s{U}$, denoted by $\cone^*(\s{U})\subseteq\spa{U}^*$, is defined as
\begin{align}
	\cone^*(\s{U})&\coloneq\left\{\alpha\in\spa{U}^*\colon\left\langle\alpha,\gamma\right\rangle\in\spa{R}_+\;\;\forall\gamma\in\s{U}\right\}.
\end{align}
Let $\spa{V}$ be another finite-dimensional vector space over $\spa{R}$.  We denote by $\spa{T}^{\spa{U}\to\spa{V}}$ the set of all linear transformations from $\spa{U}$ to $\spa{V}$.  The \emph{adjoint transformation} of a linear transformation $\ob{T}\in\spa{T}^{\spa{U}\to\spa{V}}$ is defined as the unique transformation $\ob{T}^\dagger\in\spa{T}^{\spa{V}^*\to\spa{U}^*}$ such that
\begin{align}
\label{aeq:adjoint}
	\left\langle\alpha,\ob{T}\left[\gamma\right]\right\rangle&=\left\langle\ob{T}^\dagger\left[\alpha\right],\gamma\right\rangle\quad\forall\gamma\in\spa{U},\;\alpha\in\spa{V}^*.
\end{align}

\begin{definition}[\cite{Boyd-2004a}]
\label{adef:program}
Let $\s{U}\subseteq\spa{U}$ and $\s{V}\subseteq\spa{V}$ be two closed and convex sets.  Given $\ob{C}\in\spa{U}^*$, $\ob{A}\in\spa{V}$, and $\ob{T}\in\spa{T}^{\spa{U}\to\spa{V}}$, a \textbf{convex conic program} is a pair of optimization problems, one called the \textbf{primal problem}, defined as
\begin{subequations}
\label{aeq:program-primal}
\begin{align}
	\abb{minimize}\quad&\left\langle\ob{C},\gamma\right\rangle \\
	\abb{over}\quad&\gamma\in\cone(\s{U}) \\
	\textnormal{subject to}\quad&\ob{T}\left[\gamma\right]-\ob{A}\in\cone(\s{V}),
\end{align}
\end{subequations}
and the other called the \textbf{dual problem}, defined as
\begin{subequations}
\label{aeq:program-dual}
\begin{align}
	\abb{maximize}\quad&\left\langle\alpha,\ob{A}\right\rangle \\
	\abb{over}\quad&\alpha\in\cone^*(\s{V}) \\
	\textnormal{subject to}\quad&\ob{C}-\ob{T}^\dagger\left[\alpha\right]\in\cone^*(\s{U}).
\end{align}
\end{subequations}
\end{definition}

\begin{lemma}[Slater's condition for strong duality \cite{Boyd-2004a}]
\label{alem:duality}
Consider a convex conic program whose primal problem~\eqref{aeq:program-primal} and dual problem~\eqref{aeq:program-dual} such that both the primal and the dual problems have a feasible solution.  Then the following statements hold.
\begin{enumerate}
	\item[(1)] If the primal problem has a strictly feasible solution (i.e., a feasible solution in the relative interior of the feasible region), then the primal and dual problems yield the same optimal value, and there exists an optimal solution to the dual problem, i.e.,
	\begin{align}
		&\inf_{\scriptsize\left\{\begin{array}{c}
			\gamma\in\cone(\s{U})\colon \\
			\ob{T}\left[\gamma\right]-\ob{A}\in\cone(\s{V})
		\end{array}\right\}}\left\langle\ob{C},\gamma\right\rangle \notag\\
		&\quad=\max_{\scriptsize\left\{\begin{array}{c}
			\alpha\in\cone^*(\s{V})\colon \\
			\ob{C}-\ob{T}^\dagger\left[\alpha\right]\in\cone^*(\s{U})
		\end{array}\right\}}\left\langle\alpha,\ob{A}\right\rangle. \label{aeq:slater-dual}
	\end{align}
	\item[(2)] If the dual problem has a strictly feasible solution, then the primal and dual problems yield the same optimal value, and there exists an optimal solution to the primal problem, i.e.,
	\begin{align}
		&\min_{\scriptsize\left\{\begin{array}{c}
			\gamma\in\cone(\s{U})\colon \\
			\ob{T}\left[\gamma\right]-\ob{A}\in\cone(\s{V})
		\end{array}\right\}}\left\langle\ob{C},\gamma\right\rangle \notag\\
		&\quad=\sup_{\scriptsize\left\{\begin{array}{c}
			\alpha\in\cone^*(\s{V})\colon \\
			\ob{C}-\ob{T}^\dagger\left[\alpha\right]\in\cone^*(\s{U})
		\end{array}\right\}}\left\langle\alpha,\ob{A}\right\rangle. \label{aeq:slater-primal}
	\end{align}
\end{enumerate}
\end{lemma}

\section{Adjoint, conjugate, and transpose transformations}
\label{app:transformation}

Let $\spa{U}$ be equipped with an involution onto itself, called \emph{conjugation}, which maps an element $\gamma\in\spa{U}$ to its \emph{conjugate} $\cc{\gamma}\in\spa{U}$.  Let $\spa{V}$ be also equipped with conjugation.  The \emph{conjugate transformation} of a linear transformation $\ob{T}\in\spa{T}^{\spa{U}\to\spa{V}}$, denoted by $\cc{\ob{T}}\in\spa{T}^{\spa{U}\to\spa{V}}$, is defined as $\cc{\ob{T}}[\cdot]\coloneq\cc{\ob{T}[\cc{(\cdot)}]}$.  The \emph{transpose transformation} of $\ob{T}\in\spa{T}^{\spa{U}\to\spa{V}}$, denoted by $\ob{T}^\top\in\spa{T}^{\spa{V}^*\to\spa{U}^*}$, is defined as $\ob{T}^\top\coloneq(\cc{\ob{T}})^\dagger$, where the adjoint transformation $\ob{T}^\dagger$ of $\ob{T}\in\spa{T}^{\spa{U}\to\spa{V}}$ is defined in Eq.~\eqref{aeq:adjoint}.

The definitions of adjoint, conjugate, and transpose transformations can be demonstrated in quantum theory in both the static and the dynamic settings.  In the static setting, we set the underlying vector spaces $\spa{U}$ and $\spa{V}$ to be the spaces of Hermitian operators $\spa{H}^\AO$ and $\spa{H}^\AI$ associated with the static systems $\AO$ and $\AI$, respectively.  We may as well consider the space $\spa{H}^\AO$ to be self-dual, i.e., $(\spa{H}^\AO)^*=\spa{H}^\AO$, so that the Hilbert--Schmidt inner product between Hermitian operators is naturally employed, i.e., $\langle\tau,\rho\rangle\coloneq\tr[\tau\rho]$ for any $\rho,\tau\in\spa{H}^\AO$.  The space of linear transformations $\spa{T}^{\spa{H}^\AO\to\spa{H}^\AI}$ is the space of HP linear maps $\spa{L}^\A$ from $\AO$ to $\AI$.  The conjugation in $\spa{H}^\AO$ is the complex conjugation on Hermitian operators.  Then given an HP linear map $\Lambda\in\spa{L}^\A$, its adjoint map $\Lambda^\dagger\in\spa{L}^{\AI\to\AO}$, conjugate map $\cc{\Lambda}\in\spa{L}^\A$, and transpose map $\Lambda^\top\in\spa{L}^{\AI\to\AO}$ can all be defined accordingly.

In the dynamic setting, we set the underlying vector spaces $\spa{U}$ and $\spa{V}$ to be the spaces of HP linear maps $\spa{L}^\A$ and $\spa{L}^\B$ associated with the dynamic systems $\A$ and $\B$, respectively.  We may as well consider the space $\spa{L}^\A$ to be self-dual, i.e., $(\spa{L}^\A)^*=\spa{L}^\A$, so that the inner product can be defined between HP linear maps as $\langle\Psi,\Lambda\rangle\coloneq\tr[J_\Psi J_\Lambda]$ for any $\Lambda,\Psi\in\spa{L}^\A$.  The space of linear transformations $\spa{T}^{\spa{L}^\A\to\spa{L}^\B}$ is the space of HP-preserving supermaps $\spa{S}^{\A\To\B}$ from $\A$ to $\B$.  The conjugation in $\spa{L}^\A$ is the conjugation on HP linear maps defined in the static setting.  Then given an HP-preserving supermap $\sm{\Theta}\in\spa{S}^{\A\To\B}$, its adjoint supermap $\sm{\Theta}^\dagger\in\spa{S}^{\B\To\A}$, conjugate supermap $\cc{\sm{\Theta}}\in\spa{S}^{\A\To\B}$, and transpose map $\sm{\Theta}^\top\in\spa{S}^{\A\To\B}$ can all be defined accordingly.

\begin{lemma}[\cite{Gour-2019a}]
\label{alem:transpose}
Let $\Lambda\in\spa{L}^\A$ be an HP linear map, and let $\sm{\Theta}\in\spa{S}^{\A\To\B}$ be an HP-preserving supermap.  Then their Choi operators satisfy the following properties.
\begin{enumerate}
	\item[(1)] $J_{\Lambda^\top}=J_\Lambda$.
	\item[(2)] $J_{\sm{\Theta}^\top}=J_\sm{\Theta}$.
\end{enumerate}
\end{lemma}

\section{Dual formulation of the free min-entropies}
\label{app:dual}

\subsection{Derivation of Eq.~\eqref{eq:FME-dual}}
\label{app:FME-dual}

We note that the minimization in Eq.~\eqref{eq:FME} is exactly in the form of Eq.~\eqref{aeq:program-primal} by substituting $\spa{U}=\spa{R}$, $\spa{V}=\spa{H}^\BI$, $\s{U}=\spa{R}$, $\s{V}=\cone^*(\f{F}^\BI)$, $\ob{C}=1$, $\ob{A}=\rho^\BI$, and $\ob{T}[\cdot]=(\cdot)\spec{I}^\BI$.  Since $r=1$ is a strictly feasible solution to this minimization, by Lemma~\ref{alem:duality}(1), strong duality holds.  Therefore, the FME can be equivalently expressed in terms of the dual problem, which following Eq.~\eqref{aeq:slater-dual} is given by
\begin{align}
	2^{-H_{\min}^\f{F}(\BI)_\rho}&=\max_{\scriptsize\left\{\begin{array}{c}
		\alpha\in\cone(\f{F}^\BI)\colon \\
		1-\tr\left[\alpha\right]=0
	\end{array}\right\}}\tr\left[\alpha\rho\right] \\
	&=\max_{\tau\in\f{F}^\BI}\tr\left[\tau\rho\right],
\end{align}
which recovers Eq.~\eqref{eq:FME-dual}.

\subsection{Derivation of Eqs.~\eqref{eq:FCME-dual-1} and \eqref{eq:FCME-dual-2}}
\label{app:FCME-dual}

We note that the minimization in Eq.~\eqref{eq:FCME} is exactly in the form of Eq.~\eqref{aeq:program-primal} by substituting $\spa{U}=\spa{H}^\AI$, $\spa{V}=\spa{H}^{\AI\BI}$, $\s{U}=\spa{H}^\AI$, $\s{V}=\cone^*(\f{O}_\abb{J}^{\AI\to\BI})$, $\ob{C}=\spec{I}^\AI$, $\ob{A}=\rho^{\AI\BI}$, and $\ob{T}[\cdot]=(\cdot)^\AI\otimes\spec{I}^\BI$.  Since $\gamma=\spec{I}^\AI$ is a strictly feasible solution to this minimization, by Lemma~\ref{alem:duality}(1), strong duality holds.  Therefore, the FCME can be equivalently expressed in terms of the dual problem, which following Eq.~\eqref{aeq:slater-dual} is given by
\begin{align}
	2^{-H_{\min}^\f{O}(\BI|\AI)_\rho}&=\max_{\scriptsize\left\{\begin{array}{c}
		\alpha\in\cone(\f{O}_\abb{J}^{\AI\to\BI})\colon \\
		\spec{I}^\AI-\alpha^\AI=0
	\end{array}\right\}}\tr\left[\alpha\rho\right] \\
	&=\max_{\alpha\in\f{O}_\abb{J}^{\AI\to\BI}}\tr\left[\alpha\rho\right] \\
	&=\max_{\Psi\in\f{O}^{\AI\to\BI}}\tr\left[J_\Psi\rho\right],
\end{align}
which recovers Eq.~\eqref{eq:FCME-dual-1}.  To derive Eq.~\eqref{eq:FCME-dual-2},
\begin{align}
	&2^{-H_{\min}^\f{O}(\BI|\AI)_\rho} \notag\\
	&\quad=\max_{\Psi\in\f{O}^{\AI\to\BI}}\tr\left[J_\Psi\rho\right] \\
	&\quad=\max_{\Psi\in\f{O}^{\AI\to\BI}}\tr\left[J_{\Psi^\top}\rho\right] \label{aeq:FCME-1}\\
	&\quad=\max_{\Psi\in\f{O}^{\AI\to\BI}}\left\langle\left((\Psi^\top)^{\RBI\to\AI}\otimes\id^\BI\right)\left[\spec{\phi}_+^{\RBI\BI}\right],\rho^{\AI\BI}\right\rangle \\
	&\quad=\max_{\Psi\in\f{O}^{\AI\to\BI}}\left\langle\spec{\phi}_+^{\RBI\BI},\left(\cc{\Psi}^{\AI\to\RBI}\otimes\id^{\BI}\right)\left[\rho^{\AI\BI}\right]\right\rangle \label{aeq:FCME-2}\\
	&\quad=\max_{\Psi\in\f{O}^{\AI\to\BI}}\left\langle\spec{\phi}_+^{\RBI\BI},\left(\Psi^{\AI\to\RBI}\otimes\id^\BI\right)\left[\rho^{\AI\BI}\right]\right\rangle \label{aeq:FCME-3}\\
	&\quad=\max_{\Psi\in\f{O}^{\AI\to\BI}}\tr\left[\spec{\phi}_+^{\RBI\BI}\left(\Psi^{\AI\to\RBI}\otimes\id^\BI\right)\left[\rho^{\AI\BI}\right]\right].
\end{align}
Here Eq.~\eqref{aeq:FCME-1} uses Lemma~\ref{alem:transpose}(1); Eq.~\eqref{aeq:FCME-2} is by the definition of transpose transformations; Eq.~\eqref{aeq:FCME-3} is a consequence of the conjugation invariance principle of resource theories (see Sec.~\ref{sec:axiom}).

\subsection{Derivation of Eqs.~\eqref{eq:EFCME-dual-1} and \eqref{eq:EFCME-dual-2}}
\label{app:EFCME-dual}

Define the following two sets:
\begin{align}
	\s{U}^{\AO\AI\BO}&\coloneq\left\{\gamma\in\spa{H}^{\AO\AI\BO}\colon\gamma^{\AO\BO}=\spec{\pi}^{\AO}\otimes\gamma^\BO\right\}, \\
	\s{B}^{\AO\AI\BO}&\coloneq\left\{\beta\in\spa{H}^{\AO\AI\BO}\colon\beta^{\AI\BO}=0,\right. \notag\\
	&\quad\left.\beta^{\AO\AI\BO}=\beta^{\AO\BO}\otimes\spec{\pi}^\AI\right\}.
\end{align}

\begin{lemma}
	\label{alem:EFCME}
	$\cone^*(\s{U}^{\AO\AI\BO})=\s{B}^{\AO\AI\BO}$.
\end{lemma}

\begin{proof}
We first show $\s{B}^{\AO\AI\BO}\subseteq\cone^*(\s{U}^{\AO\AI\BO})$.  For any $\gamma\in\s{U}^{\AO\AI\BO}$ and $\beta\in\s{B}^{\AO\AI\BO}$, we have that
\begin{align}
	\left\langle\beta,\gamma\right\rangle&=\left\langle\beta^{\AO\BO}\otimes\spec{\pi}^\AI,\gamma^{\AO\AI\BO}\right\rangle \\
	&=\frac{1}{d_\AI}\left\langle\beta^{\AO\BO},\gamma^{\AO\BO}\right\rangle \\
	&=\frac{1}{d_\AI}\left\langle\beta^{\AO\BO},\spec{\pi}^\AO\otimes\gamma^\BO\right\rangle \\
	&=\frac{1}{d_\AO d_\AI}\left\langle\beta^\BO,\gamma^\BO\right\rangle \\
	&=\frac{1}{d_\AO}\left\langle\beta^{\AI\BO},\spec{\pi}^\AI\otimes\gamma^\BO\right\rangle \\
	&=\frac{1}{d_\AO}\left\langle0,\spec{\pi}^\AI\otimes\gamma^\BO\right\rangle \\
	&=0.
\end{align}
Conversely, it suffices to show that for any $\beta\in\spa{H}^{\AO\AI\BO}$ such that $\beta\notin\s{B}^{\AO\AI\BO}$, there exists an operator $\gamma_\star\in\s{U}^{\AO\AI\BO}$ such that $\langle\beta,\gamma_\star\rangle<0$.  This is done by choosing
\begin{align}
	\gamma_\star^{\AO\AI\BO}&\coloneq\beta^{\AO\BO}\otimes\spec{\pi}^\AI-\beta^{\AO\AI\BO}-\spec{\pi}^\AO\otimes\beta^{\AI\BO},
\end{align} 
and it follows that
\begin{align}
	&\left\langle\beta,\gamma_\star\right\rangle \notag\\
	&\quad=\left\langle\beta^{\AO\AI\BO},\beta^{\AO\BO}\otimes\spec{\pi}^\AI-\beta^{\AO\AI\BO}-\spec{\pi}^\AO\otimes\beta^{\AI\BO}\right\rangle \\
	&\quad=\left\langle\beta^{\AO\AI\BO},\beta^{\AO\BO}\otimes\spec{\pi}^\AI-\beta^{\AO\AI\BO}\right\rangle \notag\\
	&\quad\quad-\left\langle\beta^{\AO\AI\BO},\spec{\pi}^\AO\otimes\beta^{\AI\BO}\right\rangle \\
	&\quad=-\left\langle\beta^{\AO\AI\BO}-\beta^{\AO\BO}\otimes\spec{\pi}^\AI,\beta^{\AO\AI\BO}-\beta^{\AO\BO}\otimes\spec{\pi}^\AI\right\rangle \notag\\
	&\quad\quad-\frac{1}{d_\AO}\left\langle\beta^{\AI\BO},\beta^{\AI\BO}\right\rangle \\
	&\quad=-\left\|\beta^{\AO\AI\BO}-\beta^{\AO\BO}\otimes\spec{\pi}^\AI\right\|_2^2-\frac{1}{d_\AO}\left\|\beta^{\AI\BO}\right\|_2^2 \\
	&\quad<0. \label{aeq:EFCME-1}
\end{align}
Here Eq.~\eqref{aeq:EFCME-1} uses the assumption $\beta\notin\s{B}^{\AO\AI\BO}$.
\end{proof}

To derive Eq.~\eqref{eq:EFCME-dual-1}, we note that the minimization in Eq.~\eqref{eq:EFCME} is exactly in the form of Eq.~\eqref{aeq:program-primal} by substituting $\spa{U}=\spa{H}^{\AO\AI\BO}$, $\spa{V}=\spa{H}^{\AO\AI\BO\BI}$, $\s{U}=\s{U}^{\AO\AI\BO}$, $\s{V}=\cone^*(\f{S}_\abb{J}^{\A\To\B})$, $\ob{C}=\spec{I}^{\AO\AI\BO}$, $\ob{A}=J_\Lambda^{\AO\AI\BO\BI}$, and $\ob{T}[\cdot]=(\cdot)^{\AO\AI\BO}\otimes\spec{I}^\BI$.  Since $\gamma=\spec{I}^{\AO\AI\BO}$ is a strictly feasible solution to this minimization, by Lemma~\ref{alem:duality}(1), strong duality holds.  Therefore, the EFCME can be equivalently expressed in terms of the dual problem, which following Eq.~\eqref{aeq:slater-dual} is given by
\begin{align}
	&d_\AO d_\BO2^{-H_{\min}^\f{S}(\B|\A)_\Lambda} \notag\\
	&\quad=\max_{\scriptsize\left\{\begin{array}{c}
		\alpha\in\cone(\f{S}_\abb{J}^{\A\To\B})\colon \\
		\spec{I}^{\AO\AI\BO}-\alpha^{\AO\AI\BO}\in\cone^*(\s{U}^{\AO\AI\BO})
	\end{array}\right\}}\tr\left[\alpha J_\Lambda\right] \\
	&\quad=\max_{\scriptsize\left\{\begin{array}{c}
		\alpha\in\cone(\f{S}_\abb{J}^{\A\To\B})\colon \\
		\spec{I}^{\AO\AI\BO}-\alpha^{\AO\AI\BO}\in\s{B}^{\AO\AI\BO}
	\end{array}\right\}}\tr\left[\alpha J_\Lambda\right] \label{aeq:EFCME-2}\\
	&\quad=\max_{\scriptsize\left\{\begin{array}{c}
		\alpha\in\cone(\f{S}_\abb{J}^{\A\To\B}), \\
		\beta\in\spa{H}^{\AO\AI\BO}\colon \\
		\spec{I}^{\AO\AI\BO}-\alpha^{\AO\AI\BO}=\beta^{\AO\AI\BO}, \\
		\beta^{\AI\BO}=0, \\
		\beta^{\AO\AI\BO}=\beta^{\AO\BO}\otimes\spec{\pi}^\AI
	\end{array}\right\}}\tr\left[\alpha J_\Lambda\right] \\
	&\quad=\max_{\scriptsize\left\{\begin{array}{c}
		\alpha\in\cone(\f{S}_\abb{J}^{\A\To\B})\colon \\
		d_\AO\spec{I}^{\AI\BO}=\alpha^{\AI\BO}, \\
		\alpha^{\AO\AI\BO}=\alpha^{\AO\BO}\otimes\spec{\pi}^\AI
	\end{array}\right\}}\tr\left[\alpha J_\Lambda\right] \\
	&\quad=\max_{\scriptsize\left\{\begin{array}{c}
		\alpha\in\spa{H}^{\AO\AI\BO\BI}\colon \\
		\frac{1}{d_{\AO}}\alpha^{\AO\AI\BO\BI}\in\f{S}_\abb{J}^{\A\To\B}
	\end{array}\right\}}\tr\left[\alpha J_\Lambda\right] \\
	&\quad=d_\AO\max_{\sm{\Theta}\in\f{S}^{\A\To\B}}\tr\left[J_\sm{\Theta}J_\Lambda\right],
\end{align}
which recovers Eq.~\eqref{eq:EFCME-dual-1}.  Here Eq.~\eqref{aeq:EFCME-2} uses Lemma~\ref{alem:EFCME}.   To derive Eq.~\eqref{eq:EFCME-dual-2}, 
\begin{align}
	&d_\BO2^{-H_{\min}^\f{S}(\B|\A)_\Lambda} \notag\\
	&\quad=\max_{\sm{\Theta}\in\f{S}^{\A\To\B}}\tr\left[J_\sm{\Theta}J_\Lambda\right] \\
	&\quad=\max_{\sm{\Theta}\in\f{S}^{\A\To\B}}\tr\left[J_{\sm{\Theta}^\top}J_\Lambda\right] \label{aeq:EFCME-3}\\
	&\quad=\max_{\sm{\Theta}\in\f{S}^{\A\To\B}}\tr\left[J_{\Lambda_{\sm{\Theta}^\top}}J_\Lambda\right] \\
	&\quad=\max_{\sm{\Theta}\in\f{S}^{\A\To\B}}\left\langle\left((\sm{\Theta}^\top)^{\RB\To\A}\otimes\Id^\B\right)\left\{\spec{\Phi}_+^{\RB\B}\right\},\Lambda^{\A\B}\right\rangle \\
	&\quad=\max_{\sm{\Theta}\in\f{S}^{\A\To\B}}\left\langle\spec{\Phi}_+^{\RB\B},\left(\cc{\sm{\Theta}}^{\A\To\RB}\otimes\Id^\B\right)\left\{\Lambda^{\A\B}\right\}\right\rangle \label{aeq:EFCME-4}\\
	&\quad=\max_{\sm{\Theta}\in\f{S}^{\A\To\B}}\left\langle\spec{\Phi}_+^{\RB\B},\left(\sm{\Theta}^{\A\To\RB}\otimes\Id^\B\right)\left\{\Lambda^{\A\B}\right\}\right\rangle \label{aeq:EFCME-5}\\
	&\quad=\max_{\sm{\Theta}\in\f{S}^{\A\To\B}}\tr\left[J_{\spec{\Phi}_+^{\RB\B}}J_{\left(\sm{\Theta}^{\A\To\RB}\otimes\Id^\B\right)\left\{\Lambda^{\A\B}\right\}}\right] \\
	&\quad=\max_{\sm{\Theta}\in\f{S}^{\A\To\B}}\tr\left[\spec{\phi}_+^{\RBI\BI}\left(\sm{\Theta}^{\A\To\RB}\otimes\Id^\B\right)\left\{\Lambda^{\A\B}\right\}\left[\spec{\phi}_+^{\RBO\BO}\right]\right].
\end{align}
Here Eq.~\eqref{aeq:EFCME-3} uses Lemma~\ref{alem:transpose}(2); Eq.~\eqref{aeq:EFCME-4} is by the definition of transpose transformations; Eq.~\eqref{aeq:EFCME-5} is a consequence of the conjugation invariance principle of resource theories.

\section{Entropic and operational properties of the free min-entropies}
\label{app:properties}

\subsection{Proof of Proposition~\ref{prop:monotonicity}}
\label{app:monotoncity}

We first prove Proposition~\ref{prop:monotonicity}(3), which implies (1) by Proposition~\ref{prop:reducibility}(2).  Let $\Lambda\in\q{L}_\leqslant^\B$ be a subchannel, and let $\sm{\Theta}\in\q{S}_\leqslant^{\B\To\BP}$ be a subsuperchannel such that $\sm{\Theta}^\dagger\in\f{S}^{\BP\B}$.  Then by Eq.~\eqref{eq:EFME-dual-1},
\begin{align}
	H_{\min}^\f{C}(\BP)_{\sm{\Theta}\left\{\Lambda\right\}}&=\log d_\BO-\log\max_{\Psi\in\f{C}^\BP}\tr\left[J_\Psi J_{\sm{\Theta}\left\{\Lambda\right\}}\right] \\
	&=\log d_\BO-\log\max_{\Psi\in\f{C}^\BP}\left\langle\Psi,\sm{\Theta}\left\{\Lambda\right\}\right\rangle \\
	&=\log d_\BO-\log\max_{\Psi\in\f{C}^\BP}\left\langle\sm{\Theta}^\dagger\left\{\Psi\right\},\Lambda\right\rangle \label{aeq:monotonicity-1}\\
	&\geq\log d_\BO-\log\max_{\Psi\in\f{C}^\B}\left\langle\Psi,\Lambda\right\rangle \label{aeq:monotonicity-2}\\
	&=H_{\min}^\f{C}(\B)_\Lambda.
\end{align}
Here Eq.~\eqref{aeq:monotonicity-1} is by the definition of adjoint transformations; Eq.~\eqref{aeq:monotonicity-2} uses the fact that the free superoperation $\sm{\Theta}^\dagger$ preserves the set of free channels.

Next we prove Proposition~\ref{prop:monotonicity}(4), which implies (2) by Proposition~\ref{prop:reducibility}(4).  Let $\Lambda\in\q{L}_\leqslant^{\A\B}$ be a subchannel, and let $\sm{\Theta}\in\f{S}^{\A\To\AP}$ be a free superoperation.  Then by Eq.~\eqref{eq:EFCME-dual-2},
\begin{align}
	&H_{\min}^\f{S}(\B|\AP)_{\left(\sm{\Theta}\otimes\Id\right)\left\{\Lambda\right\}} \notag\\
	&\quad=\log d_\BO-\log\max_{\sm{\Theta}'\in\f{S}^{\AP\To\B}}\tr\left[\spec{\phi}_+^{\RBI\BI}\right. \notag\\
	&\quad\quad\left.\times\left(\sm{\Theta}'^{\AP\To\RB}\circ\sm{\Theta}^{\A\To\AP}\otimes\Id^\B\right)\left\{\Lambda^{\A\B}\right\}\left[\spec{\phi}_+^{\RBO\BO}\right]\right] \\
	&\quad\geq\log d_\BO-\log\max_{\sm{\Theta}'\in\f{S}^{\A\To\B}}\tr\left[\spec{\phi}_+^{\RBI\BI}\right. \notag\\
	&\quad\quad\left.\times\left(\sm{\Theta}'^{\A\To\RB}\otimes\Id^\B\right)\left\{\Lambda^{\A\B}\right\}\left[\spec{\phi}_+^{\RBO\BO}\right]\right] \label{aeq:monotonicity-3}\\
	&\quad=H_{\min}^\f{S}(\B|\A)_\Lambda.
\end{align}
Here Eq.~\eqref{aeq:monotonicity-3} uses the closedness of the set of free superoperations under composition.

\subsection{Proof of Proposition~\ref{prop:subadditivity}}
\label{app:subadditivity}

It suffices to prove Proposition~\ref{prop:subadditivity}(4), which implies (1)--(3) by Proposition~\ref{prop:reducibility}(2)--(4).  Suppose that $\f{S}$ is closed under tensor product.  Let $\Lambda\in\q{L}_\leqslant^{\A\B}$ and $\Psi\in\q{L}_\leqslant^{\AP\BP}$ be two subchannels.  Then by Eq.~\eqref{eq:EFCME-dual-1},
\begin{align}
	&H_{\min}^\f{S}(\B\BP|\A\AP)_{\Lambda\otimes\Psi} \notag\\
	&\quad=-\log\max_{\sm{\Theta}\in\f{S}^{\A\AP\To\B\BP}}\tr\left[J_{\sm{\Theta}}J_{\Lambda\otimes\Psi}\right] \\
	&\quad\leq-\log\max_{\sm{\Theta}\in\f{S}^{\A\To\B}}\max_{\sm{\Theta}'\in\f{S}^{\AP\To\BP}}\tr\left[J_{\sm{\Theta}\otimes\sm{\Theta}'}J_{\Lambda\otimes\Psi}\right] \\
	&\quad\leq-\log\max_{\sm{\Theta}\in\f{S}^{\A\To\B}}\max_{\sm{\Theta}'\in\f{S}^{\AP\To\BP}}\tr\left[\left(J_\sm{\Theta}^{\AO\AI\BO\BI}\otimes J_{\sm{\Theta}'}^{\APO\API\BPO\BPI}\right)\right. \notag\\
	&\quad\quad\left.\times\left(J_\Lambda^{\AO\AI\BO\BI}\otimes J_\Psi^{\APO\API\BPO\BPI}\right)\right] \\
	&\quad=-\log\max_{\sm{\Theta}\in\f{S}^{\A\To\B}}\tr\left[J_{\sm{\Theta}}J_\Lambda\right]-\log\max_{\sm{\Theta}'\in\f{S}^{\AP\To\BP}}\tr\left[J_{\sm{\Theta}'}J_\Psi\right] \\
	&\quad=H_{\min}^\f{S}(\B|\A)_\Lambda+H_{\min}^\f{S}(\BP|\AP)_\Psi.
\end{align}

\subsection{Proof of Proposition~\ref{prop:bounds}}
\label{app:bounds}

We first prove Proposition~\ref{prop:bounds}(2), which implies (1) by Proposition~\ref{prop:reducibility}(2) and (4).  Suppose $\Tr\in\f{S}^{\A\To(\#\to\#)}$, which in the static setting corresponds to $\tr\in\f{O}^\AI$.  Let $\Lambda\in\q{L}^\A$ be a channel, and let $\Omega\in\q{L}_\leqslant^\B$ be a subchannel.  Then by Eq.~\eqref{eq:EFCME-dual-2},
\begin{align}
	&H_{\min}^\f{S}(\B|\A)_{\Lambda\otimes\Omega} \notag\\
	&\quad=\log d_\BO-\log\max_{\sm{\Theta}\in\f{S}^{\A\To\B}}\tr\left[\spec{\phi}_+^{\RBI\BI}\right. \notag\\
	&\quad\quad\left.\times\left(\sm{\Theta}^{\A\To\RB}\left\{\Lambda^\A\right\}\otimes\Omega^\B\right)\left[\spec{\phi}_+^{\RBO\BO}\right]\right] \\
	&\quad\geq\log d_\BO-\log\max_{\Psi\in\q{L}^\B}\tr\left[\spec{\phi}_+^{\RBI\BI}\left(\Psi^\RB\otimes\Omega^\B\right)\left[\spec{\phi}_+^{\RBO\BO}\right]\right] \\
	&\quad=H_{\min}(\B)_\Omega. \label{aeq:bounds-1}
\end{align}
Here Eq.~\eqref{aeq:bounds-1} uses the dual formulation of extended quantum conditional min-entropy \cite{Gour-2019a}.  Conversely,
\begin{align}
	&H_{\min}^\f{S}(\B|\A)_{\Lambda\otimes\Omega} \notag\\
	&\quad=\log d_\BO-\log\max_{\sm{\Theta}\in\f{S}^{\A\To\B}}\tr\left[\spec{\phi}_+^{\RBI\BI}\right. \notag\\
	&\quad\quad\left.\times\left(\sm{\Theta}^{\A\To\RB}\left\{\Lambda^\A\right\}\otimes\Omega^\B\right)\left[\spec{\phi}_+^{\RBO\BO}\right]\right] \\
	&\quad\leq\log d_\BO-\log\max_{\sm{\Theta}\in\f{S}^{(\#\to\#)\To\B}}\tr\left[\spec{\phi}_+^{\RBI\BI}\right. \notag\\
	&\quad\quad\left.\times\left(\sm{\Theta}^{(\#\to\#)\To\RB}\circ\Tr\left\{\Lambda^\A\right\}\otimes\Omega^\B\right)\left[\spec{\phi}_+^{\RBO\BO}\right]\right] \label{aeq:bounds-2}\\
	&\quad=\log d_\BO-\log\max_{\Psi\in\f{C}^\B}\tr\left[\spec{\phi}_+^{\RBI\BI}\left(\Psi^\RB\otimes\Omega^\B\right)\left[\spec{\phi}_+^{\RBO\BO}\right]\right] \\
	&\quad=H_{\min}^\f{C}(\B)_\Omega. \label{aeq:bounds-3}
\end{align}
Here Eq.~\eqref{aeq:bounds-2} uses the closedness of the set of free superoperations under composition; Eq.~\eqref{aeq:bounds-3} follows from Eq.~\eqref{eq:EFME-dual-2}.  Let $\Psi\in\f{C}^\A$ be a free channel.  Then
\begin{align}
	&H_{\min}^\f{S}(\B|\A)_{\Psi\otimes\Omega} \notag\\
	&\quad=\log d_\BO-\log\max_{\sm{\Theta}\in\f{S}^{\A\To\B}}\tr\left[\spec{\phi}_+^{\RBI\BI}\right. \notag\\
	&\quad\quad\left.\times\left(\sm{\Theta}^{\A\To\RB}\left\{\Psi^\A\right\}\otimes\Omega^\B\right)\left[\spec{\phi}_+^{\RBO\BO}\right]\right] \\
	&\quad=\log d_\BO-\log\max_{\Psi'\in\f{C}^\B}\tr\left[\spec{\phi}_+^{\RBI\BI}\left(\Psi'^\RB\otimes\Omega^\B\right)\left[\spec{\phi}_+^{\RBO\BO}\right]\right] \\
	&\quad=H_{\min}^\f{C}(\B)_\Omega. \label{aeq:bounds-4}
\end{align}
Equations~\eqref{aeq:bounds-1}, \eqref{aeq:bounds-3}, and \eqref{aeq:bounds-4} complete the proof of Proposition~\ref{prop:bounds}(2).

Next we prove Proposition~\ref{prop:bounds}(4), which implies (2) by Proposition~\ref{prop:reducibility}(2).  Let $\Lambda\in\q{L}^\B$ be a channel.  By Definition~\ref{def:EFME} and the fact $\f{C}\subseteq\q{L}$,
\begin{align}
	H_{\min}^\f{C}(\B)_\Lambda&\geq H_{\min}(\B)_\Lambda \\
	&=H_{\min}(\BI|\BO)_{\frac{1}{d_\BO}J_\Lambda} \\
	&\geq-\log\min\left\{d_\BO,d_\BI\right\}. \label{aeq:bounds-5}
\end{align}
Here Eq.~\eqref{aeq:bounds-5} uses the lower bound on the quantum conditional min-entropy~\cite{Tomamichel-2012a}.  Let $\ob{K}\in\f{C}^\B$ be a free isometric channel (so that $d_\BO\leq d_\BI$).  Then by Eq.~\eqref{eq:EFME-dual-1},
\begin{align}
	H_{\min}^\f{C}(\B)_\ob{K}&=\log d_\BO-\log\max_{\Psi\in\f{C}^\B}\tr\left[J_\Psi J_\ob{K}\right] \\
	&=\log d_\BO-\log\tr\left[J_\ob{K}J_\ob{K}\right] \\
	&=-\log d_\BO. \label{aeq:bounds-6}
\end{align}
Suppose $\spec{\Pi}\in\f{C}^\B$, which in the static setting corresponds to $\spec{\pi}\in\f{F}^\BI$.  Then
\begin{align}
	H_{\min}^\f{C}(\B)_\Lambda&=\log d_\BO-\log\max_{\Psi\in\f{C}^\B}\tr\left[J_\Psi J_\Lambda\right] \\
	&\leq\log d_\BO-\log\tr\left[J_\spec{\Pi} J_\Lambda\right] \\
	&=\log d_\BO-\log\tr\left[\left(\spec{I}^\BO\otimes\spec{\pi}^\BI\right)J_\Lambda^{\BO\BI}\right] \\
	&=\log d_\BI. \label{aeq:bounds-7}
\end{align}
We also have that
\begin{align}
	H_{\min}^\f{C}(\B)_\spec{\Pi}&=\log d_\BO-\log\max_{\Psi\in\f{C}^\B}\tr\left[J_\Psi J_\spec{\Pi}\right] \\
	&=\log d_\BO-\log\max_{\Psi\in\f{C}^\B}\tr\left[J_\Psi^{\BO\BI}\left(\spec{I}^\BO\otimes\spec{\pi}^\BI\right)\right] \\
	&=\log d_\BI. \label{aeq:bounds-8}
\end{align}
Equations~\eqref{aeq:bounds-5}, \eqref{aeq:bounds-6}, \eqref{aeq:bounds-7}, and \eqref{aeq:bounds-8} complete the proof of Proposition~\ref{prop:bounds}(4).

\subsection{Proof of Proposition~\ref{prop:operational}}
\label{app:operational}

It suffices to prove Proposition~\ref{prop:operational}(2), which implies (1) by Proposition~\ref{prop:reducibility}(4).  Let $\povm{\Lambda}\equiv\{\Lambda_{x_1|x_0}\}_{x_0,x_1}$ be a multi-instrument in $\A$.  By Eq.~\eqref{eq:EFCME-dual-1},
\begin{align}
	&2^{-H_{\min}^\f{S}(\X|\A)_\Lambda} \notag\\
	&\quad=\frac{1}{d_\XO}\max_{\sm{\Theta}\in\f{S}^{\A\To\X}}\tr\left[J_{\sm{\Theta}}J_\Lambda\right] \\
	&\quad=\frac{1}{d_\XO}\max_{\sm{\Theta}\in\f{S}^{\A\To\X}}\sum_{x_0,x_1}\tr\left[J_{\sm{\Theta}}^{\AO\AI\XO\XI}\right. \notag\\
	&\quad\quad\left.\times\left(J_{\Lambda_{x_1|x_0}}^{\AO\AI}\otimes\op{x_0}{x_0}^{\XO}\otimes\op{x_1}{x_1}^{\XI}\right)\right] \\
	&\quad=\frac{1}{d_\XO}\max_{\sm{\Theta}\in\f{S}^{\A\To\X}}\sum_{x_0,x_1}\tr\left[\left(\spec{I}^{\AO\AI}\otimes\bra{x_0}^{\XO}\otimes\bra{x_1}^{\XI}\right)\right. \notag\\
	&\quad\quad\left.\vphantom{}\times J_{\sm{\Theta}}^{\AO\AI\XO\XI}\left(\spec{I}^{\AO\AI}\otimes\ket{x_0}^{\XO}\otimes\ket{x_1}^{\XI}\right)J_{\Lambda_{x_1|x_0}}^{\AO\AI}\right] \\
	&\quad=\frac{1}{d_\XO}\max_{\scriptsize\left\{\begin{array}{c}
		\left\{\mu_{x_1|x_0}\right\}_{x_0,x_1}\subset\spa{H}^{\AO\AI}\colon \\
		\sum_{x_0,x_1}\mu_{x_1|x_0}^{\AO\AI}\otimes\op{x_0}{x_0}^\XO\otimes\op{x_1}{x_1}^\XI\in\f{S}_\abb{J}^{\A\To\X}
	\end{array}\right\}} \notag\\
	&\quad\quad\times\sum_{x_0,x_1}\tr\left[\mu_{x_1|x_0}J_{\Lambda_{x_1|x_0}}\right] \\
	&\quad=\frac{1}{d_\XO}\max_{\{\mu_{x_1|x_0}\}_{x_0,x_1}\in\f{S}_\abb{M}^\A}\sum_{x_0,x_1}\tr\left[\mu_{x_1|x_0}J_{\Lambda_{x_1|x_0}}\right] \label{aeq:operational}\\
	&\quad=P_\abb{guess}^\f{S}(\povm{\Lambda}).
\end{align}
Here Eq.~\eqref{aeq:operational} is by Definition~\ref{def:supermeasurement-free}.

\section{Resource theories beyond quantum}
\label{app:resource}

Let $\{\spa{U}^{(i)}\}_i$ be a collection of finite-dimensional vector spaces over $\spa{R}$.  We denote by $\id^{(i)}\in\spa{T}^{(i)\to(i)}$ the identity map from $\spa{U}^{(i)}$ to itself.

\begin{definition}
\label{adef:resource}
Let $\q{Q}\subseteq\bigcup_{i}\spa{U}^{(i)}$ be a set of elements called \textbf{objects}.  Let $\f{F}\subseteq\q{Q}$ be a subset of objects, and let $\f{T}\subseteq\bigcup_{i,j}\spa{T}^{(i)\to(j)}$ be a subset of transformations.  For any $i$ and $j$, denote $\q{Q}^{(i)}\equiv\q{Q}\cap\spa{U}^{(i)}$, $\f{F}^{(i)}\equiv\f{F}\cap\spa{U}^{(i)}$, and $\f{T}^{(i)\to(j)}\equiv\f{T}\cap\spa{T}^{(i)\to(j)}$.  Then the tuple $(\f{T},\f{F};\q{Q})$ is called a \textbf{quasi-resource theory} whenever the following conditions hold:
\begin{enumerate}
	\item[(1)] for any $i$, it holds that $\id^{(i)}\in\f{T}^{(i)\to(i)}$;
	\item[(2)] for any $i$, $j$, and $k$, if $\ob{T}\in\f{T}^{(i)\to(j)}$ and $\ob{T}'\in\f{T}^{(j)\to(k)}$, then $\ob{T}'\circ\ob{T}\in\f{T}^{(i)\to(k)}$;
	\item[(3)] for any $i$ and $j$, if $\ob{Q}\in\q{Q}^{(i)}$ and $\ob{T}\in\f{T}^{(i)\to(j)}$, then $\ob{T}[\ob{Q}]\in\q{Q}^{(j)}$;
	\item[(4)] for any $i$ and $j$, if $\ob{F}\in\f{F}^{(i)}$ and $\ob{T}\in\f{T}^{(i)\to(j)}$, then $\ob{T}[\ob{F}]\in\f{F}^{(j)}$.
\end{enumerate}
In this case, transformations in $\f{T}$ are called \textbf{free transformations}, and objects in $\f{F}$ are called \textbf{free objects}.  We say that $(\f{T},\f{F};\q{Q})$ is \textbf{closed} and \textbf{convex} whenever $\q{Q}^{(i)}$, $\f{F}^{(i)}$, and $\f{T}^{(i)\to(j)}$ are closed and convex for all $i$ and $j$.
\end{definition}

\subsection{Complete sets of resource monotones}
\label{app:convertibility}

\begin{definition}
\label{adef:monotone}
A function $f\colon\q{Q}\to\spa{R}$, which maps each object to a real number, is called a \textbf{resource monotone} in a quasi-resource theory $(\f{T},\f{F};\q{Q})$ whenever $f(\ob{Q})\geq f(\ob{T}[\ob{Q}])$ for every object $\ob{Q}\in\q{Q}^{(i)}$ and every free transformation $\ob{T}\in\f{T}^{(i)\to(j)}$ for all $i$ and $j$.  A set of resource monotones $\{f_w\}_{w\in\idx{W}}$ is said to be \textbf{complete} with respect to $i$ and $j$ whenever the following condition holds: given any two objects $\ob{Q}\in\q{Q}^{(i)}$ and $\ob{Q}'\in\q{Q}^{(j)}$, there exists a free transformation $\ob{T}\in\f{T}^{(i)\to(j)}$ such that $\ob{T}[\ob{Q}]=\ob{Q}'$ if and only if $f_w(\ob{Q})\geq f_w(\ob{Q}')$ for all $w\in\idx{W}$.
\end{definition}

\begin{theorem}
\label{athm:convertibility}
Let $(\f{T},\f{F};\q{Q})$ be a closed and convex quasi-resource theory.  Let $\{f_w\}_{w\in\idx{W}}$ be a set of resource monotones in $(\f{T},\f{F};\q{Q})$ with $f_w\colon\q{Q}\to\spa{R}$ for all $w\in\idx{W}$.  Then $\{f_w\}_{w\in\idx{W}}$ is complete with respect to $i$ and $j$ if there exists a sequence of mappings $\{h_d\}_{d=2}^\infty$ with $h_d\colon(\spa{U}^{(j)})^*\to\idx{W}$ for all $d\geq2$ such that the following condition holds:
\begin{enumerate}
	\item[(1)] given any functional $\alpha\in(\spa{U}^{(j)})^*$ and two objects $\ob{Q}\in\q{Q}^{(i)}$ and $\ob{Q}'\in\q{Q}^{(j)}$,
	\begin{align}
	\label{aeq:convertibility}
		\max_{\ob{T}\in\f{T}^{(i)\to(j)}}\left\langle\alpha,\ob{T}\left[\ob{Q}\right]\right\rangle&\geq\max_{\ob{T}'\in\f{T}^{(j)\to(j)}}\left\langle\alpha,\ob{T}'\left[\ob{Q}'\right]\right\rangle
	\end{align}
	if and only if $\lim_{d\to\infty}f_{h_d(\alpha)}(\ob{Q})\geq\lim_{d\to\infty}f_{h_d(\alpha)}(\ob{Q}')$.
\end{enumerate} 
\end{theorem}

\begin{proof}
Let $\{h_d\}_{d=2}^\infty$ be a sequence of mappings with $h_d\colon(\spa{U}^{(j)})^*\to\idx{W}$ for all $d\geq2$ such that Condition~(1) holds.  Let $\ob{Q}\in\q{Q}^{(i)}$ and $\ob{Q}'\in\q{Q}^{(j)}$ be two objects such that $f_w(\ob{Q})\geq f_w(\ob{Q}')$ for all $w\in\idx{W}$.  We prove by contradiction that there exists a free transformation $\ob{T}_\star\in\f{T}^{(i)\to(j)}$ such that $\f{T}_\star(\ob{Q})=\ob{Q}'$.  Suppose that such $\ob{T}_\star$ does not exist, i.e.,
\begin{align}
\label{aeq:convertibility-1}
	\ob{Q}'\notin\left\{\ob{T}\left[\ob{Q}\right]\colon\ob{T}\in\f{T}^{(i)\to(j)}\right\}.
\end{align}
Since the set on the right-hand side of Eq.~\eqref{aeq:convertibility-1} is closed and convex, by the Hahn-Banach Separation Theorem \cite{Rudin-1991a}, there exists a functional $\alpha_\star\in(\spa{U}^{(j)})^*$ such that 
\begin{align}
	\max_{\ob{T}\in\f{T}^{(i)\to(j)}}\left\langle\alpha_\star,\ob{T}\left[\ob{Q}\right]\right\rangle&<\left\langle\alpha_\star,\ob{Q}'\right\rangle \\
	&\leq\max_{\ob{T}'\in\f{T}^{(j)\to(j)}}\left\langle\alpha_\star,\ob{T}'\left[\ob{Q}'\right]\right\rangle.
\end{align}
By Condition~(1), this implies $\lim_{d\to\infty}f_{h_d(\alpha_\star)}(\ob{Q})<\lim_{d\to\infty}f_{h_d(\alpha_\star)}(\ob{Q}')$, and thus there exists an integer $d_\star\geq2$ such that $f_{h_{d_\star}(\alpha_\star)}(\ob{Q})<f_{h_{d_\star}(\alpha_\star)}(\ob{Q}')$.  This contradicts the assumption $f_w(\ob{Q})\geq f_w(\ob{Q}')$ for all $w\in\idx{W}$.  We thus conclude that there exists a free transformation $\ob{T}_\star\in\f{T}^{(i)\to(j)}$ such that $\ob{T}_\star[\ob{Q}]=\ob{Q}'$.  Then by Definition~\ref{adef:monotone}, $\{f_w\}_w$ is complete with respect to $i$ and $j$.
\end{proof}

\begin{corollary}
\label{acor:convertibility}
Let $(\f{T},\f{F};\q{Q})$ be a closed and convex quasi-resource theory.  Let $\{f_w\}_{w\in\idx{W}}$ be a set of functions with $f_w\colon\q{Q}\to\spa{R}$ for all $w\in\idx{W}$.  Then $\{f_w\}_{w\in\idx{W}}$ is a complete set of resource monotones with respect to $i$ and $j$ in $(\f{T},\f{F};\q{Q})$ if there exists a surjective mapping $h\colon(\spa{U}^{(j)})^*\to\idx{W}$ satisfying the following condition:
\begin{enumerate}
	\item[(1)] given any functional $\alpha\in(\spa{U}^{(j)})^*$ and two objects $\ob{Q}\in\q{Q}^{(i)}$ and $\ob{Q}'\in\q{Q}^{(j)}$, Eq.~\eqref{aeq:convertibility} holds if and only if $f_{h(\alpha)}(\ob{Q})\geq f_{h(\alpha)}(\ob{Q}')$.
\end{enumerate}
\end{corollary}

\begin{proof}
Let $h\colon(\spa{U}^{(j)})^*\to\idx{W}$ be a mapping satisfying Condition~(1).  To show that $\{f_w\}_w$ is a set of resource monotones, consider any two objects $\ob{Q}\in\q{Q}^{(i)}$ and $\ob{Q}'\in\q{Q}^{(j)}$ such that $\ob{T}_\star[\ob{Q}]=\ob{Q}'$ for some free transformation $\ob{T}_\star\in\f{T}^{(i)\to(j)}$.  Then for any $\alpha\in(\spa{U}^{(j)})^*$, we have that 
\begin{align}
	\max_{\ob{T}\in\f{T}^{(i)\to(j)}}\left\langle\alpha,\ob{T}\left[\ob{Q}\right]\right\rangle&\geq\max_{\ob{T}'\in\f{T}^{(j)\to(j)}}\left\langle\alpha,\ob{T}'\circ\ob{T}_\star\left[\ob{Q}\right]\right\rangle \\
	&=\max_{\ob{T}'\in\f{T}^{(j)\to(j)}}\left\langle\alpha,\ob{T}'\left[\ob{Q}'\right]\right\rangle.
\end{align}
By Condition~(1), this implies $f_{h(\alpha)}(\ob{Q})\geq f_{h(\alpha)}(\ob{Q}')$ for all $\alpha\in(\spa{U}^{(j)})^*$.  Since $h$ is surjective, we necessarily have $f_w(\ob{Q})\geq f_w(\ob{Q}')$ for all $w\in\idx{W}$.  By Definition~\ref{adef:monotone}, $\{f_w\}_w$ is a set of resource monotones.  Since a sequence of the same mapping $h$ satisfies Condition~(1) of Theorem~\ref{athm:convertibility}, by Theorem~\ref{athm:convertibility}, $\{f_w\}_w$ is complete with respect to $i$ and $j$.
\end{proof}

\subsection{Resource robustness measures}
\label{app:robustness}

\begin{definition}
\label{adef:robustness}
The \textbf{resource robustness} of an object $\ob{Q}\in\q{Q}^{(i)}$ \textbf{against} a subset of objects $\s{K}\subseteq\q{Q}$ in a quasi-resource theory $(\f{T},\f{F};\q{Q})$ is defined as
\begin{align}
	\label{aeq:robustness}
	R^{\s{K},\f{F}}(\ob{Q})&\coloneq\min_{\scriptsize\left\{\begin{array}{c}
		r\in\spa{R}_+, \\
		\ob{K}\in\s{K}^{(i)}\colon \\
		\frac{\ob{Q}+r\ob{K}}{1+r}\in\f{F}^{(i)}
	\end{array}\right\}}r,
\end{align}
where $\s{K}^{(i)}\equiv\s{K}\cap\q{Q}^{(i)}$ for all $i$.
\end{definition}

\begin{definition}
A subset of objects $\s{K}\subseteq\q{Q}$ is said to be \textbf{contractive} under a subset of transformations $\f{T}\subseteq\bigcup_{i,j}\spa{T}^{(i)\to(j)}$ whenever the following condition holds: for any $i$ and $j$, if $\ob{Q}\in\s{K}^{(i)}$ and $\ob{T}\in\f{T}^{(i)\to(j)}$, then $\ob{T}[\ob{K}]\in\s{K}^{(j)}$.
\end{definition}

\begin{proposition}
\label{aprop:robustness-monotone}
Let $(\f{T},\f{F};\q{Q})$ be a quasi-resource theory, and let $\s{K}\subseteq\q{Q}$ be a subset of objects contractive under $\f{T}$.  Then the resource robustness $R^{\s{K},\f{F}}$ against $\s{K}$ is a resource monotone in $(\f{T},\f{F};\q{Q})$.
\end{proposition}

\begin{proof}
Let $\ob{K}_\star\in\s{K}^{(i)}$ be an optimizer of the program on the right-hand side of Eq.~\eqref{aeq:robustness}, and it satisfies $\frac{\ob{Q}+R^{\s{K},\f{F}}(\ob{Q})\ob{K}_\star}{1+R^{\s{K},\f{F}}(\ob{Q})}\eqcolon\ob{F}_\star\in\f{F}^{(i)}$.  Then for any $j$ and free transformation $\ob{T}\in\f{T}^{(i)\to(j)}$, we have that
\begin{align}
	\frac{\ob{T}\left[\ob{Q}\right]+R^{\s{K},\f{F}}(\ob{Q})\ob{T}\left[\ob{K}_\star\right]}{1+R^{\s{K},\f{F}}(\ob{Q})}&=\ob{T}\left[\ob{F}_\star\right]\in\f{F}^{(j)}. \label{aeq:robustness-1}
\end{align}
If $\s{K}$ is contractive under $\f{T}$, then $\ob{T}[\ob{K}_\star]\in\s{K}^{(j)}$.  This implies that
\begin{align}
	R^{\s{K},\f{F}}(\ob{T}\left[\ob{Q}\right])&=\min_{\scriptsize\left\{\begin{array}{c}
		r\in\spa{R}_+, \\
		\ob{K}\in\s{K}^{(j)}\colon \\
		\frac{\ob{T}\left[\ob{Q}\right]+r\ob{K}}{1+r}\in\f{F}^{(j)}
	\end{array}\right\}}r \\
	&\leq\min_{\scriptsize\left\{\begin{array}{c}
		r\in\spa{R}_+\colon \\
		\frac{\ob{T}\left[\ob{Q}\right]+r\ob{T}\left[\ob{K}_\star\right]}{1+r}\in\f{F}^{(j)}
	\end{array}\right\}}r \\
	&\leq R^{\s{K},\f{F}}(\ob{Q}). \label{aeq:robustness-2}
\end{align}
Here Eq.~\eqref{aeq:robustness-2} uses Eq.~\eqref{aeq:robustness-1}.
\end{proof}

\begin{definition}
A set of objects $\q{Q}\subset\bigcup_{i}\spa{U}^{(i)}$ is said to be \textbf{normalized} whenever the following condition holds: for any $i$, there exists a functional $\spec{\ob{I}}\in(\spa{U}^{(i)})^*$ such that $\langle\spec{\ob{I}},\ob{Q}\rangle=1$ for all $\ob{Q}\in\q{Q}^{(i)}$.
\end{definition}

\begin{proposition}
\label{aprop:robustness-program}
Let $\q{Q}$ be a normalized set of objects, and let $(\f{T},\f{F};\q{Q})$ be a closed and convex quasi-resource theory.  Let $\s{K}\subseteq\q{Q}$ be a subset of objects satisfying the following conditions:
\begin{enumerate}
	\item[(1)] for any $i$, the set $\s{K}^{(i)}$ is closed and convex;
	\item[(2)] the value of $R^{\s{K},\f{F}}(\ob{Q})$ is bounded for all $\ob{Q}\in\q{Q}$.
\end{enumerate}
Let $\ob{Q}\in\q{Q}^{(i)}$ be an object.  Then $R^{\s{K},\f{F}}(\ob{Q})$ can be formulated as the following convex conic program, with the primal form being
\begin{align}
\label{aeq:robustness-primal}
	R^{\s{K},\f{F}}(\ob{Q})=&\min_{\scriptsize\left\{\begin{array}{c}
		\gamma\in\cone(\f{F}^{(i)})\colon \\
		\gamma-\ob{Q}\in\cone(\s{K}^{(i)})
	\end{array}\right\}}\left\langle\spec{\ob{I}},\gamma\right\rangle-1,
\end{align}
and the dual form being
\begin{align}
\label{aeq:robustness-dual}
	R^{\s{K},\f{F}}(\ob{Q})=&\sup_{\scriptsize\left\{\begin{array}{c}
		\alpha\in\cone^*(\s{K}^{(i)})\colon \\
		\spec{\ob{I}}-\alpha\in\cone^*(\f{F}^{(i)})
	\end{array}\right\}}\left\langle\alpha,\ob{Q}\right\rangle-1.
\end{align}
\end{proposition}

\begin{proof}
Equation~\eqref{aeq:robustness-primal} is derived from Eq.~\eqref{aeq:robustness} by substituting $\ob{Q}+r\ob{K}=\gamma$.  We note that the minimization in Eq.~\eqref{aeq:robustness-primal} is exactly in the form of Eq.~\eqref{aeq:program-primal} by substituting $\spa{U}=\spa{V}=\spa{U}^{(i)}$, $\s{U}=\f{F}^{(i)}$, $\s{V}=\s{K}^{(i)}$, $\ob{C}=\spec{\ob{I}}$, $\ob{A}=\ob{Q}$, and $\ob{T}=\id^{(i)}$.  Then following Eq.~\eqref{aeq:program-dual}, the dual problem is exactly the one on the right-hand side of Eq.~\eqref{aeq:robustness-dual}.  Now we show that strong duality holds, so that two sides of Eq.~\eqref{aeq:robustness-dual} are equal.

Since $R^{\s{K},\f{F}}(\ob{Q})$ has a bounded value, the primal problem on the right-hand side of Eq.~\eqref{aeq:robustness-primal} has a feasible solution.  Since $\s{K}\subseteq\q{Q}$ is normalized, we have $\langle\frac{1}{2}\spec{\ob{I}},\ob{K}\rangle=\frac{1}{2}>0$ for all $\ob{K}\in\s{K}^{(i)}$, and therefore $\frac{1}{2}\spec{\ob{I}}\in\intr(\cone^*(\s{K}^{(i)}))$, which implies that $\alpha=\frac{1}{2}\spec{\ob{I}}$ is an interior point of the dual feasible region.  Then by Lemma~\ref{alem:duality}(2), strong duality holds.
\end{proof}

\begin{theorem}
\label{athm:robustness}
Let $(\f{T},\f{F};\q{Q})$ be a closed and convex quasi-resource theory with $\q{Q}$ normalized, and let $\s{K}\subseteq\q{Q}$ be a subset of objects contractive under $\f{T}$ and satisfying the conditions in Proposition~\ref{aprop:robustness-program}.  Let $\ob{Q}\in\q{Q}^{(i)}$ be an object.  Then for any $j$,
\begin{align}
\label{aeq:robustness-characterization}
	1+R^{\s{K},\f{F}}(\ob{Q})&\geq\max_{\alpha\in\cone^*(\s{K}^{(j)})}\frac{\max_{\ob{T}\in\f{T}^{(i)\to(j)}}\left\langle\alpha,\ob{T}\left[\ob{Q}\right]\right\rangle}{\max_{\ob{F}\in\f{F}^{(j)}}\left\langle\alpha,\ob{F}\right\rangle}.
\end{align}
The inequality in Eq.~\eqref{aeq:robustness-characterization} becomes equality if $j=i$.
\end{theorem}

\begin{proof}
Let $\gamma_\star\in\cone(\f{F}^{(i)})$ be an optimizer of the program on the right-hand side of Eq.~\eqref{aeq:robustness-primal}.  Then for any $j$ and $\alpha\in\cone^*(\s{K}^{(j)})$, we have that
\begin{align}
	&\max_{\ob{T}\in\f{T}^{(i)\to(j)}}\left\langle\alpha,\ob{T}\left[\ob{Q}\right]\right\rangle \notag\\
	&\quad=\max_{\ob{T}\in\f{T}^{(i)\to(j)}}\left\langle\alpha,\ob{T}\left[\gamma_\star\right]-\ob{T}\left[\gamma_\star-\ob{Q}\right]\right\rangle \\
	&\quad\leq\max_{\ob{T}\in\f{T}^{(i)\to(j)}}\left\langle\alpha,\ob{T}\left[\gamma_\star\right]\right\rangle \label{aeq:robustness-3}\\
	&\quad=\left\langle\spec{\ob{I}},\gamma_\star\right\rangle\max_{\ob{T}\in\f{T}^{(i)\to(j)}}\left\langle\alpha,\ob{T}\left[\frac{\gamma_\star}{\left\langle\spec{\ob{I}},\gamma_\star\right\rangle}\right]\right\rangle \\
	&\quad\leq\left\langle\spec{\ob{I}},\gamma_\star\right\rangle\max_{\ob{F}\in\f{F}^{(j)}}\left\langle\alpha,\ob{F}\right\rangle \label{aeq:robustness-4} \\
	&\quad=\left(1+R^{\s{K},\f{F}}(\ob{Q})\right)\max_{\ob{F}\in\f{F}^{(j)}}\left\langle\alpha,\ob{F}\right\rangle. \label{aeq:robustness-5}
\end{align}
Here Eq.~\eqref{aeq:robustness-3} is due to $\gamma_\star-\ob{Q}\in\cone(\s{K}^{(i)})$ and that $\s{K}$ is contractive under $\f{T}$; Eq.~\eqref{aeq:robustness-4} is due to $\frac{\gamma_\star}{\langle\spec{\ob{I}},\gamma_\star\rangle}\in\f{F}^{(i)}$ and that $\f{F}$ is contractive under $\f{T}$; Eq.~\eqref{aeq:robustness-5} follows from Eq.~\eqref{aeq:robustness-primal}.  Given any $\ob{Q}'\in\q{Q}^{(j)}$, since $R^{\s{K},\f{F}}(\ob{Q}')$ has a bounded value, by Definition~\ref{adef:robustness}, there exists $\ob{K}_{\ob{Q}'}\in\s{K}^{(j)}$ such that $\frac{\ob{Q}'+R^{\s{K},\f{F}}(\ob{Q}')\ob{K}_{\ob{Q}'}}{1+R^{\s{K},\f{F}}(\ob{Q}')}\eqcolon\ob{F}_{\ob{Q}'}\in\f{F}^{(j)}$.  This implies that
\begin{align}
	\max_{\ob{F}\in\f{F}^{(j)}}\left\langle\alpha,\ob{F}\right\rangle&\geq\max_{\ob{Q}'\in\q{Q}^{(j)}}\left\langle\alpha,\ob{F}_{\ob{Q}'}\right\rangle \\
	&=\max_{\ob{Q}'\in\q{Q}^{(j)}}\left\langle\alpha,\frac{\ob{Q}'+R^{\s{K},\f{F}}(\ob{Q}')\ob{K}_{\ob{Q}'}}{1+R^{\s{K},\f{F}}(\ob{Q}')}\right\rangle \label{aeq:robustness-6}\\
	&\geq\max_{\ob{Q}'\in\q{Q}^{(j)}}\frac{1}{1+R^{\s{K},\f{F}}(\ob{Q}')}\left\langle\alpha,\ob{Q}'\right\rangle \\
	&\geq\max_{\ob{K}'\in\s{K}^{(j)}}\frac{1}{1+R^{\s{K},\f{F}}(\ob{K}')}\left\langle\alpha,\ob{K}'\right\rangle \label{aeq:robustness-7}\\
	&\geq0. \label{aeq:robustness-8}
\end{align}
Here Eq.~\eqref{aeq:robustness-6} is due to $\alpha\in\cone^*(\s{K}^{(j)})$ and $R^{\s{K},\f{Q}}(\ob{Q}')\geq0$; Eq.~\eqref{aeq:robustness-7} is due to $\s{K}^{(j)}\subseteq\q{Q}^{(j)}$.  Combining Eqs.~\eqref{aeq:robustness-5} and \eqref{aeq:robustness-8}, which hold for all $\alpha\in\cone^*(\s{K}^{(j)})$, we have that
\begin{align}
	1+R^{\s{K},\f{F}}(\ob{Q})&\geq\sup_{\alpha\in\cone^*(\s{K}^{(j)})}\frac{\max_{\ob{T}\in\f{T}^{(i)\to(j)}}\left\langle\alpha,\ob{T}\left[\ob{Q}\right]\right\rangle}{\max_{\ob{F}\in\f{F}^{(j)}}\left\langle\alpha,\ob{F}\right\rangle}. \label{aeq:robustness-9}
\end{align}
Now we show that the inequality in Eq.~\eqref{aeq:robustness-9} becomes equality if $j=i$.  For any feasible solution $\alpha\in\cone^*(\s{K}^{(i)})$ to the program on the right-hand side of Eq.~\eqref{aeq:robustness-dual}, we have that
\begin{align}
	\max_{\ob{F}\in\f{F}^{(i)}}\langle\alpha,\ob{F}\rangle&=\max_{\ob{F}\in\f{F}^{(i)}}\left\langle\spec{\ob{I}}-\left(\spec{\ob{I}}-\alpha\right),\ob{F}\right\rangle \\
	&\leq\max_{\ob{F}\in\f{F}^{(i)}}\left\langle\spec{\ob{I}},\ob{F}\right\rangle \label{aeq:robustness-10}\\
	&=1. \label{aeq:robustness-11}
\end{align}
Here Eq.~\eqref{aeq:robustness-10} is due to $\spec{\ob{I}}-\alpha\in\cone^*(\f{F}^{(i)})$; Eq.~\eqref{aeq:robustness-11} follows from the fact that $\f{F}\subseteq\q{Q}$ is normalized.  Then it follows that
\begin{align}
	&\sup_{\alpha\in\cone^*(\s{K}^{(i)})}\frac{\max_{\ob{T}\in\f{T}^{(i)\to(i)}}\left\langle\alpha,\ob{T}\left[\ob{Q}\right]\right\rangle}{\max_{\ob{F}\in\f{F}^{(i)}}\left\langle\alpha,\ob{F}\right\rangle} \notag\\
	&\quad\geq\sup_{\scriptsize\left\{\begin{array}{c}
		\alpha\in\cone^*(\s{K}^{(i)})\colon \\
		\spec{\ob{I}}-\alpha\in\cone^*(\f{F}^{(i)})
	\end{array}\right\}}\frac{\max_{\ob{T}\in\f{T}^{(i)\to(i)}}\left\langle\alpha,\ob{T}\left[\ob{Q}\right]\right\rangle}{\max_{\ob{F}\in\f{F}^{(i)}}\left\langle\alpha,\ob{F}\right\rangle} \\
	&\quad\geq\sup_{\scriptsize\left\{\begin{array}{c}
		\alpha\in\cone^*(\s{K}^{(i)})\colon \\
		\spec{\ob{I}}-\alpha\in\cone^*(\f{F}^{(i)})
	\end{array}\right\}}\max_{\ob{T}\in\f{T}^{(i)\to(i)}}\left\langle\alpha,\ob{T}\left[\ob{Q}\right]\right\rangle \label{aeq:robustness-12}\\
	&\quad\geq\sup_{\scriptsize\left\{\begin{array}{c}
		\alpha\in\cone^*(\s{K}^{(i)})\colon \\
		\spec{\ob{I}}-\alpha\in\cone^*(\f{F}^{(i)})
	\end{array}\right\}}\left\langle\alpha,\ob{Q}\right\rangle \label{aeq:robustness-13}\\
	&\quad=1+R^{\s{K},\f{F}}(\ob{Q}). \label{aeq:robustness-14}
\end{align}
Here Eq.~\eqref{aeq:robustness-12} uses Eq.~\eqref{aeq:robustness-11}; Eq.~\eqref{aeq:robustness-13} uses the property $\id^{(i)}\in\f{T}^{(i)\to(i)}$; Eq.~\eqref{aeq:robustness-14} uses Eq.~\eqref{aeq:robustness-dual}.  Setting $j=i$ in Eq.~\eqref{aeq:robustness-9} and combining it with Eq.~\eqref{aeq:robustness-14}, we have that
\begin{align}
	1+R^{\s{K},\f{F}}(\ob{Q})&=\sup_{\alpha\in\cone^*(\s{K}^{(i)})}\frac{\max_{\ob{T}\in\f{T}^{(i)\to(i)}}\left\langle\alpha,\ob{T}\left[\ob{Q}\right]\right\rangle}{\max_{\ob{F}\in\f{F}^{(i)}}\left\langle\alpha,\ob{F}\right\rangle}. \label{aeq:robustness-15}
\end{align}
The supremums on the right-hand sides of Eqs.~\eqref{aeq:robustness-9} and \eqref{aeq:robustness-15} can be taken as maximums since they are reachable by functionals from a bounded and closed set.
\end{proof}

\begin{corollary}
\label{acor:robustness}
Let $(\f{T},\f{F};\q{Q})$ be a closed and convex quasi-resource theory with $\q{Q}$ normalized, and let $\s{K}\subseteq\q{Q}$ be a subset of objects satisfying the conditions in Proposition~\ref{aprop:robustness-program}.  Let $\ob{Q}\in\q{Q}^{(i)}$ be an object.  Then
\begin{align}
	1+R^{\s{K},\f{F}}(\ob{Q})&=\max_{\alpha\in\cone^*(\s{K}^{(i)})}\frac{\left\langle\alpha,\ob{Q}\right\rangle}{\max_{\ob{F}\in\f{F}^{(i)}}\left\langle\alpha,\ob{F}\right\rangle}.
\end{align}
\end{corollary}

\begin{proof}
Denote $\f{I}\equiv\{\id^{(i)}\}_i\subseteq\bigcup_{i,j}\spa{T}^{(i)\to(j)}$.  Since $(\f{T},\f{F};\q{Q})$ is a closed and convex quasi-resource theory, so is $(\f{I},\f{F};\q{Q})$.  In addition, $\s{K}$ is contractive under $\f{I}$.  Applying Theorem~\ref{athm:robustness} to $(\f{I},\f{F};\q{Q})$, we have that
\begin{align}
	1+R^{\s{K},\f{F}}(\ob{Q})&=\max_{\alpha\in\cone^*(\s{K}^{(i)})}\frac{\max_{\ob{T}\in\f{I}^{(i)}}\left\langle\alpha,\ob{T}\left[\ob{Q}\right]\right\rangle}{\max_{\ob{F}\in\f{F}^{(i)}}\left\langle\alpha,\ob{F}\right\rangle} \\
	&=\max_{\alpha\in\cone^*(\s{K}^{(i)})}\frac{\left\langle\alpha,\ob{Q}\right\rangle}{\max_{\ob{F}\in\f{F}^{(i)}}\left\langle\alpha,\ob{F}\right\rangle}. \label{aeq:robustness-16}
\end{align}
Here Eq.~\eqref{aeq:robustness-16} is due to $\f{I}^{(i)}=\{\id^{(i)}\}$.
\end{proof}

\begin{corollary}
\label{acor:robustness-limit}
Let $(\f{T},\f{F};\q{Q})$ be a closed and convex quasi-resource theory with $\q{Q}$ normalized such that the following condition holds:
\begin{enumerate}
	\item[(1)] there exists $k$ and an object $\spec{\ob{P}}\in\q{Q}^{(k)}$ such that $\f{F}^{(i)}=\{\ob{T}[\spec{\ob{P}}]\colon\ob{T}\in\f{T}^{(k)\to(i)}\}$ for all $i$.
\end{enumerate}
Let $\s{K}\subseteq\q{Q}$ be a subset of objects contractive under $\f{T}$ and satisfying the conditions in Proposition~\ref{aprop:robustness-program}.  Let $\{f_w\}_{w\in\idx{W}}$ be a set of functions with $f_w\colon\q{Q}\to\spa{R}$ for all $w\in\idx{W}$ such that the following condition holds:
\begin{enumerate}
	\item[(2)] for any $w\in\idx{W}$, there exists $j_w$ and $\alpha_w\in\cone^*(\s{K}^{(j_w)})$ such that $f_w(\ob{Q}')=\max_{\ob{T}\in\f{T}^{(i')\to(j_w)}}\langle\alpha_w,\ob{T}[\ob{Q}']\rangle$ for all $\ob{Q}'\in\q{Q}^{(i')}$ and all $i'$.
\end{enumerate}
Let $\ob{Q}\in\q{Q}^{(i)}$ be an object.  If there exists a sequence of mappings $\{h_d\}_{d=2}^\infty$ with $h_d\colon\cone^*(\s{K}^{(i)})\to\idx{W}$ for all $d\geq2$ such that the following condition holds:
\begin{enumerate}
	\item[(3)] given any functional $\alpha\in\cone^*(\s{K}^{(i)})$, there exists a real number $c_\alpha\in\spa{R}$ such that for any $i'$ and object $\ob{Q}'\in\q{Q}^{(i')}$,
	\begin{align}
		\lim_{d\to\infty}f_{h_d(\alpha)}(\ob{Q}')&=c_\alpha\max_{\ob{T}\in\f{T}^{(i')\to(i)}}\left\langle\alpha,\ob{T}\left[\ob{Q}'\right]\right\rangle,
	\end{align} 
\end{enumerate}
then 
\begin{align}
	1+R^{\s{K},\f{F}}(\ob{Q})&=\sup_{w\in\idx{W}}\frac{f_w(\ob{Q})}{\max_{\ob{F}\in\f{F}}f_w(\ob{F})}.
\end{align}
\end{corollary}

\begin{proof}
Let $\{h_d\}_{d=2}^\infty$ be a sequence of mappings with $h_d\colon\cone^*(\s{K}^{(i)})\to\idx{W}$ for all $d\geq2$ such that Condition~(3) holds.  Then for any $\alpha\in\cone^*(\s{K}^{(i)})$,
\begin{align}
	&\lim_{d\to\infty}\max_{\ob{F}\in\f{F}}f_{h_d(\alpha)}(\ob{F}) \notag\\
	&\quad=\lim_{d\to\infty}\sup_{i'}\max_{\ob{F}\in\f{F}^{(i')}}\max_{\ob{T}\in\f{T}^{(i')\to(j_{h_d(\alpha)})}}\left\langle\alpha_{h_d(\alpha)},\ob{T}\left[\ob{F}\right]\right\rangle \label{aeq:robustness-limit-1}\\
	&\quad=\lim_{d\to\infty}\max_{\ob{T}\in\f{T}^{(k)\to(j_{h_d(\alpha)})}}\left\langle\alpha_{h_d(\alpha)},\ob{T}\left[\spec{\ob{P}}\right]\right\rangle \label{aeq:robustness-limit-2}\\
	&\quad=\lim_{d\to\infty}f_{h_d(\alpha)}(\spec{\ob{P}}) \label{aeq:robustness-limit-3}\\
	&\quad=c_\alpha\max_{\ob{T}\in\f{T}^{(k)\to(i)}}\left\langle\alpha,\ob{T}\left[\spec{\ob{P}}\right]\right\rangle \label{aeq:robustness-limit-4}\\
	&\quad=c_\alpha\max_{\ob{F}\in\f{F}^{(i)}}\left\langle\alpha,\ob{F}\right\rangle. \label{aeq:robustness-limit-5}
\end{align}
Here Eqs.~\eqref{aeq:robustness-limit-1} and \eqref{aeq:robustness-limit-3} use Condition~(2); Eq.~\eqref{aeq:robustness-limit-2} and \eqref{aeq:robustness-limit-5} use Condition~(1); Eq.~\eqref{aeq:robustness-limit-4} uses Condition~(3).  By Theorem~\ref{athm:robustness}, we have that
\begin{align}
	&1+R^{\s{K},\f{F}}(\ob{Q}) \notag\\
	&\quad=\max_{\alpha\in\cone^*(\s{K}^{(i)})}\frac{\max_{\ob{T}\in\f{T}^{(i)\to(i)}}\left\langle\alpha,\ob{T}\left[\ob{Q}\right]\right\rangle}{\max_{\ob{F}\in\f{F}^{(i)}}\left\langle\alpha,\ob{F}\right\rangle} \\
	&\quad=\max_{\alpha\in\cone^*(\s{K}^{(i)})}\lim_{d\to\infty}\frac{f_{h_d(\alpha)}(\ob{Q})}{\max_{\ob{F}\in\f{F}}f_{h_d(\alpha)}(\ob{F})} \label{aeq:robustness-limit-6}\\
	&\quad\leq\lim_{d\to\infty}\max_{\alpha\in\cone^*(\s{K}^{(i)})}\frac{f_{h_d(\alpha)}(\ob{Q})}{\max_{\ob{F}\in\f{F}}f_{h_d(\alpha)}(\ob{F})} \\
	&\quad\leq\sup_{w\in\idx{W}}\frac{f_w(\ob{Q})}{\max_{\ob{F}\in\f{F}}f_w(\ob{F})}. \label{aeq:robustness-limit-7}
\end{align}
Here Eq.~\eqref{aeq:robustness-limit-6} uses Condition~(3) and Eq.~\eqref{aeq:robustness-limit-5}.  Conversely, by Theorem~\ref{athm:robustness},
\begin{align}
	&1+R^{\s{K},\f{F}}(\ob{Q}) \notag\\
	&\quad\geq\sup_{j}\max_{\alpha\in\cone^*(\s{K}^{(j)})}\frac{\max_{\ob{T}\in\f{T}^{(i)\to(j)}}\left\langle\alpha,\ob{T}\left[\ob{Q}\right]\right\rangle}{\max_{\ob{F}\in\f{F}^{(j)}}\left\langle\alpha,\ob{F}\right\rangle} \\
	&\quad=\sup_{j}\max_{\alpha\in\cone^*(\s{K}^{(j)})}\frac{\max_{\ob{T}\in\f{T}^{(i)\to(j)}}\left\langle\alpha,\ob{T}\left[\ob{Q}\right]\right\rangle}{\max_{\ob{T}\in\f{T}^{(k)\to(j)}}\left\langle\alpha,\ob{T}\left[\spec{\ob{P}}\right]\right\rangle} \label{aeq:robustness-limit-8}\\
	&\quad\geq\sup_{w\in\idx{W}}\frac{\max_{\ob{T}\in\f{T}^{(i)\to(j_w)}}\left\langle\alpha_w,\ob{T}\left[\ob{Q}\right]\right\rangle}{\max_{\ob{T}\in\f{T}^{(k)\to(j_w)}}\left\langle\alpha_w,\ob{T}\left[\spec{\ob{P}}\right]\right\rangle} \label{aeq:robustness-limit-9}\\
	&\quad=\sup_{w\in\idx{W}}\frac{f_w(\ob{Q})}{\max_{\ob{T}\in\f{T}^{(k)\to(j_w)}}f_w(\ob{T}\left[\spec{\ob{P}}\right])} \label{aeq:robustness-limit-10}\\
	&\quad=\sup_{w\in\idx{W}}\frac{f_w(\ob{Q})}{\max_{\ob{F}\in\f{F}^{(j_w)}}f_w(\ob{F})} \label{aeq:robustness-limit-11}\\
	&\quad\geq\sup_{w\in\idx{W}}\frac{f_w(\ob{Q})}{\max_{\ob{F}\in\f{F}}f_w(\ob{F})}. \label{aeq:robustness-limit-12}
\end{align}
Here Eqs.~\eqref{aeq:robustness-limit-8} and \eqref{aeq:robustness-limit-11} use Condition~(1); Eqs.~\eqref{aeq:robustness-limit-9} and \eqref{aeq:robustness-limit-10} use Condition~(2).  Combining Eqs.~\eqref{aeq:robustness-limit-7} and \eqref{aeq:robustness-limit-12}, we have that
\begin{align}
	1+R^{\s{K},\f{F}}(\ob{Q})&=\sup_{w\in\idx{W}}\frac{f_w(\ob{Q})}{\max_{\ob{F}\in\f{F}}f_w(\ob{F})}.
\end{align}
\end{proof}

\section{Entropic characterization of quantum resources}
\label{app:characterization-entropy}

Let $(\f{S},\f{C})$ be a closed and convex dynamic QRT\@.  By Definition~\ref{adef:resource}, this implies that $(\f{S},\f{C};\q{L})$ is a closed and convex quasi-resource theory.

\subsection{Proof of Theorem~\ref{thm:convertibility-deterministic}}
\label{app:convertibility-deterministic}

The implication (1) $\Rightarrow$ (2) follows from Proposition~\ref{prop:monotonicity}(4).  The implication (2) $\Rightarrow$ (3) is self-evident.  We now show (3) $\Rightarrow$ (1), and then (1) $\Rightarrow$ (4) and (4) $\Rightarrow$ (1).

Let $\pmt{D}$ denote the set of collections of states with an index range $\idx{N}$ in $\BI$:
\begin{align}
	\label{aeq:convertibility-deterministic-1}
	\pmt{D}&\coloneq\left\{\left\{\omega_n\right\}_{n\in\idx{N}}\colon\omega_n\in\q{D}^\BI\;\forall n\in\idx{N}\right\}.
\end{align}
For every $\{\omega_n\}_{n\in\idx{N}}\in\pmt{D}$, define a measure-and-prepare channel $\Omega\in\q{L}^\B$ according to Eq.~\eqref{eq:convertibility-measure-and-prepare}.  Let $\{f_{\{\omega_n\}_n}\}_{\{\omega_n\}_n\in\pmt{D}}$ be a set of functions with $f_{\{\omega_n\}_n}\colon\q{L}\to\spa{R}$ for all $\{\omega_n\}_n\in\pmt{D}$, defined as follows:
\begin{align}
	f_{\{\omega_n\}_n}(\Lambda)&\coloneq d_\BO2^{-H_{\min}^\f{S}(\RB|\A)_{\Lambda^\A\otimes\Omega^\RB}} \\
	&=\max_{\sm{\Theta}\in\f{S}^{\A\To\B}}\tr\left[\spec{\phi}_+^{\BI\RBI}\left(\sm{\Theta}^{\A\To\B}\left\{\Lambda^\A\right\}\otimes\Omega^\RB\right)\left[\spec{\phi}_+^{\BO\RBO}\right]\right] \label{aeq:convertibility-deterministic-2}\\
	&=\max_{\sm{\Theta}\in\f{S}^{\A\To\B}}\tr\left[J_\cc{\Omega}J_{\sm{\Theta}\left\{\Lambda\right\}}\right] \\
	&=\max_{\sm{\Theta}\in\f{S}^{\A\To\B}}\tr\left[\left(\sum_{n}N_n^\BO\otimes\cc{\omega}_n^\BI\right)J_{\sm{\Theta}\left\{\Lambda\right\}}\right]\quad\forall\Lambda\in\q{L}. \label{aeq:convertibility-deterministic-3}
\end{align}
Here Eq.~\eqref{aeq:convertibility-deterministic-2} follows from Eq.~\eqref{eq:EFCME-dual-2}; Eq.~\eqref{aeq:convertibility-deterministic-3} follows from Eq.~\eqref{eq:convertibility-measure-and-prepare}.  By Proposition~\ref{prop:monotonicity}(4), $f_{\{\omega_n\}_n}$ is a resource monotone for all $\{\omega_n\}_n\in\pmt{D}$.  In what follows, we define a mapping $h\colon\aff(\q{L}^\B)\to\pmt{D}$.  Since the POVM $\povm{N}\equiv\{N_n\}_{n\in\idx{N}}$ is informationally complete and has linearly independent elements, it forms a basis of $\spa{H}^\BO$.  Let $\{K_n\}_{n\in\idx{N}}$ be the dual basis of $\povm{N}$ with $K_n\in\spa{H}^\BO$ for all $n\in\idx{N}$, and it satisfies that
\begin{align}
	\tr\left[K_nN_{n'}\right]&=\delta_{n,n'}\quad\forall n,n'\in\idx{N},
\end{align}
where $\delta_{n,n'}\coloneq1$ if $n=n'$ and $0$ otherwise.  The affine space generated by the set of channels $\q{L}^\B$ is given by
\begin{align}
	\aff(\q{L}^\B)&=\left\{\Xi\in\spa{L}^\B\colon J_\Xi^\BO=c\spec{I}^\BO,\;c\in\spa{R}\right\}.
\end{align}
For any linear map $\Xi\in\aff(\q{L}^\B)$, there exists a real number $c\in\spa{R}$ such that $J_\Xi^\BO=c\spec{I}^\BO$, and there exists a set of operators $\{\xi_n\}_{n\in\idx{N}}$ with $\xi_n\in\spa{H}^\BI$ for all $n\in\idx{N}$ such that
\begin{align}
\label{aeq:convertibility-deterministic-4}
	J_\Xi^{\BO\BI}&=\sum_{n}N_n^\BO\otimes\xi_n^\BI.
\end{align}
Then we have that
\begin{align}
	\tr\left[\xi_n\right]&=\sum_{n'}\tr\left[K_nN_{n'}\right]\tr\left[\xi_{n'}\right] \\
	&=\tr\left[\left(K_n^\BO\otimes\spec{I}^\BI\right)\left(\sum_{n'}N_{n'}^\BO\otimes\xi_{n'}^\BI\right)\right] \\
	&=\tr\left[\left(K_n^\BO\otimes\spec{I}^\BI\right)J_\Xi^{\BO\BI}\right] \\
	&=\tr\left[K_nJ_\Xi^\BO\right] \\
	&=c\tr\left[K_n\right] \\
	&=c\sum_{n'}\tr\left[K_nN_{n'}\right] \label{aeq:convertibility-deterministic-5}\\
	&=c\quad\forall n\in\idx{N}.
\end{align}
Here Eq.~\eqref{aeq:convertibility-deterministic-5} is due to $\sum_{n'}N_{n'}=\spec{I}$.  Define $b\coloneq\max_n\|\xi_n\|_\infty$.  Define $h(\Xi)\coloneq\{\eta_n\}_{n\in\idx{N}}$ with $\eta_n\in\q{D}^\BI$ for all $n\in\idx{N}$ as follows:
\begin{align}
\label{aeq:convertibility-deterministic-6}
	\eta_n&\coloneq\frac{1}{c+d_\BI b }\left(\cc{\xi}_n+b\spec{I}\right)\quad\forall n\in\idx{N}.
\end{align}
It can be verified that $\eta_n$ is a valid state for all $n\in\idx{N}$, as $\eta_n\in\spa{H}_+^\BI$ and $\tr[\eta_n]=1$ for all $n\in\idx{N}$.  Then by Eqs.~\eqref{aeq:convertibility-deterministic-3} and \eqref{aeq:convertibility-deterministic-6},
\begin{align}
	&f_{h(\Xi)}(\Lambda) \notag\\
	&\quad=\frac{1}{c+d_\BI b}\max_{\sm{\Theta}\in\f{S}^{\A\To\B}}\tr\left[\left(\sum_{n}N_n^\BO\otimes\xi_n^\BI+b\spec{I}^{\BO\BI}\right)J_{\sm{\Theta}\left\{\Lambda\right\}}\right] \label{aeq:convertibility-deterministic-7}\\
	&\quad=\frac{1}{c+d_\BI b}\left(\max_{\sm{\Theta}\in\f{S}^{\A\To\B}}\tr\left[J_\Xi J_{\sm{\Theta}\left\{\Lambda\right\}}\right]+d_\BO b\right) \label{aeq:convertibility-deterministic-8}\\
	&\quad=\frac{1}{c+d_\BI b}\left(\max_{\sm{\Theta}\in\f{S}^{\A\To\B}}\left\langle\Xi,\sm{\Theta}\left\{\Lambda\right\}\right\rangle+d_\BO b\right)\quad\forall\Lambda\in\q{L}. \label{aeq:convertibility-deterministic-9}
\end{align}
Here Eq.~\eqref{aeq:convertibility-deterministic-7} is due to $\sum_{n}N_n=\spec{I}$; Eq.~\eqref{aeq:convertibility-deterministic-8} follows from Eq.~\eqref{aeq:convertibility-deterministic-3}.  By Eq.~\eqref{aeq:convertibility-deterministic-9}, a sequence of the same mapping $h$ satisfies Condition~(1) of Theorem~\ref{athm:convertibility} given the set of resource monotones $\{f_{\{\omega_n\}_n}\}_{\{\omega_n\}_n\in\pmt{D}}$ by substituting $\spa{U}^{(i)}=\aff(\q{L}^\A)$, $\spa{U}^{(j)}=\aff(\q{L}^\B)$, $\q{Q}=\q{L}$, $\f{T}=\f{S}$, $\f{F}=\f{C}$, and $\idx{W}=\pmt{D}$.  By Theorem~\ref{athm:convertibility}, $\{f_{\{\omega_n\}_n}\}_{\{\omega_n\}_n\in\pmt{D}}$ is complete with respect to $\A$ and $\B$, and consequently the implication (3) $\Rightarrow$ (1) holds.

To show (1) $\Rightarrow$ (4), we note that if there exists a free superoperation $\sm{\Theta}\in\f{S}^{\A\To\B}$ such that $\sm{\Theta}\{\Lambda\}=\Psi$, then $r=1$, $\alpha=(J_\sm{\Theta}^\top)^{\AO\AI\BO\BI}$ is a feasible solution to the program in Eq.~\eqref{eq:convertibility-program}, which implies $g(\Lambda,\Psi)\geq1$.

To show (4) $\Rightarrow$ (1), we define a new dynamic QRT $(\f{S}',\f{C}')$ as follows.  Specifically, the new set of free superoperations $\f{S}'$ is defined as
\begin{align}
	\f{S}'&\coloneq\f{S}^{\A\To\B}\cup\bigcup_{\BPO,\BPI}\left\{\Id^\BP\right\},
\end{align}
and the new set of free channels $\f{C}'$ is reduced from $\f{S}'$ by Definition~\ref{def:QRT-dynamic}.  Since $(\f{S},\f{C})$ is a closed and convex dynamic QRT, so is $(\f{S}',\f{C}')$.  Applying the equivalence (1) $\Leftrightarrow$ (3) in Theorem~\ref{athm:convertibility} (which has just been proved) to $(\f{S}',\f{C}')$ with respect to the POVM $\{\cc{N}_n\}_{n\in\idx{N}}$, we know that there exists a free superoperation $\sm{\Theta}\in(\f{S}')^{\A\To\B}=\f{S}^{\A\To\B}$ such that $\sm{\Theta}\{\Lambda\}\in\Psi$ if and only if
\begin{align}
	\label{aeq:convertibility-deterministic-10}
	H_{\min}^{\f{S}'}(\RB|\A)_{\Lambda^\A\otimes\Omega^\RB}&\leq H_{\min}^{\f{S}'}(\RB|\B)_{\Psi^\B\otimes\Omega^\RB}
\end{align}
for every measure-and-prepare channel $\Omega\in\q{L}^\B$ of the form
\begin{align}
	\Omega^\B\left[\cdot\right]&=\sum_{n}\tr\left[\cc{N}_n\left(\cdot\right)\right]\omega_n.
\end{align}
By Eq.~\eqref{eq:EFCME-dual-1}, and due to $(\f{S}')^{\A\To\B}=\f{S}^{\A\To\B}$ and $(\f{S}')^{\B\To\B}=\{\Id^\B\}$, this is equivalent to
\begin{align}
	&\max_{\sm{\Theta}\in\f{S}^{\A\To\B}}\sum_{n}\tr\left[J_\sm{\Theta}^{\AO\AI\BO\BO}\left(J_\Lambda^{\AO\AI}\otimes N_n^\BO\otimes\omega_n^\BI\right)\right] \notag\\
	&\quad\geq\sum_{n}\tr\left[\cc{J}_\Psi^{\BO\BI}\left(N_n^\BO\otimes\omega_n^\BI\right)\right]
\end{align}
for all $\{\omega_n\}_{n\in\idx{N}}\in\pmt{D}$.  This condition can be reformulated as $\bar{g}(\Lambda,\Psi)\geq1$, where
\begin{subequations}
\begin{align}
	\bar{g}(\Lambda,\Psi)&\coloneq\max r \\
	\textnormal{s.t.:}&\quad r\in\spa{R}, \\
	&\quad\sm{\Theta}\in\f{S}^{\A\To\B}, \\
	&\quad\tr_{\AO\AI\BO}\left[J_\sm{\Theta}^{\AO\AI\BO\BI}\left(J_\Lambda^{\AO\AI}\otimes N_n^\BO\otimes\omega_n^\BI\right)\right] \notag\\
	&\quad\quad\geq r\tr_\BO\left[(J_\Psi^\top)^{\BO\BI}\left(N_n^\BO\otimes\omega_n^\BI\right)\right] \notag\\
	&\quad\quad\quad\forall\left\{\omega_n\right\}_{n\in\idx{N}}\in\pmt{D}. \label{aeq:convertibility-deterministic-11}
\end{align}
\end{subequations}
By Lemma~\ref{lem:superchannel}, $\sm{\Theta}\in\f{S}^{\A\To\B}$ is equivalent to Eqs.~\eqref{eq:convertibility-program-2}--\eqref{eq:convertibility-program-3} with $\alpha\in\cone(\f{S}_\abb{J}^{\A\To\B})$; Eq.~\eqref{aeq:convertibility-deterministic-11} is implied by Eq.~\eqref{eq:convertibility-program-1}.  Consequently, $\bar{g}(\Lambda,\Psi)\geq g(\Lambda,\Psi)$, which is defined in Eq.~\eqref{eq:convertibility-program}.  Since Condition~(1) is equivalent to $\bar{g}(\Lambda,\Psi)\geq1$, we conclude that if $g(\Lambda,\Psi)\geq1$, then Condition~(1) holds.

\subsection{Proof of Theorem~\ref{thm:convertibility-probabilistic}}
\label{app:convertibility-probabilistic}

For every subchannel $\Gamma_0\in\q{L}_\leqslant^\B$, define a subchannel $\Omega_0\in\q{L}_\leqslant^{\B^\times}$ according to Eq.~\eqref{eq:convertibility-QC}.  Let $\{f_{\Gamma_0}\}_{\Gamma_0\in\q{L}_\leqslant^\B}$ be a set of functions with $f_{\Gamma_0}\colon\q{L}\to\spa{R}$ for all $\Gamma_0\in\q{L}_\leqslant^\B$, defined as follows:
\begin{align}
	&f_{\Gamma_0}(\Lambda_0) \notag\\
	&\quad\coloneq d_\BO2^{-H_{\min}^\f{S}(\RB^\times|\A)_{\Lambda_0^\A\otimes\Omega_0^{\RB^\times}+\Lambda_1^\A\otimes\frac{2}{3}\spec{\Pi}^{\RB^\times}}} \\
	&\quad=\max_{\sm{\Theta}\in\f{S}^{\A\To\B^\times}}\tr\left[\left(\spec{\phi}_+^{\BI\RBI}\otimes\spec{\phi}_+^{\XI\RXI}\right)\left((\sm{\Theta}\left\{\Lambda_0\right\})^{\B^\times}\otimes\Omega_0^{\RB^\times}\phantom{\frac{1}{1}}\right.\right. \notag\\
	&\quad\quad\left.\left.\vphantom{}+(\sm{\Theta}\left\{\Lambda_1\right\})^{\B^\times}\otimes\frac{2}{3}\spec{\Pi}^{\RB^\times}\right)\left[\spec{\phi}_+^{\BO\RBO}\right]\right] \label{aeq:convertibility-probabilistic-1}\\
	&\quad=\max_{\sm{\Theta}\in\f{S}^{\A\To\B^\times}}\tr\left[J_{\cc{\Omega}_0}^{\BO\BI\XI}J_{\sm{\Theta}\left\{\Lambda_0\right\}}^{\BO\BI\XI}+\frac{2}{3}J_\spec{\Pi}^{\BO\BI\XI}J_{\sm{\Theta}\left\{\Lambda_1\right\}}^{\BO\BI\XI}\right] \\
	&\quad=\frac{1}{3}\max_{\sm{\Theta}\in\f{S}^{\A\To\B^\times}}\tr\left[\left(\left(2J_{\cc{\Gamma}_0}^{\BO\BI}-J_\spec{\Pi}^{\BO\BI}\right)\otimes\op{0}{0}^\XI\right)J_{\sm{\Theta}\left\{\Lambda_0\right\}}^{\BO\BI\XI}\right. \notag\\
	&\quad\quad\left.\vphantom{}+2J_\spec{\Pi}^{\BO\BI\XI}J_{\sm{\Theta}\left\{\Lambda_0+\Lambda_1\right\}}^{\BO\BI\XI}\right] \label{aeq:convertibility-probabilistic-2} \\
	&\quad=\frac{1}{3}\max_{\sm{\Theta}\in\f{S}^{\A\To\B^\times}}\tr\left[\left(\left(2J_{\cc{\Gamma}_0}^{\BO\BI}-J_\spec{\Pi}^{\BO\BI}\right)\otimes\op{0}{0}^\XI\right)J_{\sm{\Theta}\left\{\Lambda_0\right\}}^{\BO\BI\XI}\right] \notag\\
	&\quad\quad+\frac{d_\BO}{3d_\BI}\quad\forall\Lambda_0\in\q{L}_\leqslant. \label{aeq:convertibility-probabilistic-3}
\end{align}
Here Eq.~\eqref{aeq:convertibility-probabilistic-1} follows from Eq.~\eqref{eq:EFCME-dual-2}; Eq.~\eqref{aeq:convertibility-probabilistic-2} follows from Eq.~\eqref{eq:convertibility-QC}; Eq.~\eqref{aeq:convertibility-probabilistic-3} is due to $\sm{\Theta}\{\Lambda_0+\Lambda_1\}\in\q{L}^\B$.  Define a mapping $h\colon\spa{L}^\B\to\q{L}_\leqslant^\B$ as follows.  For any linear map $\Xi\in\spa{L}^\B$, define
\begin{align}
\label{aeq:convertibility-probabilistic-4}
	h(\Xi)&\coloneq\frac{1}{2d_\BI\left\|J_\Xi\right\|_\infty}\cc{\Xi}+\frac{1}{2}\spec{\Pi}.
\end{align}
It can be verified that $h(\Xi)$ is a valid subchannel, as $J_{h(\Xi)}\in\spa{H}_+^{\BO\BI}$ and $\spec{I}^\BO-J_{h(\Xi)}^\BO\in\spa{H}_+^\BO$.  By Eqs.~\eqref{aeq:convertibility-probabilistic-3} and \eqref{aeq:convertibility-probabilistic-4},
\begin{align}
	&f_{h(\Xi)}(\Lambda_0) \notag\\
	&\quad=\frac{1}{3d_\BI\left\|J_\Xi\right\|_\infty}\max_{\sm{\Theta}\in\f{S}^{\A\To\B^\times}}\tr\left[\left(J_\Xi^{\BO\BI}\otimes\op{0}{0}^\XI\right)J_{\sm{\Theta}\left\{\Lambda_0\right\}}^{\BO\BI\XI}\right] \notag\\
	&\quad\quad+\frac{d_\BO}{3d_\BI} \label{aeq:convertibility-probabilistic-5}\\
	&\quad=\frac{1}{3d_\BI\left\|J_\Xi\right\|_\infty}\max_{\sm{\Theta}_0\in\f{S}_\leqslant^{\A\To\B}}\tr\left[J_\Xi^{\BO\BI}J_{\sm{\Theta}_0\left\{\Lambda_0\right\}}^{\BO\BI}\right]+\frac{d_\BO}{3d_\BI} \label{aeq:convertibility-probabilistic-6}\\
	&\quad=\frac{1}{3d_\BI\left\|J_\Xi\right\|_\infty}\max_{\sm{\Theta}_0\in\f{S}_\leqslant^{\A\To\B}}\left\langle\Xi,\sm{\Theta}_0\left\{\Lambda_0\right\}\right\rangle+\frac{d_\BO}{3d_\BI} \notag\\
	&\quad\quad\forall\Lambda_0\in\q{L}_\leqslant. \label{aeq:convertibility-probabilistic-7}
\end{align}
Here Eq.~\eqref{aeq:convertibility-probabilistic-6} is by Definition~\ref{def:probabilistic-dynamic}.  By Eq.~\eqref{aeq:convertibility-probabilistic-7}, the mapping $h$ satisfies Condition~(1) of Corollary~\ref{acor:convertibility} given the set of functions $\{f_{\Gamma_0}\}_{\Gamma_0\in\q{L}_\leqslant^\B}$ by substituting $\spa{U}^{(i)}=\spa{L}^\A$, $\spa{U}^{(j)}=\spa{L}^\B$, $\q{Q}=\q{L}_\leqslant$, $\f{T}=\f{S}_\leqslant$, $\f{F}=\emptyset$, and $\idx{W}=\q{L}_\leqslant^\B$.  Since $(\f{S}_\leqslant,\emptyset;\q{L})$ is a closed and convex quasi-resource theory, and since the mapping $h$ is surjective by definition, by Corollary~\ref{acor:convertibility}, $\{f_{\Gamma_0}\}_{\Gamma_0\in\q{L}_\leqslant^\B}$ is a complete set of resource monotones with respect to $\A$ and $\B$ in $(\f{S}_\leqslant,\emptyset;\q{Q})$, and therefore it is a complete set of resource monotones in $(\f{S},\f{C})$ in the probabilistic sense.

\subsection{Proof of Theorem~\ref{thm:robustness-entropy}}
\label{app:robustness-entropy}

The dynamic QRT $(\f{S},\f{C})$ satisfies Condition~(1) of Corollary~\ref{acor:robustness-limit} by substituting $\spa{U}^{(i)}=\spa{L}^\A$, $\q{Q}=\q{L}$, $\f{T}=\f{S}$, $\f{F}=\f{C}$, $\spa{U}^{(k)}=\spa{L}^{\#\to\#}$, and $\spec{\ob{P}}=1$.  Let $\{f_\Omega\}_{\Omega\in\q{L}}$ be a set of functions with $f_\Omega\colon\q{L}\to\spa{R}$ for all $\Omega\in\q{L}$, defined as follows:
\begin{align}
	f_\Omega(\Lambda)&\coloneq d_\BO2^{-H_{\min}^\f{S}(\B|\AP)_{\Lambda^\AP\otimes\Omega^\B}} \\
	&=\max_{\sm{\Theta}\in\f{S}^{\AP\To\B}}\tr\left[\spec{\phi}_+^{\RBI\BI}\left(\sm{\Theta}^{\AP\To\RB}\left\{\Lambda^\AP\right\}\otimes\Omega^\B\right)\left[\spec{\phi}_+^{\RBO\BO}\right]\right] \label{aeq:robustness-entropy-1}\\
	&=\max_{\sm{\Theta}\in\f{S}^{\AP\To\B}}\tr\left[J_\cc{\Omega}J_{\sm{\Theta}\left\{\Lambda\right\}}\right] \label{aeq:robustness-entropy-2}\\
	&=\max_{\sm{\Theta}\in\f{S}^{\AP\To\B}}\left\langle\cc{\Omega},\sm{\Theta}\left\{\Lambda\right\}\right\rangle\quad\forall\Lambda\in\q{L}. \label{aeq:robustness-entropy-3}
\end{align}
Here Eq.~\eqref{aeq:robustness-entropy-1} follows from Eq.~\eqref{eq:EFCME-dual-2}.  By Eq.~\eqref{aeq:robustness-entropy-3}, the set of functions $\{f_\Omega\}_{\Omega\in\q{L}}$ satisfies Condition~(2) of Corollary~\ref{acor:robustness-limit} by substituting $\spa{U}^{(i')}=\spa{L}^\AP$, $\q{Q}=\q{L}$, $\s{K}=\q{L}$, $\idx{W}=\q{L}$, $\spa{U}^{(j_\Omega)}=\spa{L}^\B$, and $\alpha_\Omega=\cc{\Omega}$.  In what follows, we define a sequence of mappings $\{h_d\}_{d=2}^\infty$ with $h_d\colon\cone^*(\q{L}^\A)\to\q{L}$ for all $d\geq2$.  For any linear map $\Xi\in\cone^*(\q{L}^\A)$, there exists a positive semidefinite operator $\xi\in\spa{H}_+^{\AO\AI}$ and an operator $\beta\in\spa{H}^\AO$ with $\tr[\beta]=0$ such that
\begin{align}
	\label{aeq:robustness-entropy-4}
	J_\Xi^{\AO\AI}&=\xi^{\AO\AI}+\beta^\AO\otimes\spec{I}^\AI.
\end{align}
For any $d\geq2$,  define $h_d(\Xi)\in\q{L}^{\A^\times}$ with $\A^\times\equiv\AO\to\AI\XI$ and $d_\XI\coloneq d$ as follows:
\begin{align}
	J_{h_d(\Xi)}^{\AO\AI\XI}&\coloneq\frac{1}{\left\|\xi^\AO\right\|_\infty}\cc{\xi}^{\AO\AI}\otimes\op{0}{0}^\XI \notag\\
	&\quad+\sum_{x_1=1}^{d-1}\frac{1}{d-1}\left(\spec{I}^\AO-\frac{1}{\left\|\xi^\AO\right\|_\infty}\cc{\xi}^\AO\right)\otimes\pi^\AI \notag\\
	&\quad\otimes\op{x_1}{x_1}^\XI. \label{aeq:robustness-entropy-5}
\end{align}
It can be verified that $h_d(\Xi)$ is a valid channel for all $d\geq2$, as $J_{h_d(\Xi)}\in\spa{H}_+^{\AO\AI\XI}$ and $J_{h_d(\Xi)}^\AO=\spec{I}^\AO$.  By Eq.~\eqref{aeq:robustness-entropy-5},
\begin{align}
	J_{h_d(\Xi)}^{\AO\AI\XI}-\frac{1}{\left\|\xi^\AO\right\|_\infty}\cc{\xi}^{\AO\AI}\otimes\op{0}{0}^\XI&\in\spa{H}_+^{\AO\AI\XI},
\end{align}
which implies that
\begin{align}
	f_{h_d(\Xi)}(\Lambda)&\geq\frac{1}{\left\|\xi^\AO\right\|_\infty}\max_{\sm{\Theta}\in\f{S}^{\AP\To\A^\times}}\tr\left[\left(\xi^{\AO\AI}\otimes\op{0}{0}^\XI\right)J_{\sm{\Theta}\left\{\Lambda\right\}}^{\AO\AI\XI}\right] \notag\\
	&\quad\forall\Lambda\in\q{L}. \label{aeq:robustness-entropy-6}
\end{align}
Also by Eq.~\eqref{aeq:robustness-entropy-5},
\begin{align}
	&\frac{1}{\left\|\xi^\AO\right\|_\infty}\cc{\xi}^{\AO\AI}\otimes\op{0}{0}^\XI+\frac{1}{d_\AI\left(d-1\right)}\spec{I}^{\AO\AI\XI}-J_{h_d(\Xi)}^{\AO\AI\XI} \notag\\
	&\quad\in\spa{H}_+^{\AO\AI\XI},
\end{align}
which implies that
\begin{align}
	f_{h_d(\Xi)}(\Lambda)&\leq\frac{1}{\left\|\xi^\AO\right\|_\infty}\max_{\sm{\Theta}\in\f{S}^{\AP\To\A^\times}}\tr\left[\left(\xi^{\AO\AI}\otimes\op{0}{0}^\XI\right)J_{\sm{\Theta}\left\{\Lambda\right\}}^{\AO\AI\XI}\right] \notag\\
	&\quad+\frac{d_\AO}{d_\AI(d-1)}\quad\forall\Lambda\in\q{L}. \label{aeq:robustness-entropy-7}
\end{align}
Equations~\eqref{aeq:robustness-entropy-6} and \eqref{aeq:robustness-entropy-7} imply that
\begin{align}
	&\lim_{d\to\infty}f_{h_d(\Xi)}(\Lambda) \notag\\
	&\quad=\frac{1}{\left\|\xi^\AO\right\|_\infty}\max_{\sm{\Theta}\in\f{S}^{\AP\To\A^\times}}\tr\left[\left(\xi^{\AO\AI}\otimes\op{0}{0}^\XI\right)J_{\sm{\Theta}\left\{\Lambda\right\}}^{\AO\AI\XI}\right] \\
	&\quad=\frac{1}{\left\|\xi^\AO\right\|_\infty}\max_{\sm{\Theta}_0\in\f{S}^{\AP\To\A}}\tr\left[\xi J_{\sm{\Theta}_0\left\{\Lambda\right\}}\right] \label{aeq:robustness-entropy-8}\\
	&\quad=\frac{1}{\left\|\xi^\AO\right\|_\infty}\max_{\sm{\Theta}\in\f{S}^{\AP\To\A}}\tr\left[\xi J_{\sm{\Theta}\left\{\Lambda\right\}}\right]  \label{aeq:robustness-entropy-9}\\
	&\quad=\frac{1}{\left\|\xi^\AO\right\|_\infty}\max_{\sm{\Theta}\in\f{S}^{\AP\To\A}}\tr\left[\left(\xi^{\AO\AI}+\beta^\AO\otimes\spec{I}^\AI\right)J_{\sm{\Theta}\left\{\Lambda\right\}}^{\AO\AI}\right] \label{aeq:robustness-entropy-10}\\
	&\quad=\frac{1}{\left\|\xi^\AO\right\|_\infty}\max_{\sm{\Theta}\in\f{S}^{\AP\To\A}}\tr\left[J_\Xi J_{\sm{\Theta}\left\{\Lambda\right\}}\right] \label{aeq:robustness-entropy-11}\\
	&\quad=\frac{1}{\left\|\xi^\AO\right\|_\infty}\max_{\sm{\Theta}\in\f{S}^{\AP\To\A}}\left\langle\Xi,\sm{\Theta}\left\{\Lambda\right\}\right\rangle \label{aeq:robustness-entropy-12}\quad\forall\Lambda\in\q{L}.
\end{align}
Here Eq.~\eqref{aeq:robustness-entropy-8} is by Definition~\ref{def:probabilistic-dynamic}; Eq.~\eqref{aeq:robustness-entropy-9} is due to $\f{S}_\leqslant^{\AP\To\A}\subseteq\f{S}^{\AP\To\A}$ and $\xi\in\spa{H}_+^{\AO\AI}$; Eqs.~\eqref{aeq:robustness-entropy-10} and \eqref{aeq:robustness-entropy-11} follow from Eq.~\eqref{aeq:robustness-entropy-4}; Eq.~\eqref{aeq:robustness-entropy-10} is due to $\tr[\beta]=0$.  By Eq.~\eqref{aeq:robustness-entropy-12}, the sequence of mappings $\{h_d\}_{d=2}^\infty$ satisfies Condition~(2) of Corollary~\ref{acor:robustness-limit} given the set of functions $\{f_\Omega\}_{\Omega\in\q{L}}$ by substituting $\spa{U}^{(i)}=\spa{L}^\A$, $\spa{U}^{(i')}=\spa{L}^\AP$, $\ob{Q}'=\Lambda$, and $c_\Xi=\frac{1}{\|\xi^\AO\|_\infty}$.  By Corollary~\ref{acor:robustness-limit}, we have that
\begin{align}
	1+R_\abb{glob}^\f{C}(\Lambda)&=1+R^{\q{L},\f{C}}(\Lambda) \\
	&=\sup_{\Omega\in\q{L}}\frac{f_\Omega(\Lambda)}{\max_{\Psi\in\f{C}}f_\Omega(\Psi)} \\
	&=\sup_{\Omega\in\q{L}}\frac{2^{-H_{\min}^\f{S}(\B|\A)_{\Lambda^\A\otimes\Omega^\B}}}{\max_{\Psi\in\f{C}}2^{-H_{\min}^\f{S}(\B|\AP)_{\Psi^\AP\otimes\Omega^\B}}} \\
	&=\sup_{\Omega\in\q{L}}\frac{2^{-H_{\min}^\f{S}(\B|\A)_{\Lambda^\A\otimes\Omega^\B}}}{2^{-H_{\min}^\f{C}(\B)_\Omega}} \label{aeq:robustness-entropy-13}\\
	&=\sup_{\Omega\in\q{L}}2^{I_{\min}^\f{S}(\A;\B)_{\Lambda^\A\otimes\Omega^\B}}.
\end{align}
Here Eq.~\eqref{aeq:robustness-entropy-13} is by Proposition~\ref{prop:reducibility}(3).

\section{Operational characterization of quantum resources}
\label{app:characterization-tasks}

As before, let $(\f{S},\f{C})$ be a closed and convex dynamic QRT, which implies that $(\f{S},\f{C};\q{L})$ is a closed and convex quasi-resource theory.

\subsection{Proof of Theorem~\ref{thm:convertibility-EOP}}
\label{app:convertibility-EOP}

Let $\pmt{M}$ denote the set of POVMs on $\BO\BI$.  Let $\{f_\povm{M}\}_{\povm{M}\in\pmt{M}}$ be a set of functions with $f_\povm{M}\colon\q{L}\to\spa{R}$ for all $\povm{M}\equiv\{M_m\}_m\in\pmt{M}$, defined as follows:
\begin{align}
	f_\povm{M}(\Lambda)&\coloneq d_\BO P_\abb{EOP}^\f{S}(\Lambda;\povm{M}) \\
	&=\max_{\sm{\Theta}\in\f{S}^{\A\To\B^\times}}\sum_{m}\tr\left[\left(M_m^{\BO\BI}\otimes\op{m}{m}^\XI\right)J_{\sm{\Theta}\left\{\Lambda\right\}}^{\BO\BI\XI}\right] \notag\\
	&\quad\forall\Lambda\in\q{L}. \label{aeq:convertibility-EOP-1}
\end{align}
Here Eq.~\eqref{aeq:convertibility-EOP-1} follows from Eq.~\eqref{eq:EOP}.  By analysis in Sec.~\ref{sec:monotones}, $f_{\povm{M}}$ is a resource monotone for all $\povm{M}\in\pmt{M}$.  In what follows, we define a sequence of mappings $\{h_d\}_{d=2}^\infty$ with $h_d\colon\spa{L}^\B\to\pmt{M}$ for all $d\geq2$.  For any linear map $\Xi\in\spa{L}^\B$ and $d\geq2$, define $h_d(\Xi)\coloneq\{L_m\}_{m\in\idx{M}}$ with $|\idx{M}|\coloneq d$ and $L_m\in\spa{H}_+^{\BO\BI}$ for all $m\in\idx{M}$ as follows:
\begin{align}
\label{aeq:convertibility-EOP-2}
	\begin{cases}
		L_0\coloneq\frac{1}{2\left\|J_\Xi\right\|_\infty}J_\Xi+\frac{1}{2}\spec{I}, \\
		L_m\coloneq\frac{1}{d-1}\left(\spec{I}-L_0\right)\quad\forall m\in\idx{M}\setminus\{0\}.
	\end{cases}
\end{align}
It can be verified that $\{L_m\}_{m\in\idx{M}}$ is a valid POVM, as $L_m\in\spa{H}_+^{\BO\BI}$ for all $m\in\idx{M}$ and $\sum_{m}L_m=\spec{I}$.  By Eq.~\eqref{aeq:convertibility-EOP-2},
\begin{align}
	\sum_{m}L_m^{\BO\BI}\otimes\op{m}{m}^\XI-L_0^{\BO\BI}\otimes\op{0}{0}^\XI&\in\spa{H}_+^{\BO\BI\XI},
\end{align}
which implies that
\begin{align}
	f_{h_d(\Xi)}(\Lambda)&\geq\max_{\sm{\Theta}\in\f{S}^{\A\To\B^\times}}\tr\left[\left(L_0^{\BO\BI}\otimes\op{0}{0}^\XI\right)J_{\sm{\Theta}\left\{\Lambda\right\}}^{\BO\BI\XI}\right] \notag\\
	&\quad\forall\Lambda\in\q{L}. \label{aeq:convertibility-EOP-3}
\end{align}
Also by Eq.~\eqref{aeq:convertibility-EOP-2},
\begin{align}
	&L_0^{\BO\BI}\otimes\op{0}{0}^\XI+\frac{1}{d-1}\spec{I}^{\BO\BI\XI}-\sum_{m}L_m^{\BO\BI}\otimes\op{m}{m}^\XI \notag\\
	&\quad\in\spa{H}_+^{\BO\BI\XI},
\end{align}
which implies that
\begin{align}
	f_{h_d(\Xi)}(\Lambda)&\leq\max_{\sm{\Theta}\in\f{S}^{\A\To\B^\times}}\tr\left[\left(L_0^{\BO\BI}\otimes\op{0}{0}^\XI\right)J_{\sm{\Theta}\left\{\Lambda\right\}}^{\BO\BI\XI}\right] \notag\\
	&\quad+\frac{d_\BO}{d-1}\quad\forall\Lambda\in\q{L}. \label{aeq:convertibility-EOP-4}
\end{align}
Equations~\eqref{aeq:convertibility-EOP-3} and \eqref{aeq:convertibility-EOP-4} imply that
\begin{align}
	&\lim_{d\to\infty}f_{h_d(\Xi)}(\Lambda) \notag\\
	&\quad=\max_{\sm{\Theta}\in\f{S}^{\A\To\B^\times}}\tr\left[\left(L_0^{\BO\BI}\otimes\op{0}{0}^\XI\right)J_{\sm{\Theta}\left\{\Lambda\right\}}^{\BO\BI\XI}\right] \\
	&\quad=\max_{\sm{\Theta}_0\in\f{S}_\leqslant^{\A\To\B}}\tr\left[L_0J_{\sm{\Theta}_0\left\{\Lambda\right\}}\right] \label{aeq:convertibility-EOP-5}\\
	&\quad=\max_{\sm{\Theta}\in\f{S}^{\A\To\B}}\tr\left[L_0J_{\sm{\Theta}\left\{\Lambda\right\}}\right] \label{aeq:convertibility-EOP-6}\\
	&\quad=\frac{1}{2\left\|J_\Xi\right\|_\infty}\max_{\sm{\Theta}\in\f{S}^{\A\To\B}}\tr\left[J_\Xi J_{\sm{\Theta}\left\{\Lambda\right\}}\right]+\frac{1}{2}d_\BO \label{aeq:convertibility-EOP-7}\\
	&\quad=\frac{1}{2\left\|J_\Xi\right\|_\infty}\max_{\sm{\Theta}\in\f{S}^{\A\To\B}}\left\langle\Xi,\sm{\Theta}\left\{\Lambda\right\}\right\rangle+\frac{1}{2}d_\BO\quad\forall\Lambda\in\q{L}. \label{aeq:convertibility-EOP-8}
\end{align}
Here Eq.~\eqref{aeq:convertibility-EOP-5} is by Definition~\ref{def:probabilistic-dynamic}; Eq.~\eqref{aeq:convertibility-EOP-6} is due to $\f{S}_\leqslant^{\A\To\B}\subseteq\f{S}^{\A\To\B}$ and $M_0\in\spa{H}_+^{\BO\BI}$; Eq.~\eqref{aeq:convertibility-EOP-7} follows from Eq.~\eqref{aeq:convertibility-EOP-2}.   By Eq.~\eqref{aeq:convertibility-EOP-8}, the sequence of mappings $\{h_d\}_{d=2}^\infty$ satisfies Condition~(1) of Theorem~\ref{athm:convertibility} given the set of resource monotones $\{f_\povm{M}\}_{\povm{M}\in\pmt{M}}$ by substituting $\spa{U}^{(i)}=\spa{L}^\A$, $\spa{U}^{(j)}=\spa{L}^\B$, $\q{Q}=\q{L}$, $\f{T}=\f{S}$, $\f{F}=\f{C}$, and $\idx{W}=\pmt{M}$.  By Theorem~\ref{athm:convertibility}, $\{f_\povm{M}\}_{\povm{M}\in\pmt{M}}$ is complete with respect to $\A$ and $\B$, which yields the desired statement.

\subsection{Proof of Theorem~\ref{thm:convertibility-OCC}}
\label{app:convertibility-OCC}

Let $\pmt{E}$ denote the set of state ensembles with an index range $\idx{N}$ in $\BO$.  Let $\{f_\ens{\upsilon}\}_{\ens{\upsilon}\in\pmt{E}}$ be a set of functions with $f_\ens{\upsilon}\colon\q{L}\to\spa{R}$ for all $\ens{\upsilon}\equiv\{\nu_n\}_{n\in\idx{N}}\in\pmt{E}$, defined as follows:
\begin{align}
	f_\ens{\upsilon}(\Lambda)&\coloneq P_\abb{OCC}^\f{S}(\Lambda;\ens{\upsilon},\povm{N}) \\
	&=\max_{\sm{\Theta}\in\f{S}^{\A\To\B}}\tr\left[\left(\sum_{n}\cc{\nu}_n^\BO\otimes N_n^\BI\right)J_{\sm{\Theta}\left\{\Lambda\right\}}^{\BO\BI}\right]\quad\forall\Lambda\in\q{L}. \label{aeq:convertibility-OCC-1}
\end{align}
Here Eq.~\eqref{aeq:convertibility-OCC-1} follows from Eq.~\eqref{eq:OCC}.  By analysis in Sec.~\ref{sec:monotones}, $f_\ens{\upsilon}$ is a resource monotone for all $\ens{\upsilon}\in\pmt{E}$.  In what follows, we define a mapping $h\colon\spa{L}^\B\to\pmt{E}$.  Since the POVM $\povm{N}\equiv\{N_n\}_{n\in\idx{N}}$ is informationally complete, for any linear map $\Xi\in\spa{L}^\B$, there exists a set of operators $\{\xi_n\}_{n\in\idx{N}}$ with $\xi_n\in\spa{H}^\BO$ for all $n\in\idx{N}$ such that
\begin{align}
\label{aeq:convertibility-OCC-2}
	J_\Xi^{\BO\BI}&=\sum_{n}\xi_n^\BO\otimes N_n^\BI.
\end{align}
Define $a\coloneq\sum_n\tr[\xi_n]$ and $b\coloneq\max_n\|\xi_n\|_\infty$.  Define $h(\Xi)\coloneq\{\eta_n\}_{n\in\idx{N}}$ with $\eta_n\in\q{D}_\leqslant^\BO$ for all $n\in\idx{N}$ as follows:
\begin{align}
\label{aeq:convertibility-OCC-3}
	\eta_n&\coloneq\frac{1}{a+d_\BO\left|\idx{N}\right|b}\left(\cc{\xi}_n+b\spec{I}\right)\quad\forall n\in\idx{N}.
\end{align}
It can be verified that $\{\eta_n\}_{n\in\idx{N}}$ is a valid state ensemble, as $\eta_n\in\spa{H}_+^\BO$ for all $n\in\idx{N}$ and $\sum_{n}\tr[\eta_n]=1$.  By Eqs.~\eqref{aeq:convertibility-OCC-1} and \eqref{aeq:convertibility-OCC-3},
\begin{align}
	&f_{h(\Xi)}(\Lambda) \notag\\
	&\quad=\frac{1}{a+d_\BO\left|\idx{N}\right|b}\max_{\sm{\Theta}\in\f{S}^{\A\To\B}}\tr\left[\left(\sum_{n}\xi_n^\BO\otimes N_n^\BI+b\spec{I}^{\BO\BI}\right)J_{\sm{\Theta}\left\{\Lambda\right\}}^{\BO\BI}\right] \label{aeq:convertibility-OCC-4}\\
	&\quad=\frac{1}{a+d_\BO\left|\idx{N}\right|b}\left(\max_{\sm{\Theta}\in\f{S}^{\A\To\B}}\tr\left[J_\Xi J_{\sm{\Theta}\left\{\Lambda\right\}}\right]+d_\BO b\right) \label{aeq:convertibility-OCC-5}\\
	&\quad=\frac{1}{a+d_\BO\left|\idx{N}\right|b}\left(\max_{\sm{\Theta}\in\f{S}^{\A\To\B}}\left\langle\Xi,\sm{\Theta}\left\{\Lambda\right\}\right\rangle+d_\BO b\right)\quad\forall\Lambda\in\q{L}. \label{aeq:convertibility-OCC-6}
\end{align}
Here Eq.~\eqref{aeq:convertibility-OCC-4} is due to $\sum_{n}N_n=\spec{I}$; Eq.~\eqref{aeq:convertibility-OCC-5} follows from Eq.~\eqref{aeq:convertibility-OCC-2}.  By Eq.~\eqref{aeq:convertibility-OCC-6}, a sequence of the same mapping $h$ satisfies Condition~(1) of Theorem~\ref{athm:convertibility} given the set of resource monotones $\{f_\ens{\upsilon}\}_{\ens{\upsilon}\in\pmt{E}}$ by substituting $\spa{U}^{(i)}=\aff(\q{L}^\A)$, $\spa{U}^{(j)}=\aff(\q{L}^\B)$, $\q{Q}=\q{L}$, $\f{T}=\f{S}$, $\f{F}=\f{C}$, and $\idx{W}=\pmt{E}$.  By Theorem~\ref{athm:convertibility}, $\{f_\ens{\upsilon}\}_{\ens{\upsilon}\in\pmt{E}}$ is complete with respect to $\A$ and $\B$, which yields the desired statement.

\subsection{Proof of Theorem~\ref{thm:robustness-EOP}}
\label{app:robustness-EOP}

The dynamic QRT $(\f{S},\f{C})$ satisfies Condition~(1) of Corollary~\ref{acor:robustness-limit} by substituting $\spa{U}^{(i)}=\spa{L}^\A$, $\q{Q}=\q{L}$, $\f{T}=\f{S}$, $\f{F}=\f{C}$, $\spa{U}^{(k)}=\spa{L}^{\#\to\#}$, and $\spec{\ob{P}}=1$.  Let $\pmt{M}$ denote the set of POVMs on $\AO\AI$.  Let $\{f_\povm{M}\}_{\povm{M}\in\pmt{M}}$ be a set of functions with $f_\povm{M}\colon\q{L}\to\spa{R}$ for all $\povm{M}\equiv\{M_m\}_{m\in\idx{M}}\in\pmt{M}$, defined as follows:
\begin{align}
	f_\povm{M}(\Lambda)&\coloneq d_\AO P_\abb{EOP}^\f{S}(\Lambda;\povm{M}) \\
	&=\max_{\sm{\Theta}\in\f{S}^{\AP\To\A^\times}}\sum_{m}\tr\left[\left(M_m^{\AO\AI}\otimes\op{m}{m}^\XI\right)J_{\sm{\Theta}\left\{\Lambda\right\}}^{\AO\AI\XI}\right] \label{aeq:robustness-EOP-1}\\
	&=\max_{\sm{\Theta}\in\f{S}^{\AP\To\A^\times}}\left\langle\Xi_\povm{M},\sm{\Theta}\left\{\Lambda\right\}\right\rangle\quad\forall\Lambda\in\q{L}, \label{aeq:robustness-EOP-2}
\end{align}
where $\Xi_\povm{M}\in\spa{L}^{\A^\times}$ is defined such that
\begin{align}
	J_{\Xi_\povm{M}}^{\AO\AI\XI}&=\sum_{m}M_m^{\AO\AI}\otimes\op{m}{m}^\XI.
\end{align}
Here Eq.~\eqref{aeq:robustness-EOP-1} follows from Eq.~\eqref{eq:EOP}.  Since $\Xi_\povm{M}\in\cone^*(\q{L}^{\A^\times})$, by Eq.~\eqref{aeq:robustness-EOP-1}, the set of functions $\{f_\povm{M}\}_{\povm{M}\in\pmt{M}}$ satisfies Condition~(2) of Corollary~\ref{acor:robustness-limit} by substituting $\spa{U}^{(i')}=\spa{L}^\AP$, $\q{Q}=\q{L}$, $\s{K}=\q{L}$, $\idx{W}=\q{L}$, $\spa{U}^{(j_\Omega)}=\spa{L}^{\A^\times}$, and $\alpha_\povm{M}=\Xi_\povm{M}$.  In what follows, we define a sequence of mappings $\{h_d\}_{d=2}^\infty$ with $h_d\colon\cone^*(\q{L}^\A)\to\pmt{M}$ for all $d\geq2$.  For any linear map $\Xi\in\cone^*(\q{L}^\A)$, there exists a positive semidefinite operator $\xi\in\spa{H}_+^{\AO\AI}$ and an operator $\beta\in\spa{H}^\AO$ with $\tr[\beta]=0$ such that
\begin{align}
	\label{aeq:robustness-EOP-3}
	J_\Xi^{\AO\AI}&=\xi^{\AO\AI}+\beta^\AO\otimes\spec{I}^\AI.
\end{align}
For any $d\geq2$, define $h_d(\Xi)\coloneq\{L_m\}_{m\in\idx{M}}\in\pmt{M}$ with $|\idx{M}|\coloneq d$ and $L_m\in\spa{H}_+^{\AO\AI}$ for all $m\in\idx{M}$ as follows:
\begin{align}
\label{aeq:robustness-EOP-4}
	\begin{cases}
		L_0\coloneq\frac{1}{\left\|\xi\right\|_\infty}\xi, \\
		L_m\coloneq\frac{1}{d-1}\left(\spec{I}-L_0\right)\quad\forall m\in\idx{M}\setminus\{0\}.
	\end{cases}
\end{align}
It can be verified that $\{L_m\}_{m\in\idx{M}}$ is a valid POVM, as $L_m\in\spa{H}_+^{\AO\AI}$ for all $m\in\idx{M}$ and $\sum_{m}L_m=\spec{I}$.  By Eq.~\eqref{aeq:robustness-EOP-4},
\begin{align}
	\sum_{m}L_m^{\AO\AI}\otimes\op{m}{m}^\XI-L_0^{\AO\AI}\otimes\op{0}{0}^\XI&\in\spa{H}_+^{\AO\AI\XI},
\end{align}
which implies that
\begin{align}
	f_{h_d(\Xi)}(\Lambda)&\geq\max_{\sm{\Theta}\in\f{S}^{\AP\To\A^\times}}\tr\left[\left(L_0^{\AO\AI}\otimes\op{0}{0}^\XI\right)J_{\sm{\Theta}\left\{\Lambda\right\}}^{\AO\AI\XI}\right] \notag\\
	&\quad\forall\Lambda\in\q{L}. \label{aeq:robustness-EOP-5}
\end{align}
Also by Eq.~\eqref{aeq:robustness-EOP-4},
\begin{align}
	&L_0^{\AO\AI}\otimes\op{0}{0}^\XI+\frac{1}{d-1}\spec{I}^{\AO\AI\XI}-\sum_{m}L_m^{\AO\AI}\otimes\op{m}{m}^\XI \notag\\
	&\quad\in\spa{H}_+^{\AO\AI\XI},
\end{align}
which implies that
\begin{align}
	f_{h_d(\Xi)}(\Lambda)&\leq\max_{\sm{\Theta}\in\f{S}^{\AP\To\A^\times}}\tr\left[\left(L_0^{\AO\AI}\otimes\op{0}{0}^\XI\right)J_{\sm{\Theta}\left\{\Lambda\right\}}^{\AO\AI\XI}\right] \notag\\
	&\quad+\frac{d_\AO}{d-1}\quad\forall\Lambda\in\q{L}. \label{aeq:robustness-EOP-6}
\end{align}
Equations~\eqref{aeq:robustness-EOP-5} and \eqref{aeq:robustness-EOP-6} imply that
\begin{align}
	&\lim_{d\to\infty}f_{h_d(\Xi)}(\Lambda) \notag\\
	&\quad=\max_{\sm{\Theta}\in\f{S}^{\AP\To\A^\times}}\tr\left[\left(L_0^{\AO\AI}\otimes\op{0}{0}^\XI\right)J_{\sm{\Theta}\left\{\Lambda\right\}}^{\AO\AI\XI}\right] \\
	&\quad=\max_{\sm{\Theta}_0\in\f{S}_\leqslant^{\AP\To\A}}\tr\left[L_0J_{\sm{\Theta}_0\left\{\Lambda\right\}}\right] \label{aeq:robustness-EOP-7}\\
	&\quad=\max_{\sm{\Theta}\in\f{S}^{\AP\To\A}}\tr\left[L_0J_{\sm{\Theta}\left\{\Lambda\right\}}\right] \label{aeq:robustness-EOP-8}\\
	&\quad=\frac{1}{\left\|\xi\right\|_\infty}\max_{\sm{\Theta}\in\f{S}^{\AP\To\A}}\tr\left[\xi J_{\sm{\Theta}\left\{\Lambda\right\}}\right] \label{aeq:robustness-EOP-9}\\
	&\quad=\frac{1}{\left\|\xi\right\|_\infty}\max_{\sm{\Theta}\in\f{S}^{\AP\To\A}}\tr\left[\left(\xi^{\AO\AI}+\beta^\AO\otimes\spec{I}^\AI\right)J_{\sm{\Theta}\left\{\Lambda\right\}}^{\AO\AI}\right] \label{aeq:robustness-EOP-10}\\
	&\quad=\frac{1}{\left\|\xi\right\|_\infty}\max_{\sm{\Theta}\in\f{S}^{\AP\To\A}}\tr\left[J_\Xi J_{\sm{\Theta}\left\{\Lambda\right\}}\right] \label{aeq:robustness-EOP-11}\\
	&\quad=\frac{1}{\left\|\xi\right\|_\infty}\max_{\sm{\Theta}\in\f{S}^{\AP\To\A}}\left\langle\Xi,\sm{\Theta}\left\{\Lambda\right\}\right\rangle\quad\forall\Lambda\in\q{L}. \label{aeq:robustness-EOP-12}
\end{align}
Here Eq.~\eqref{aeq:robustness-EOP-7} is by Definition~\ref{def:probabilistic-dynamic}; Eq.~\eqref{aeq:robustness-EOP-8} is due to $\f{S}_\leqslant^{\AP\To\A}\subseteq\f{S}^{\AP\To\A}$ and $M_0\in\spa{H}_+^{\AO\AI}$; Eq.~\eqref{aeq:robustness-EOP-9} follows from Eq.~\eqref{aeq:robustness-EOP-4}; Eqs.~\eqref{aeq:robustness-EOP-10} and \eqref{aeq:robustness-EOP-11} follows from Eq.~\eqref{aeq:robustness-EOP-3}.  By Eq.~\eqref{aeq:robustness-EOP-12}, the sequence of mappings $\{h_d\}_{d=2}^\infty$ satisfies Condition~(2) of Corollary~\ref{acor:robustness-limit} given the set of functions $\{f_\povm{M}\}_{\povm{M}\in\pmt{M}}$ by substituting $\spa{U}^{(i)}=\spa{L}^\A$, $\spa{U}^{(i')}=\spa{L}^\AP$, $\ob{Q}'=\Lambda$, and $c_\Xi=\frac{1}{\|\xi\|_\infty}$.  By Corollary~\ref{acor:robustness-limit}, we have that
\begin{align}
	1+R_\abb{glob}^\f{C}(\Lambda)&=1+R^{\q{L},\f{C}}(\Lambda) \\
	&=\sup_{\povm{M}\in\pmt{M}}\frac{f_\povm{M}(\Lambda)}{\max_{\Psi\in\f{C}}f_\povm{M}(\Psi)} \\
	&=\sup_{\povm{M}\in\pmt{M}}\frac{P_\abb{EOP}^\f{S}(\Lambda;\povm{M})}{\max_{\Psi\in\f{C}}P_\abb{EOP}^\f{S}(\Psi;\povm{M})} \\
	&=\sup_{\povm{M}\in\pmt{M}}\frac{P_\abb{EOP}^\f{S}(\Lambda;\povm{M})}{P_\abb{EOP}^\f{C}(\povm{M})} \label{aeq:robustness-EOP-13}.
\end{align}
Here Eq.~\eqref{aeq:robustness-EOP-13} follows from Eq.~\eqref{eq:EOP-free}.

\subsection{Proof of Theorem~\ref{thm:entropy-OCC}}
\label{app:entropy-OCC}

Let $\{f_\Omega\}_{\Omega\in\q{L}_\abb{MP}}$ be a set of functions with $f_\Omega\colon\q{L}\to\spa{R}$ for all $\Omega\in\q{L}_\abb{MP}$, defined as follows:
\begin{align}
	f_\Omega(\Lambda)&\coloneq d_\BO2^{-H_{\min}^\f{S}(\B|\AP)_{\Lambda^\AP\otimes\Omega^\B}} \\
	&=\max_{\sm{\Theta}\in\f{S}^{\AP\To\B}}\tr\left[J_\cc{\Omega}J_{\sm{\Theta}\left\{\Lambda\right\}}\right]\quad\forall\Lambda\in\q{L}. \label{aeq:entropy-OCC-1}
\end{align}
Here Eq.~\eqref{aeq:entropy-OCC-1} follows from Eq.~\eqref{aeq:robustness-entropy-2}.  For any dynamic system $B$ and measure-and-prepare channel $\Omega\in\q{L}_\abb{MP}^\B\equiv\q{L}_\abb{MP}\cap\q{L}^\B$, there exists a POVM $\{L_n\}_{n\in\idx{N}}$ and a set of states $\{\eta_n\}_{n\in\idx{N}}$ with $L_n\in\spa{H}_+^\BO$ and $\eta_n\in\q{D}^\BI$ for all $n\in\idx{N}$ such that
\begin{align}
\label{aeq:entropy-OCC-2}
	J_\Omega^{\BO\BI}&=\sum_{n}L_n^\BO\otimes\eta_n^\BI.
\end{align}
Define $b\coloneq\|\sum_{n}\eta_n\|_\infty$.  Define a POVM $\povm{N}_\Omega\equiv\{N_n\}_{n\in\idx{N}\cup\{\perp\}}$ with $\perp\notin\idx{N}$ and $N_n\in\spa{H}_+^\BI$ for all $n\in\idx{N}\cup\{\perp\}$ as follows:
\begin{align}
\label{aeq:entropy-OCC-3}
	\begin{cases}
		N_n\coloneq\frac{1}{b}\cc{\eta}_n\quad\forall n\in\idx{N}, \\
		N_\perp\coloneq\spec{I}-\sum_{n\in\idx{N}}N_n.
	\end{cases}
\end{align}
It can be verified that $\povm{N}_\Omega$ is a valid POVM, as $N_n\in\spa{H}_+^\BI$ for all $n\in\idx{N}\cup\{\perp\}$ and $\sum_{n}N_n=\spec{I}$.  Define a state ensemble $\ens{\upsilon}_\Omega\equiv\{\nu_n\}_{n\in\idx{N}\cup\{\perp\}}$ with $\nu_n\in\spa{H}_+^\BO$ for all $n\in\idx{N}\cup\{\perp\}$ as follows:
\begin{align}
\label{aeq:entropy-OCC-4}
	\begin{cases}
		\nu_n\coloneq\frac{1}{d_\BO}L_n\quad\forall n\in\idx{N}, \\
		\nu_\perp\coloneq0.
	\end{cases}
\end{align}
It can be verified that $\ens{\upsilon}_\Omega$ is a valid state ensemble, as $\nu_n\in\spa{H}_+^\BO$ for all $n\in\idx{N}\cup\{\perp\}$ and $\sum_{n}\tr[\nu_n]=1$.  By Eqs.~\eqref{aeq:entropy-OCC-1}--\eqref{aeq:entropy-OCC-4},
\begin{align}
	f_\Omega(\Lambda)&=\max_{\sm{\Theta}\in\f{S}^{\AP\To\B}}\tr\left[J_\cc{\Omega}J_{\sm{\Theta}\left\{\Lambda\right\}}\right] \\
	&=\max_{\sm{\Theta}\in\f{S}^{\AP\To\B}}\tr\left[\sum_{n}\left(\cc{L}_n^\BO\otimes\cc{\eta}_n^\BI\right)J_{\sm{\Theta}\left\{\Lambda\right\}}^{\BO\BI}\right] \\
	&=d_\BO b\max_{\sm{\Theta}\in\f{S}^{\AP\To\B}}\tr\left[\sum_{n}\left(\cc{\nu}_n^\BO\otimes N_n^\BI\right)J_{\sm{\Theta}\left\{\Lambda\right\}}^{\BO\BI}\right] \\
	&=d_\BO bP_\abb{OCC}^\f{S}(\Lambda;\ens{\upsilon}_\Omega,\povm{N}_\Omega)\quad\forall\Lambda\in\q{L}. \label{aeq:entropy-OCC-5}
\end{align}
Here Eq.~\eqref{aeq:entropy-OCC-5} follows from Eq.~\eqref{eq:OCC}.  Then we have that
\begin{align}
	&\sup_{\Omega\in\q{L}_\abb{MP}}2^{I_{\min}^\f{S}(\A;\B)_{\Lambda^\A\otimes\Omega^\B}} \notag\\
	&\quad=\sup_{\Omega\in\q{L}_\abb{MP}}\frac{2^{-H_{\min}^\f{S}(\B|\A)_{\Lambda^\A\otimes\Omega^\B}}}{2^{-H_{\min}^\f{C}(\B)_\Omega}} \\
	&\quad=\sup_{\Omega\in\q{L}_\abb{MP}}\frac{2^{-H_{\min}^\f{S}(\B|\A)_{\Lambda^\A\otimes\Omega^\B}}}{\max_{\Psi\in\f{C}}2^{-H_{\min}^\f{S}(\B|\AP)_{\Psi^\AP\otimes\Omega^\B}}} \label{aeq:entropy-OCC-6}\\
	&\quad=\sup_{\Omega\in\q{L}_\abb{MP}}\frac{f_\Omega(\Lambda)}{\max_{\Psi\in\f{C}}f_\Omega(\Psi)} \\
	&\quad=\sup_{\Omega\in\q{L}_\abb{MP}}\frac{P_\abb{OCC}^\f{S}(\Lambda;\ens{\upsilon}_\Omega,\povm{N}_\Omega)}{\max_{\Psi\in\f{C}}P_\abb{OCC}^\f{S}(\Psi;\ens{\upsilon}_\Omega,\povm{N}_\Omega)} \\
	&\quad=\sup_{\Omega\in\q{L}_\abb{MP}}\frac{P_\abb{OCC}^\f{S}(\Lambda;\ens{\upsilon}_\Omega,\povm{N}_\Omega)}{P_\abb{OCC}^\f{C}(\ens{\upsilon}_\Omega,\povm{N}_\Omega)} \label{aeq:entropy-OCC-7}\\
	&\quad\leq\sup_{\ens{\upsilon},\povm{N}}\frac{P_\abb{OCC}^\f{S}(\Lambda;\ens{\upsilon},\povm{N})}{P_\abb{OCC}^\f{C}(\ens{\upsilon},\povm{N})}. \label{aeq:entropy-OCC-8}
\end{align}
Here Eq.~\eqref{aeq:entropy-OCC-6} is by Proposition~\ref{prop:reducibility}(3); Eq.~\eqref{aeq:entropy-OCC-7} follows from Eq.~\eqref{eq:OCC-free}.  To prove the converse, let $\pmt{E}$ denote the set of state ensembles, and let $\pmt{M}$ denote the set of POVMs.  It what follows, we define a sequence of mappings $\{h_d\}_{d=2}^\infty$ with $h_d\colon\pmt{E}\times\pmt{M}\to\q{L}_\abb{MP}$ for all $d\geq2$.  For any state ensemble $\ens{\upsilon}\equiv\{\nu_n\}_{n\in\idx{N}}$ in $\BO$, POVM $\povm{N}\equiv\{N_n\}_{n\in\idx{N}}$ on $\BI$, and $d\geq2$, define $c\coloneq\|\sum_{n}\tr[N_n]\nu_n\|_\infty$, and define $h_d(\ens{\upsilon},\povm{N})\in\q{L}_\abb{MP}^{\B^\times}$ with $\B^\times\equiv\BO\to\BI\XI$ and $d_\XI\coloneq d$ as follows:
\begin{align}
\label{aeq:entropy-OCC-9}
	J_{h_d(\ens{\upsilon},\povm{N})}^{\BO\BI\XI}&\coloneq\frac{1}{c}\sum_{n}\nu_n^\BO\otimes N_n^\BI\otimes\op{0}{0}^\XI \notag\\
	&\quad+\sum_{x_1=1}^{d-1}\frac{1}{d-1}\left(\spec{I}^\BO-\frac{1}{c}\tr\left[N_n\right]\nu_n^\BO\right) \notag\\
	&\quad\otimes\pi^\BI\otimes\op{x_1}{x_1}^\XI.
\end{align}
It can be verified that $h_d(\ens{\upsilon},\povm{N})$ is a valid prepare-and-measure channel for all $d\geq2$, as $J_{h_d(\ens{\upsilon},\povm{N})}\in\spa{H}_+^{\BO\BI\XI}$, $J_{h_d(\ens{\upsilon},\povm{N})}^\BO=\spec{I}^\BO$, and $J_{h_d(\ens{\upsilon},\povm{N})}$ is separable between $\BO$ and $\BI\XI$.  By Eq.~\eqref{aeq:entropy-OCC-9},
\begin{align}
	J_{h_d(\ens{\upsilon},\povm{N})}^{\BO\BI\XI}-\frac{1}{c}\sum_{n\in\idx{N}}\nu_n^\BO\otimes N_n^\BI\otimes\op{0}{0}^\XI\in\spa{H}_+^{\BO\BI\XI},
\end{align}
which implies that
\begin{align}
	&f_{h_d(\ens{\upsilon},\povm{N})^{\B^\times}}(\Lambda) \notag\\
	&\quad=\max_{\sm{\Theta}\in\f{S}^{\AP\To\B^\times}}\tr\left[J_\cc{h_d(\ens{\upsilon},\povm{N})}J_{\sm{\Theta}\left\{\Lambda\right\}}\right] \\
	&\quad\geq\frac{1}{c}\max_{\sm{\Theta}\in\f{S}^{\AP\To\B^\times}}\tr\left[\left(\sum_{n}\nu_n^\BO\otimes N_n^\BI\otimes\op{0}{0}^\XI\right)J_{\sm{\Theta}\left\{\Lambda\right\}}^{\BO\BI\XI}\right] \notag\\
	&\quad\quad\forall\Lambda\in\q{L}. \label{aeq:entropy-OCC-10}
\end{align}
Also by Eq.~\eqref{aeq:entropy-OCC-9},
\begin{align}
	&\frac{1}{c}\sum_{n\in\idx{N}}\nu_n^\BO\otimes N_n^\BI\otimes\op{0}{0}^\XI+\frac{1}{d_\BI\left(d-1\right)}\spec{I}^{\BO\BI\XI}-J_{h_d(\ens{\upsilon},\povm{N})}^{\BO\BI\XI} \notag\\
	&\quad\in\spa{H}_+^{\BO\BI\XI},
\end{align}
which implies that
\begin{align}
	&f_{h_d(\ens{\upsilon},\povm{N})^{\B^\times}}(\Lambda) \notag\\
	&\quad\leq\frac{1}{c}\max_{\sm{\Theta}\in\f{S}^{\AP\To\B^\times}}\tr\left[\left(\sum_{n}\nu_n^\BO\otimes N_n^\BI\otimes\op{0}{0}^\XI\right)J_{\sm{\Theta}\left\{\Lambda\right\}}^{\BO\BI\XI}\right] \notag\\
	&\quad\quad+\frac{d_\BO}{d_\BI\left(d-1\right)}\quad\forall\Lambda\in\q{L}. \label{aeq:entropy-OCC-11}
\end{align}
Equations~\eqref{aeq:entropy-OCC-10} and \eqref{aeq:entropy-OCC-11} imply that
\begin{align}
	&\lim_{d\to\infty}f_{h_d(\ens{\upsilon},\povm{N})}(\Lambda) \notag\\
	&\quad=\frac{1}{c}\max_{\sm{\Theta}\in\f{S}^{\AP\To\B^\times}}\tr\left[\left(\sum_{n}\nu_n^\BO\otimes N_n^\BI\otimes\op{0}{0}^\XI\right)J_{\sm{\Theta}\left\{\Lambda\right\}}^{\BO\BI\XI}\right] \\
	&\quad=\frac{1}{c}\max_{\sm{\Theta}_0\in\f{S}_\leqslant^{\AP\To\B}}\tr\left[\left(\sum_{n}\nu_n^\BO\otimes N_n^\BI\right)J_{\sm{\Theta}_0\left\{\Lambda\right\}}^{\BO\BI\XI}\right] \label{aeq:entropy-OCC-12}\\
	&\quad=\frac{1}{c}\max_{\sm{\Theta}\in\f{S}^{\AP\To\B}}\tr\left[\left(\sum_{n}\nu_n^\BO\otimes N_n^\BI\right)J_{\sm{\Theta}\left\{\Lambda\right\}}^{\BO\BI\XI}\right] \label{aeq:entropy-OCC-13}\\
	&\quad=\frac{1}{c}P_\abb{OCC}^\f{S}(\Lambda;\ens{\upsilon},\povm{N})\quad\forall\Lambda\in\q{L}. \label{aeq:entropy-OCC-14}
\end{align}
Here Eq.~\eqref{aeq:entropy-OCC-12} is by Definition~\ref{def:probabilistic-dynamic}; Eq.~\eqref{aeq:entropy-OCC-13} is due to $\f{S}_\leqslant^{\AP\To\B}\subseteq\f{S}^{\AP\To\B}$ and $\sum_{n}\nu_n\otimes N_n\in\spa{H}_+^{\BO\BI}$; Eq.~\eqref{aeq:entropy-OCC-14} follows from Eq.~\eqref{eq:OCC}.  Then we have that
\begin{align}
	&\sup_{\ens{\upsilon},\povm{N}}\frac{P_\abb{OCC}^\f{S}(\Lambda;\ens{\upsilon},\povm{N})}{P_\abb{OCC}^\f{C}(\ens{\upsilon},\povm{N})} \notag\\
	&\quad=\sup_{\ens{\upsilon},\povm{N}}\frac{P_\abb{OCC}^\f{S}(\Lambda;\ens{\upsilon},\povm{N})}{\max_{\Psi\in\f{C}}P_\abb{OCC}^\f{S}(\Psi;\ens{\upsilon},\povm{N})} \\
	&\quad=\sup_{\ens{\upsilon},\povm{N}}\lim_{d\to\infty}\frac{f_{h_d(\ens{\upsilon},\povm{N})}(\Lambda)}{\max_{\Psi\in\f{C}}f_{h_d(\ens{\upsilon},\povm{N})}(\Psi)} \\
	&\quad\leq\lim_{d\to\infty}\sup_{\ens{\upsilon},\povm{N}}\frac{f_{h_d(\ens{\upsilon},\povm{N})}(\Lambda)}{\max_{\Psi\in\f{C}}f_{h_d(\ens{\upsilon},\povm{N})}(\Psi)} \\
	&\quad\leq\sup_{\Omega\in\q{L}_\abb{MP}}\frac{f_\Omega(\Lambda)}{\max_{\Psi\in\f{C}}f_\Omega(\Psi)} \\
	&\quad=\sup_{\Omega\in\q{L}_\abb{MP}}\frac{2^{-H_{\min}^\f{S}(\B|\A)_{\Lambda^\A\otimes\Omega^\B}}}{\max_{\Psi\in\f{C}}2^{-H_{\min}^\f{S}(\B|\AP)_{\Psi^\AP\otimes\Omega^\B}}} \\
	&\quad=\sup_{\Omega\in\q{L}_\abb{MP}}\frac{2^{-H_{\min}^\f{S}(\B|\A)_{\Lambda^\A\otimes\Omega^\B}}}{2^{-H_{\min}^\f{C}(\B)_\Omega}} \\
	&\quad=\sup_{\Omega\in\q{L}_\abb{MP}}2^{I_{\min}^\f{S}(\A;\B)_{\Lambda^\A\otimes\Omega^\B}}. \label{aeq:entropy-OCC-15}
\end{align}
Combining Eqs.~\eqref{aeq:entropy-OCC-7} and \eqref{aeq:entropy-OCC-15}, we have that
\begin{align}
	\sup_{\Omega\in\q{L}_\abb{MP}}I_{\min}^\f{S}(\A;\B)_{\Lambda^\A\otimes\Omega^\B}&=\log\sup_{\ens{\upsilon},\povm{N}}\frac{P_\abb{OCC}^\f{S}(\Lambda;\ens{\upsilon},\povm{N})}{P_\abb{OCC}^\f{C}(\ens{\upsilon},\povm{N})}.
\end{align}

\subsection{Proof of Theorem~\ref{thm:robustness-OCC}}
\label{app:robustness-OCC}

For any score rule $s\equiv\{s_{n,n'}\}_{n,n'\in\idx{N}}$ with $s_{n,n'}\in\spa{R}$ for all $n,n'\in\idx{N}$, it follows from Eq.~\eqref{eq:OCC-Id} that
\begin{align}
	S_\abb{OCC}^{\{\Id\}}(\Lambda;s,\ens{\upsilon},\povm{N})&=\tr\left[\sum_{n,n'}s_{n,n'}\left(\cc{\nu}_n^\AO\otimes N_{n'}^\AI\right)J_{\sm{\Theta}\left\{\Lambda\right\}}^{\AO\AI}\right] \\
	&=\left\langle\Xi_s,\sm{\Theta}\left\{\Lambda\right\}\right\rangle\quad\forall\Lambda\in\q{L}, \label{aeq:robustness-OCC-1}
\end{align}
where $\Xi_s\in\spa{L}^\A$ is defined such that
\begin{align}
	J_{\Xi_s}^{\AO\AI}&=\sum_{n,n'}s_{n,n'}\cc{\nu}_n^\AO\otimes N_{n'}^\AI.
\end{align}
Since the POVM $\povm{N}\equiv\{N_{n'}\}_{n'\in\idx{N}}$ is informationally complete and the state ensemble $\ens{\upsilon}\equiv\{\nu_n\}_{n\in\idx{N}}$ is tomographically complete, the mapping $s\mapsto\Xi_s$ is a bijection between $\spa{R}^{\idx{N}\times\idx{N}}$ and $\spa{L}^\A$.  Likewise, it follows from Eq.~\eqref{eq:OCC-score} that
\begin{align}
	S_\abb{OCC}^\f{S}(\Lambda;s,\ens{\upsilon},\povm{N})&=\max_{\sm{\Theta}\in\f{S}^{\AP\To\A}}\tr\left[\sum_{n,n'}s_{n,n'}\left(\cc{\nu}_n^\AO\otimes N_{n'}^\AI\right)J_{\sm{\Theta}\left\{\Lambda\right\}}^{\AO\AI}\right] \\
	&=\max_{\sm{\Theta}\in\f{S}^{\AP\To\A}}\left\langle\Xi_s,\sm{\Theta}\left\{\Lambda\right\}\right\rangle\quad\forall\Lambda\in\q{L}. \label{aeq:robustness-OCC-2}
\end{align}
Substituting $\spa{U}^{(i)}=\spa{L}^\A$, $\q{Q}=\q{L}$, $\f{T}=\f{S}$, and $\f{F}=\f{C}$, and by Eq.~\eqref{eq:OCC-K}, we have that $\Xi_s\in\cone^*(\s{K}^\A)$ if and only if $\bar{S}_\abb{OCC}^\s{K}(s,\ens{\upsilon},\povm{N})\geq0$.  Then by Corollary~\ref{acor:robustness},
\begin{align}
	&1+R^{\s{K},\f{C}}(\Lambda) \notag\\
	&\quad=\max_{\Xi\in\cone^*(\s{K}^\A)}\frac{\left\langle\Xi,\Lambda\right\rangle}{\max_{\Psi\in\f{C}^\A}\left\langle\Xi,\Psi\right\rangle} \\
	&\quad=\max_{\scriptsize\left\{\begin{array}{c}
		s\in\spa{R}^{\idx{N}\times\idx{N}}\colon \\
		\Xi_s\in\cone^*(\s{K}^\A)
	\end{array}\right\}}\frac{\left\langle\Xi_s,\Lambda\right\rangle}{\max_{\Psi\in\f{C}^\A}\left\langle\Xi_s,\Psi\right\rangle} \\
	&\quad=\max_{\scriptsize\left\{\begin{array}{c}
		s\in[-1,1]^{\idx{N}\times\idx{N}}\colon \\
		\Xi_s\in\cone^*(\s{K}^\A)
	\end{array}\right\}}\frac{\left\langle\Xi_s,\Lambda\right\rangle}{\max_{\Psi\in\f{C}^\A}\left\langle\Xi_s,\Psi\right\rangle} \label{aeq:robustness-OCC-3}\\
	&\quad=\max_{\scriptsize\left\{\begin{array}{c}
		s\in[-1,1]^{\idx{N}\times\idx{N}}\colon \\
		\bar{S}_\abb{OCC}^\s{K}(s,\ens{\upsilon},\povm{N})\geq0
	\end{array}\right\}}\frac{S_\abb{OCC}^{\{\Id\}}(\Lambda;s,\ens{\upsilon},\povm{N})}{\max_{\Psi\in\f{C}^\A}S_\abb{OCC}^{\{\Id\}}(\Psi;s,\ens{\upsilon},\povm{N})} \label{aeq:robustness-OCC-4}\\
	&\quad=\max_{\scriptsize\left\{\begin{array}{c}
		s\in[-1,1]^{\idx{N}\times\idx{N}}\colon \\
		\bar{S}_\abb{OCC}^\s{K}(s,\ens{\upsilon},\povm{N})\geq0
	\end{array}\right\}}\frac{S_\abb{OCC}^{\{\Id\}}(\Lambda;s,\ens{\upsilon},\povm{N})}{S_\abb{OCC}^\f{C}(s,\ens{\upsilon},\povm{N})}, \label{aeq:robustness-OCC-5}
\end{align}
which recovers Eq.~\eqref{eq:robustness-OCC-1}.  Here Eq.~\eqref{aeq:robustness-OCC-3} is due to the fact that the objective ratio is invariant under overall scaling of $s$; Eq.~\eqref{aeq:robustness-OCC-4} follows from Eq.~\eqref{aeq:robustness-OCC-1}; Eq.~\eqref{aeq:robustness-OCC-5} follows from Eq.~\eqref{eq:OCC-Id-free-score}.  Likewise, if $\s{K}$ is contractive under $\f{S}$, by Theorem~\ref{athm:robustness},
\begin{align}
	1+R^{\s{K},\f{C}}(\Lambda)&=\max_{\Xi\in\cone^*(\s{K}^\A)}\frac{\max_{\sm{\Theta}\in\f{S}^{\A\To\A}}\left\langle\Xi,\Lambda\right\rangle}{\max_{\Psi\in\f{C}^\A}\left\langle\Xi,\Psi\right\rangle} \\
	&=\max_{\scriptsize\left\{\begin{array}{c}
		s\in\spa{R}^{\idx{N}\times\idx{N}}\colon \\
		\Xi_s\in\cone^*(\s{K}^\A)
	\end{array}\right\}}\frac{\max_{\sm{\Theta}\in\f{S}^{\A\To\A}}\left\langle\Xi_s,\Lambda\right\rangle}{\max_{\Psi\in\f{C}^\A}\left\langle\Xi_s,\Psi\right\rangle} \\
	&=\max_{\scriptsize\left\{\begin{array}{c}
		s\in[-1,1]^{\idx{N}\times\idx{N}}\colon \\
		\Xi_s\in\cone^*(\s{K}^\A)
	\end{array}\right\}}\frac{\max_{\sm{\Theta}\in\f{S}^{\A\To\A}}\left\langle\Xi_s,\Lambda\right\rangle}{\max_{\Psi\in\f{C}^\A}\left\langle\Xi_s,\Psi\right\rangle} \\
	&=\max_{\scriptsize\left\{\begin{array}{c}
		s\in[-1,1]^{\idx{N}\times\idx{N}}\colon \\
		\bar{S}_\abb{OCC}^\s{K}(s,\ens{\upsilon},\povm{N})\geq0
	\end{array}\right\}}\frac{S_\abb{OCC}^\f{S}(\Lambda;s,\ens{\upsilon},\povm{N})}{S_\abb{OCC}^\f{C}(s,\ens{\upsilon},\povm{N})},
\end{align}
which recovers Eq.~\eqref{eq:robustness-OCC-2}.



\bibliographystyle{IEEEtran}
\bibliography{references}

\end{document}